\documentclass[ALICE,manyauthors]{cernphprep}
\usepackage[english]{babel}
\usepackage{graphicx}
\usepackage{verbatim}
\usepackage{amsfonts}
\usepackage{makeidx}
\usepackage{color}
\usepackage{amsmath}
\usepackage{lineno}
\usepackage{epsfig}
\usepackage{epstopdf}
\usepackage{hyperref}
\usepackage{cite}
\usepackage{multirow}
\usepackage{subfigure}

\usepackage[comma,square,numbers,sort&compress]{natbib}

\newcommand{\sqrts}{\sqrt{s}}
\newcommand{\sqrtsNN}{\sqrt{s_{\scriptscriptstyle \rm NN}}}

\newcommand{\av}[1]{\left\langle #1 \right\rangle}

\newcommand{\GeV}{\mathrm{GeV}}
\newcommand{\TeV}{\mathrm{TeV}}

\newcommand{\gev}{\mathrm{GeV}}
\newcommand{\gevc}{\mathrm{GeV}/c}
\newcommand{\tev}{\mathrm{TeV}}

\newcommand{\mum}{\mathrm{\mu m}}

\newcommand{\mub}{\mathrm{\mu b}}

\newcommand{\PbPb}{\mbox{Pb--Pb}}
\newcommand{\pPb}{\mbox{p--Pb}}
\newcommand{\AuAu}{\mbox{Au--Au}}

\newcommand{\pt}{p_{\rm T}}

\newcommand{\DtoKpi}{{\rm D}^0 \to {\rm K}^-\pi^+}
\newcommand{\DtoKpipi}{{\rm D}^+\to {\rm K}^-\pi^+\pi^+}
\newcommand{\DstartoDpi}{{\rm D}^{*+} \to {\rm D}^0 \pi^+}

\newcommand{\Dzero}{{\rm D^0}}

\newcommand{\Dstar}{{\rm D^{*+}}}

\newcommand{\Dplus}{{\rm D^+}}

\renewcommand{\d}{\mathrm{d}}

\renewcommand{\PbPb}{\mbox{Pb--Pb}}

\newcommand{\Raa}{R_{\rm AA}}
\newcommand{\RpPb}{R_{\rm pPb}}
\newcommand{\RAA}{R_{\rm AA}}
\newcommand{\TAA}{T_{\rm AA}}

\begin{document}

\begin{titlepage}

\PHyear{2015}
\PHnumber{252}                 
\PHdate{14 September}              

\title{Transverse momentum dependence of D-meson production\\ in $\PbPb$ collisions at 
$\mathbf{\sqrt{{\textit s}_ {NN}}}=$2.76~TeV}
\Collaboration{ALICE Collaboration%
         \thanks{See Appendix~\ref{app:collab} for the list of collaboration 
                      members}}
                     
\ShortTitle{Transverse momentum dependence of D-meson production in $\PbPb$ collisions}

\begin{abstract} 
The production of prompt charmed mesons $\Dzero$, $\Dplus$ and $\Dstar$, and their antiparticles, was measured with the ALICE detector in $\PbPb$ collisions at the centre-of-mass energy per nucleon pair, $\sqrtsNN$, of~$2.76~\TeV$. The production yields for rapidity $|y|<0.5$ are presented as a function of transverse momentum, $\pt$, in the interval 1--36~$\gevc$ for the centrality class 0--10\% and in the interval 1--16~GeV/$c$ for the centrality class 30--50\%. The nuclear modification factor $\Raa$ was computed using a proton--proton reference at $\sqrts = 2.76~\TeV$, based on measurements at $\sqrts = 7~\TeV$ and on theoretical calculations.  A maximum suppression by a factor of 5--6 with respect to binary-scaled pp yields is observed for the most central collisions at $\pt$ of about $10~\gevc$. A suppression by a factor of about 2--3 persists 
at the highest $\pt$ covered by the measurements. At low $\pt$ (1--3~GeV/$c$), the $\Raa$ has large uncertainties that span the range 0.35 (factor of about 3 suppression) to 1 (no suppression).
In all $\pt$ intervals, the $\Raa$ is larger in 
the 30--50\% centrality class compared to central collisions.
The D-meson $\Raa$ is also compared with that of charged pions and, at large $\pt$, charged hadrons, and with model calculations.
\end{abstract}

\end{titlepage}
\setcounter{page}{2}

\newpage 

\section{Introduction}
\label{sec:intro}
A state of strongly-interacting matter characterised by high energy density and temperature is predicted to be formed in ultra-relativistic collisions of heavy nuclei. According to calculations using Quantum Chromodynamics (QCD) on the lattice, these extreme conditions lead to the formation of a Quark--Gluon Plasma (QGP) state, in which quarks and gluons are deconfined, 
and chiral symmetry is partially restored (see e.g.~\cite{Karsch:2006xs,Borsanyi:2010bp,Borsanyi:2013bia,Bazavov:2011nk}).

Heavy quarks are produced in the hard scattering processes that occur in the early stage of the collision between partons of the incoming nuclei. Their production is characterised by a timescale $\Delta t < 1/(2\,m_{\rm c,b})$, $\sim 0.1$~fm/$c$ for charm and $\sim 0.01$~fm/$c$ for beauty quarks, that is shorter than the formation time of the QGP medium, about $0.3$~fm/$c$ at Large Hadron Collider (LHC) energies~\cite{Liu:2012ax}.
They can successively interact with the constituents of the medium and lose part of their energy,
via inelastic processes (gluon radiation)~\cite{Gyulassy:1990ye,Baier:1996sk} or elastic scatterings (collisional processes)~\cite{Thoma:1990fm,Braaten:1991jj,Braaten:1991we}. 
Energy loss can be studied using the nuclear modification factor $\Raa$, which compares
 the transverse-momentum ($\pt$) differential production yields in nucleus--nucleus collisions (${\rm d} N_{\rm AA}/{\rm d}\pt$)
with the cross section in proton--proton collisions (${\rm d}\sigma_{\rm pp}/{\rm d}\pt$) scaled by the average nuclear overlap function
($\av{T_{\rm AA}}$)
\begin{equation}
\label{eq:Raa}
R_{\rm AA}(\pt)=
{1\over \av{T_{\rm AA}}} \cdot 
{{\rm d} N_{\rm AA}/{\rm d}\pt \over 
{\rm d}\sigma_{\rm pp}/{\rm d}\pt}\,.
\end{equation}
The average nuclear overlap function $\av{T_{\rm AA}}$ over a centrality class is proportional to the number of binary nucleon-nucleon collisions per A--A collision in that class and it can be estimated via Glauber model calculations~\cite{Glauber:1970jm,Miller:2007ri}. 

According to QCD calculations, quarks are expected to lose less energy than gluons because their coupling to the medium is smaller~\cite{Gyulassy:1990ye,Baier:1996sk}. In the energy regime of the LHC, light-flavour hadrons with $\pt$ ranging from 5 to $20~\gev/c$ originate predominantly from gluons produced in hard scattering processes, while for larger $\pt$ they originate mainly from light quarks  (see e.g.~\cite{Djordjevic:2013pba}). Charmed mesons, instead, provide an experimental tag for a quark parent at all momenta. Therefore, the comparison of the heavy-flavour hadron $\RAA$ with that of pions is expected to be sensitive to the colour-charge dependence of energy loss. 
However, other aspects than the energy loss, like the parton $\pt$ spectrum and fragmentation into hadrons, influence the nuclear modification factor (see e.g.\,\cite{Armesto:2005iq,Djordjevic:2013pba}).  The effect of the colour-charge dependence of the energy loss should be then studied via the comparison with model calculations, that include the description of the aforementioned aspects.

In addition, several mass-dependent effects are predicted to influence the energy loss for quarks (see~\cite{Andronic:2015wma}
for a recent review).
The dead-cone effect should reduce small-angle gluon radiation for quarks that have moderate energy-over-mass values, i.e.\,for c and b quarks 
with momenta up to about 10 and $30~\gev/c$, respectively~\cite{Dokshitzer:2001zm,Armesto:2003jh,Djordjevic:2003zk,Zhang:2003wk,Wicks:2005gt,Horowitz:2011gd,Horowitz:2011wm}.
Likewise, collisional energy loss is predicted to be reduced for heavier quarks, because the spatial diffusion coefficient, which regulates
the momentum transfers with the medium, scales with the inverse of the quark mass for a given quark momentum~\cite{vanHees:2005wb}.
In particular, the study of D mesons from low-$\pt$ to high-$\pt$ allows to study the variation of the energy loss for different charm quark velocity: from a non-relativistic regime to an highly relativistic one.
Low-momentum heavy quarks, including those shifted to low momentum  
by parton energy loss, could participate in the collective expansion of the system as a consequence of multiple interactions~\cite{Batsouli:2002qf,Greco:2003vf}. 
It was also suggested that low-momentum heavy quarks could hadronise
not only via fragmentation in the vacuum, but also via the mechanism of recombination with other quarks from the 
medium~\cite{Greco:2003vf,Andronic:2003zv}.

The nuclear modification factor of heavy-flavour production was first studied at the 
Relativistic Heavy Ion Collider (RHIC). 
The PHENIX and STAR Collaborations reported measurements using heavy-flavour decay electrons and muons in Au--Au and Cu--Cu collisions at centre-of-mass energy per nucleon pair, $\sqrtsNN=200$~GeV~\cite{Adler:2005xv,Adare:2010de,Adare:2012px,Abelev:2006db}.
A suppression with respect to binary scaling 
was observed for $\pt$ larger than about 3~GeV/$c$, 
reaching a minimum $\Raa$ of about 0.2--0.3 in the interval $4<\pt<6$~GeV/$c$. 
The STAR Collaboration recently measured the $\RAA$ of $\rm D^0$ mesons 
in Au--Au collisions for the interval $0<\pt<6$~GeV/$c$~\cite{Adamczyk:2014uip}. At $\pt$ of about 5--6~GeV/$c$ the 
$\RAA$ value is similar to that observed for electrons from heavy-flavour decays and the
$\RAA$ increases towards low $\pt$, reaching a maximum value of about 1.5 at 1--2~GeV/$c$.
This feature is described by heavy-flavour transport calculations that include radial flow and a contribution due to recombination 
in the charm hadronisation process~\cite{Adamczyk:2014uip}.

A first measurement of the production of prompt D mesons at mid-rapidity in the $\pt$ interval 2--16~$\gev/c$ was published, using the
$\PbPb$ data at $\sqrtsNN=2.76~\TeV$ collected in 2010 during LHC Run 1~\cite{ALICE:2012ab}. 
A minimum $\Raa$ of about 0.2--0.3 was measured at $\pt$ of about 10~GeV/$c$ for the 20\% most central collisions.
The measurement of D-meson production in p--Pb collisions at $\sqrtsNN=5.02$~TeV, showing an $\RpPb$ compatible with unity, has provided clear 
evidence that the suppression with respect to 
binary-scaled pp cross sections, observed in Pb--Pb collisions, 
cannot be attributed to cold nuclear matter effects for $\pt$ larger than 2~GeV/$c$ and 
is, thus, caused by final-state interactions in the hot and dense medium~\cite{Abelev:2014hha}. 

In $\PbPb$ collisions, the nuclear modification factor at low $\pt$ results from the interplay of different effects occurring in the initial and in the final state. The measured D-meson nuclear modification factor in p--Pb collisions, although consistent with unity, is also described within uncertainties by calculations that include substantial initial-state effects, such as parton shadowing or saturation~\cite{Abelev:2014hha}, 
that could manifest as a reduction of the yields in Pb--Pb (and thus of the $\RAA$) by up to 50\% at low $\pt$.
In addition,  the measurement of
a significant azimuthal anisotropy of D-meson production, with respect to the estimated direction of the reaction plane in 
non-central Pb--Pb collisions, indicates that charm quarks participate in the collective expansion of the medium~\cite{Abelev:2013lca,Abelev:2014ipa}. 
Therefore, radial flow could play a relevant role as well.
In order to investigate these aspects, it is important to have a precise measurement of $\RAA$ down to low $\pt$.
In the high-$\pt$ region, where parton energy loss is expected to be dominated by radiative processes, 
the extension of the D-meson $\RAA$ beyond 20~$\gev/c$ would provide the first measurement of identified-hadron $\RAA$ at such high $\pt$. 

In this article we present the measurement of $\pt$-differential yields and nuclear modification factors of prompt $\Dzero$, $\Dplus$ and $\Dstar$ mesons (including their antiparticles), reconstructed via their hadronic decays in $\PbPb$ collisions at $\sqrtsNN =2.76~\TeV$, using the data sample recorded in 2011. For central collisions, the integrated luminosity is larger by a factor of about 10 than that used for the previously published results~\cite{ALICE:2012ab}.
This allows us to extend the measurement of $\RAA$ to lower and higher $\pt$
(from 2--16~$\gev/c$ to 1--36~$\gev/c$), to improve its precision, and to perform the
study in a narrower class of the most central collisions (10\% most central instead of 20\% most central).

The article is organised as follows:
the experimental apparatus is described in Section~\ref{sec:detector}, together with the data sample. In Section~\ref{sec:Dmesons}, the D-meson decay reconstruction and all corrections applied to the yields are presented, along with the procedure used to obtain the pp reference at $\sqrt s=2.76~\tev$. 
In Section~\ref{sec:Syst} the systematic uncertainties are discussed. 
The results for the 0--10\% (central) and 30--50\% (semi-peripheral) centrality classes are presented in Section~\ref{sec:DRAA}.  In the same Section results obtained in $\PbPb$ collisions are compared with the nuclear modification factor measured in $\pPb$ collisions at $\sqrtsNN=5.02~\tev$.
A comparison with charged pions, charged particles ($ch$) and with theoretical model predictions is also reported. These comparisons are presented in terms of the ratio $\RAA^{\rm D}/\RAA^{\pi,\,ch}$ as well.
Conclusions are drawn in Section~\ref{sec:conclusions}.

\section{Experimental apparatus and data sample}
\label{sec:detector}
The ALICE experimental apparatus~\cite{Aamodt:2008zz} is composed of various detectors for particle reconstruction and identification at mid-rapidity ($|\eta|<0.9$), a forward muon spectrometer ($-4<\eta<-2.5$) and a set of forward-backward detectors for triggering and event characterization. The detector performance for measurements in
pp, p--Pb and Pb--Pb collisions from the LHC Run~1 is presented in~\cite{Abelev:2014ffa}.

The main detector components used in this analysis are the V0 detector, the Inner Tracking System (ITS), the Time Projection Chamber (TPC) and the Time Of Flight (TOF) detector, which are located inside a large solenoidal magnet providing a uniform magnetic field of 0.5 T parallel to the LHC beam direction ($z$ axis in the ALICE reference system) and the Zero Degree Calorimeter (ZDC), located at $\pm 114$~m from the interaction point. 

Pb--Pb collision data were recorded with a minimum-bias interaction trigger based on information from the V0 detector, which
consists of two scintillator arrays covering the full azimuth in the pseudorapidity intervals $-3.7< \eta <-1.7$
and $2.8< \eta <5.1$~\cite{Abbas:2013taa}. The trigger logic required the coincidence of signals on both sides of the detector.
An online selection based on the V0 signal amplitudes was used to enhance the sample of central and mid-central collisions through two separate trigger classes.
The scintillator arrays have an intrinsic time resolution better than 0.5 ns, and their timing information was used 
together with that from the ZDCs for offline rejection of events produced by the interaction of the beams with residual gas in the vacuum pipe. 
Only events with a reconstructed interaction point (primary vertex) within $\pm10$~cm from the centre of the detector along the beam line were used in the analysis.

Collisions were divided into centrality classes, determined from the sum of the V0 signal amplitudes and defined in terms of percentiles of the total hadronic Pb--Pb cross section. In order to relate the centrality classes to the collision geometry, the distribution of the V0 summed amplitudes was fitted with a function based on the Glauber model~\cite{Glauber:1970jm,Miller:2007ri} combined with a two-component model for particle production~\cite{Abelev:2013qoq}. 
The centrality classes used in the analysis are reported in Tab.~\ref{tab:events}, together with the 
average of the nuclear overlap function $T_{\rm AA}$, the 
number of events in each class ($N_{\rm events}$) and the integrated luminosity. 

\begin{table}[!b]
 \begin{center}
  \begin{tabular}{|cccc|}
\hline
Centrality class & $\av{\TAA}$ (mb$^{-1}$) & $N_{\rm events}$ & $L_{\rm int}$ ($\rm \mub^{-1})$ \\
\hline
$\phantom{0}0$--10$\%$  & $23.44\pm 0.76$ & $16.4 \times 10^{6}$& $21.3 \pm 0.7$ \\
30--50$\%$ & $\phantom{0}3.87\pm 0.18$ & $\phantom{0}9.0 \times 10^{6} $& $\phantom{0}5.8 \pm 0.2$ \\
\hline
  \end{tabular}
 \caption{Average of the nuclear overlap function, number of events and integrated luminosity for the two centrality classes used in the analysis. The uncertainty on the integrated luminosity stems from the uncertainty of the hadronic Pb--Pb cross section from the Glauber model~\cite{Abelev:2013qoq}.}
 \label{tab:events}
 \end{center}
\end{table}

The charged-particle tracks used to reconstruct the decay of D mesons were measured in the TPC and ITS.
The tracking algorithm, based on a Kalman filter~\cite{Fruhwirth:1987fm}, starts from three-dimensional space points in the TPC, 
a large cylindrical drift detector with both total length and diameter of about 5~m, covering the pseudorapidity range $\rm |\eta| <$ 0.9 with full azimuthal acceptance~\cite{Alme2010316}.
Tracks are reconstructed in the TPC with up to 159 space points and with a measurement of the specific ionisation energy loss d$E$/d$x$ with a resolution of about 6\%.

Hits in the ITS are associated to the prolongation of the TPC tracks, forming the global tracks.
The ITS consists of six cylindrical layers of silicon detectors~\cite{Aamodt:2010aa}. 
The two innermost layers, placed at 3.9 and 7.6 cm from the beam line, 
consist of Silicon Pixel Detectors (SPD).
The third and fourth layers use Silicon Drift Detectors (SDD) and the two outermost layers contain double-sided Silicon Strip Detectors (SSD).
The effective spatial resolutions, including the intrinsic detector resolutions and residual mis-alignments, are about 14, 40 and 25~$\mum$, for SPD, SDD and SSD, respectively, along the most precise direction ($r\varphi$)~\cite{Aamodt:2010aa}.

Global tracks are used to reconstruct the primary interaction vertex and the secondary vertices of D-meson decays. 
The transverse momentum resolution for global tracks ranges from about 1\% at $\pt = 1~\gev/c$ 
to about 2\% at $10~\gev/c$, both in pp and Pb--Pb collisions. 
The spatial precision of global tracks is quantified by the resolution on the impact parameter $d_0$, 
which is the signed distance of closest approach between the track and the primary vertex in the $xy$-plane transverse to the beam direction.
In Pb--Pb collisions, the $d_0$ resolution is better than $65~\mum$ for tracks with a transverse momentum larger than $1~\gev/c$ and reaches $20~\mum$ 
for $\pt>20~\gev/c$~\cite{Abelev:2014ffa}. 

The TOF detector is an array of Multi-Gap Resistive Plate Chambers positioned at a distance of about 370~cm from the beam line and covering the full azimuth over the pseudorapidity interval $|\eta| < 0.9$.
TOF particle identification is based on the difference between the particle arrival time at the TOF detector and a start time determined using the arrival time of all particles of the event with a $\chi^2$ minimization~\cite{Akindinov:2013tea}.
The resolution ($\sigma$) of the time-of-flight measurement is about 80~ps for pions at $\pt = 1~\gev/c$ in the Pb--Pb collision centrality intervals used in this analysis. TOF provides charged-particle identification in the intermediate momentum range, with a 3$\sigma$ separation up to about $2.5~\gev/c$ for pions and kaons, and up to about $4~\gev/c$ for kaons and protons~\cite{Abelev:2014ffa}.

\section{Data analysis}
\label{sec:Dmesons}
\subsection{D-meson reconstruction}

$\Dzero$, $\Dplus$ and $\Dstar$ mesons, and their antiparticles,
were reconstructed via their hadronic decay channels $\DtoKpi$ (weak decay with
branching ratio, BR, of $3.88 \pm 0.05\%$),
$\DtoKpipi$ (weak decay, BR of $9.13\pm0.19\%$) and $\DstartoDpi$ (strong decay, BR of $67.7\pm0.05\%$) 
followed by $\DtoKpi$~\cite{Agashe:2014kda}. 
 $\Dzero$ and $\Dplus$ mesons have mean proper decay lengths ($c\tau$)
 of 123 and 312~$\mum$, respectively~\cite{Agashe:2014kda}.
 In the case of the $\Dstar$, the decay topology of the produced $\Dzero$ was exploited.  
The transverse momentum of the soft pions produced in the $\Dstar$
decays typically ranges from 0.1 to 1.5 $\gev/c$,
depending on the $\Dstar$ $\pt$.

$\Dzero$ and $\Dplus$ candidates were formed using pairs and triplets
of tracks with the correct charge-sign combination, requiring $|\eta|<0.8$,
$\pt>0.4~\gev/c$, at least 70 associated space points (out of a
maximum of 159) and fit quality $\chi^2/{\rm ndf} < 2$ in the TPC, and at least
two hits (out of six) in the ITS, out of which at least one in either of the two SPD layers.
$\Dstar$ candidates were formed by combining $\Dzero$ candidates with
tracks with $\pt>0.1~\gev/c$ and at least three hits in the ITS, out of which at least one in the SPD.

The aforementioned track selection limits the D-meson acceptance in rapidity. 
The acceptance drops steeply to zero for $|y|> 0.5$ at low $\pt$ and
$|y|> 0.8$ for $\pt> 5~\gev/c$.
A $\pt$-dependent fiducial acceptance cut, $|y_{\rm D}|<y_{\rm fid}(\pt)$, was therefore applied to the D-meson rapidity.
The cut value, $y_{\rm fid}(\pt)$, increases from 0.5 to 0.8 in the range $0<\pt<5~\gev/c$ according to a second-order
polynomial function and with a constant value of 0.8 for $\pt>5~\gev/c$. 

The selection of the decay topology was based on the displacement of
the decay tracks from the interaction vertex (via their impact
parameter, $d_0$), the separation between the secondary and primary
vertices (decay length, $L$) and the pointing angle of the
reconstructed D-meson momentum to the primary vertex.
This pointing condition was applied via a selection on the angle $\theta_{\rm pointing}$ between the direction of the reconstructed momentum of the candidate and the straight line connecting the primary and secondary vertices.
The projections of the pointing angle and of the decay length onto the transverse plane ($\theta^{xy}_{\rm pointing}$ and $L^{xy}$) were also used. 
The selection requirements were tuned to provide a large statistical significance for the signal and to keep the selection efficiency as high as possible. 
The chosen selection values depend on the $\pt$ of the D meson and become tighter from peripheral to central collisions.
A detailed description of the selection criteria was reported in~\cite{ALICE:2012ab,Abelev:2014ipa}.

In order to further reduce the combinatorial background,  pions and kaons were identified using the TPC and TOF detectors.
A 3$\sigma$ compatibility cut was applied to the difference between the measured and expected signals (for pions and kaons) for 
the TPC d$E$/d$x$ and TOF time-of-flight.
Tracks that are not matched with a hit in the TOF detector were identified using only the TPC information. 
Particle identification (PID) was not applied to the pion track from the $\Dstar$ decay.
This PID selection provides a reduction of the background by a factor
of 2--3 at low $\pt$ with respect to the case without applying the selection, while having an efficiency of about $95\%$ for the signal.

\begin{figure}[!t]
\begin{center}
\includegraphics[width=1\columnwidth]{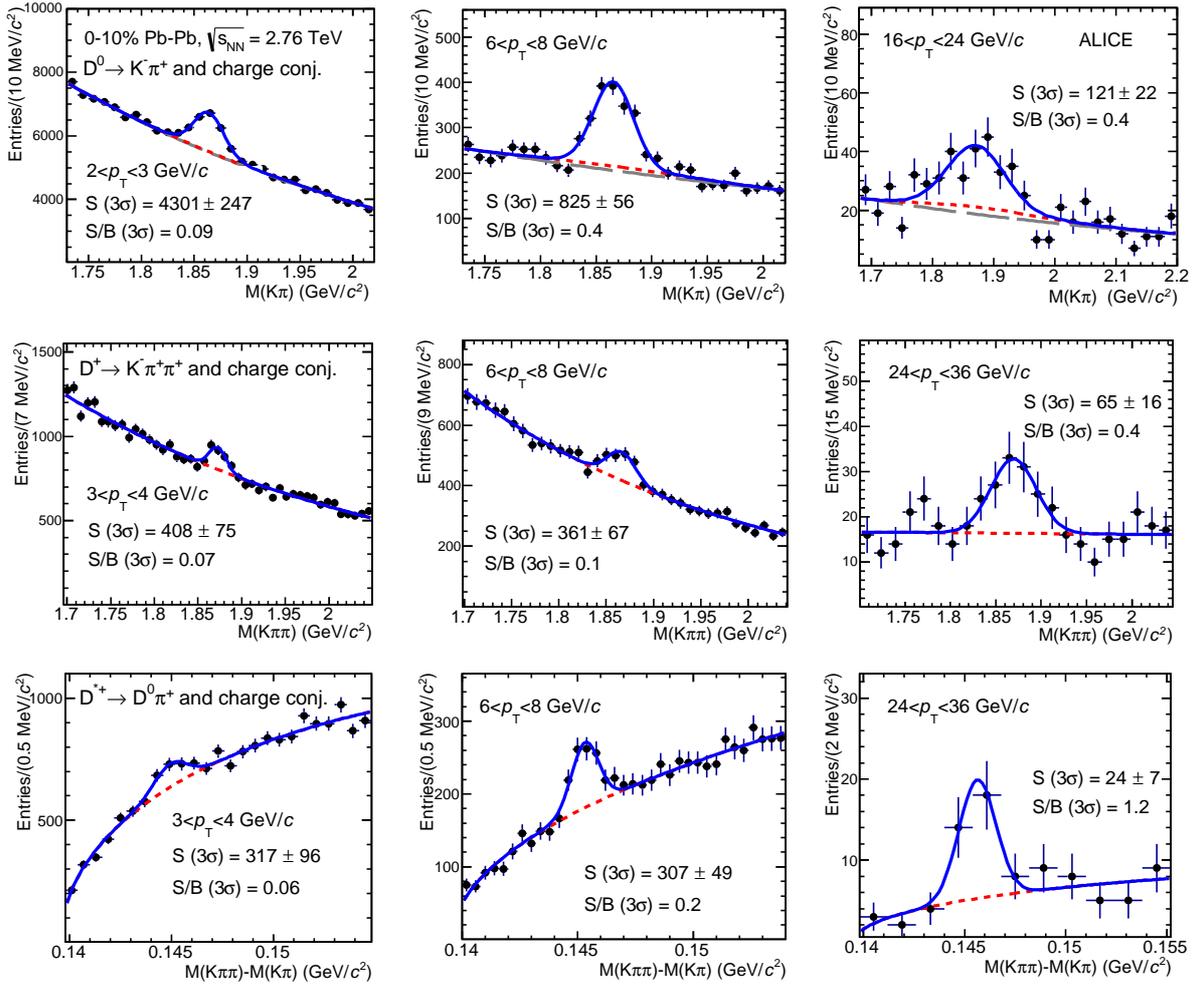}
\caption{ (K, $\pi$) (top row) and (K, $\pi$, $\pi$) (central row) invariant-mass distributions for the centrality class 0--10\%. 
Bottom row: Distribution of the mass difference $\Delta M=M({\rm
  K}\pi\pi)-M({\rm K}\pi)$ for the centrality class 0--10\%. The distributions are reported in three
$\pt$ intervals for each meson (left, middle and right column). The fit functions
used to describe the background (dash), the background without signal
reflections (only for $\Dzero$, long-dash) and the total distribution including the signal (solid) are shown.}
\label{fig:DInvMass}
\end{center}
\end{figure}

The raw D-meson yields were obtained from fits to the candidate invariant-mass distributions $M({\rm K}\pi)$
for $\Dzero$, $M({\rm K} \pi\pi)$ for $\Dplus$, and the mass difference $\Delta M = M({\rm K} \pi\pi) -M({\rm K} \pi)$ for $\Dstar$. 
The $\Dzero$ and $\Dplus$ candidate invariant-mass distributions were fitted with a function composed of a Gaussian for the signal and an 
exponential term to describe the background shape. In the 0--10\% centrality class, the background in the $M({\rm K}\pi)$ distribution 
for the interval $1<\pt<2~\gev/c$ could not be accounted for by an
exponential shape and was instead modelled with a fourth-order polynomial function. The $\Delta M$ 
distribution of $\Dstar$ candidates was fitted with a Gaussian function for the signal and a threshold 
function multiplied by an exponential for the background:
$a\,\sqrt{\Delta M - m_{\pi} } \cdot {\rm e}^{b (\Delta M-m_{\pi})}$. 

In the case of  $\Dzero$ mesons, an additional term was included in
the fit function to account for the background from  ``reflections", 
i.e.\,signal candidates that remain in the invariant-mass distribution when the 
${\rm (K,\pi)}$ mass hypotheses for the two decay tracks are swapped.
A study of simulations showed that about 70$\%$ of these reflections are rejected by the PID selection, 
while the residual contribution results in a broad invariant-mass
distribution, which can be described using a sum of two Gaussians. 
In order to account for the contribution of reflections in the data
(2--5\% at low $\pt$, about 10\% at high $\pt$), a
template consisting of two Gaussians was included in the fit. 
The centroids and widths, as well as the ratios of the integrals of these Gaussians to the signal integral, were fixed to the values obtained in the simulations (see also~\cite{Abelev:2014ipa}).

\begin{figure}[!tpb]
 \begin{center}
\includegraphics[angle=0, width=7.5cm]{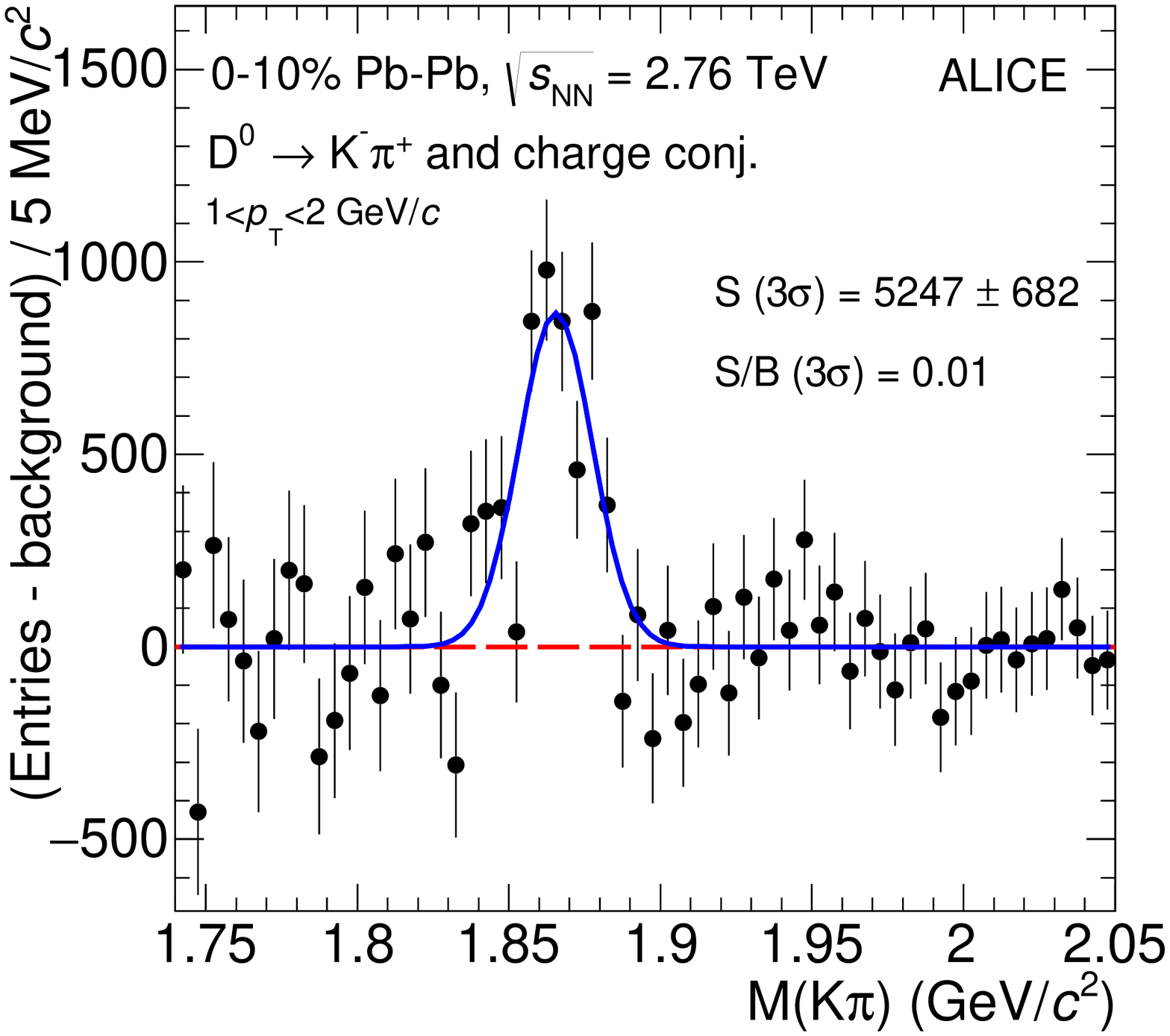}
 \includegraphics[angle=0, width=7.5cm]{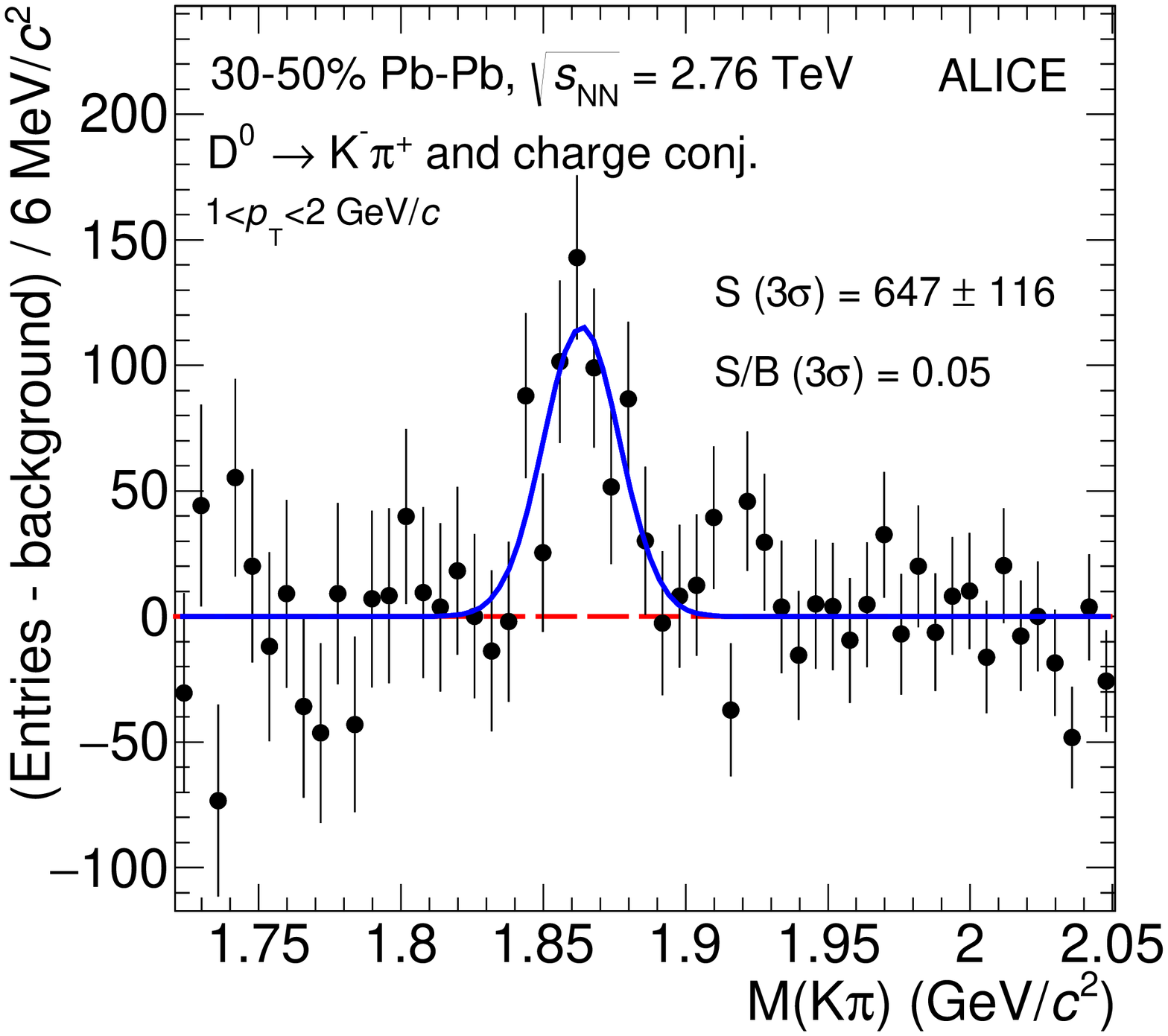}
 \end{center}
 \caption{(K, $\pi$) invariant-mass distribution for the interval
   $1<\pt<2~\gev/c$ for the 0--10\% (left) and 30--50\% (right) centrality
   classes, obtained after the subtraction of the background estimated
   by a fourth-order polynomial function for the most central
   collisions and an exponential for the 30--50\% centrality class. The
   contribution of reflections is also included in the fit. 
   The fit function used to describe the signal (solid line) is shown.}
 \label{fig:DInvMass1to2} 
\end{figure} 

In the centrality class 0--10\%, the signal extraction was 
performed in the interval $1<\pt<24~\gev/c$ for $\Dzero$ mesons,
divided in 9 $\pt$ bins, and in the interval $3<\pt<36~\gev/c$ for
$\Dplus$ and $\Dstar$ mesons, divided in 8 $\pt$ bins. In the centrality class 30--50\%, the signal extraction was 
possible in the interval $1<\pt<16~\gev/c$ for $\Dzero$ mesons and in the interval $2<\pt<16~\gev/c$ for $\Dplus$ and $\Dstar$
mesons. 
Beyond these intervals, the signal extraction was prevented by
the low signal-over-background ratio at low $\pt$, and by the low signal yield at high $\pt$. 
Figure~\ref{fig:DInvMass} shows the $\Dzero$ and $\Dplus$ invariant-mass
distributions and $\Dstar$ mass difference distributions in three
$\pt$ intervals for the centrality class 0--10\%. 
In the interval $16<\pt<24~\gev/c$ the fit range for the $\Dzero$ case is asymmetric. 
The range was limited to values larger than $1.68~\gev/c^{2}$ because 
the invariant-mass distribution of $({\rm K},\pi)$ pairs from D mesons decaying in three or more prongs 
produces a wide structure below about $1.72~\gev/c^{2}$, which cannot be accounted for 
by the background terms of the fit function.

Figure~\ref{fig:DInvMass1to2} shows the background-subtracted $\Dzero$ invariant-mass distribution for the
interval $1< \pt < 2~\gev/c$ for the 0--10\% (left
panel) and 30--50\% (right panel) centrality classes.

For all three D-meson species, the position of the 
signal peak was found to be compatible with the world average value and its $\pt$-dependent 
width with the values observed in the simulation. The statistical significance of the observed 
signals $\rm S/\sqrt{S+B}$ varies from 3 to 18, while the signal-over-background ratio S/B 
ranges from 0.01 to 1.8, depending on
the meson species, $\pt$ interval and centrality class.

\label{sec:EffppRef}
\subsection{${\rm d}N/{\rm d}\pt$  spectra corrections}
\label{Dcorr} 

The D-meson raw yields were corrected in order to obtain the
$\pt$-differential yields of prompt D mesons 

\begin{equation}
  \label{eq:dNdpt}
  \left.\frac{{\rm d} N^{\rm D}}{{\rm d}\pt}\right|_{|y|<0.5}=
  \frac{\left.f_{\rm prompt}(\pt)\cdot \frac{1}{2} N_{\rm raw}^{\rm
        D+\overline{D}}(\pt)\right|_{|y|<y_{\rm fid}}}{\Delta\pt \cdot
    \alpha_y \cdot ({\rm Acc}\times\epsilon)_{\rm prompt}(\pt)
    \cdot{\rm BR} \cdot N_{\rm events}}\,, 
\end{equation}

where prompt refers to mesons not coming from weak decays of B
hadrons. 
The raw yields $N_{\rm raw}^{\rm D+\overline{D}}$ were divided by a factor of two to obtain the charge-averaged 
 (particle and 
antiparticle)
yields.
To correct for the contribution of B-meson decay feed-down, the raw yields 
were  multiplied by the fraction of promptly produced D mesons, 
$f_{\rm prompt}$ (discussed in details later in this section). 
Furthermore, they were divided by the product of prompt 
D-meson acceptance and efficiency $({\rm Acc}\times\epsilon)_{\rm prompt}$, 
by the decay channel branching ratio ({\rm BR}), by the transverse momentum interval width ($\Delta \pt$)
and by the number of events ($N_{\rm events}$).
The factor $\alpha_y=y_{\rm fid}/0.5$ normalises the corrected yields
measured in $|y|<y_{\rm fid}$ to one unit of rapidity $|y|<0.5$,
assuming a uniform rapidity distribution for D mesons in the measured
range. This assumption was validated to the 1\% level with simulations~\cite{ALICE:2011aa, Skands:2009zm}.

The correction for acceptance and efficiency $({\rm
  Acc}\times\epsilon)_{\rm prompt}$ was determined using Monte Carlo simulations with a detailed description of the detector and its response, based on 
the GEANT3 transport package~\cite{Brun:1994aa}. 
The simulation was tuned to reproduce the (time-dependent) position and width of the interaction vertex distribution, as well as the number of active read-out channels and the accuracy of the detector calibration.
The underlying Pb--Pb events at $\sqrtsNN = 2.76$~TeV were simulated using the HIJING v1.383 generator~\cite{Wang:1991hta} 
and D-meson signals were added with the PYTHIA v6.421 generator~\cite{Sjostrand:2006za} with Perugia-0 tune~\cite{Skands:2010ak}.
Each simulated PYTHIA pp event contained a $\rm c\overline c$ or $\rm b\overline b$ pair, and D mesons 
were forced to decay in the hadronic channels of interest
for the analysis. 
Out of all the particles produced in these PYTHIA pp events, only the
heavy-flavour decay products were kept and transported through the
detector simulation together with the particles produced according to HIJING. In order to minimise the bias on the detector occupancy, the number of D mesons injected into each HIJING event was adjusted according to the Pb--Pb collision centrality.
In the most central event class, the $\pt$ distribution of D mesons
was weighted in order to match the shape measured for the \mbox{$\rm
  D^0$ meson}.
In the semi-peripheral centrality class, the D-meson $\pt$
distribution was weighted so as to match the shape given by
fixed-order-next-to-leading-log perturbative QCD calculations
(FONLL)~\cite{Cacciari:1998it,Cacciari:2001td} multiplied by the
$\RAA(\pt)$ computed using the BAMPS model~\cite{Uphoff:2011ad,Fochler:2011en,Uphoff:2012gb}.

\begin{figure}[!t]
\begin{center}
\includegraphics[width=0.95\columnwidth]{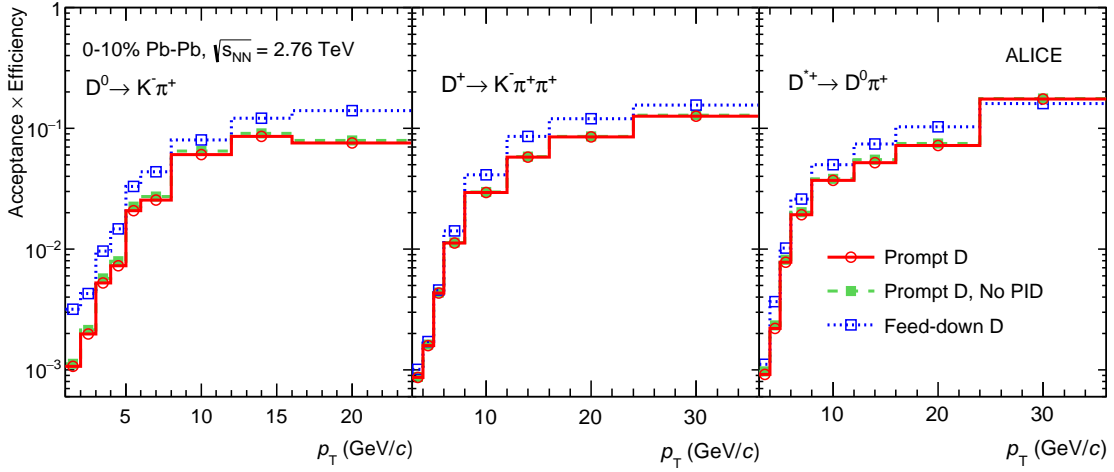}
\caption{Product of acceptance and efficiency for D mesons in Pb--Pb collisions for the 0--10\% centrality class. 
The rapidity interval is $|y| < y_{\rm fid}$ (see Section~3.1). The values for prompt (solid lines) and feed-down (dotted lines) D mesons are shown. Also displayed, for comparison, are the values for prompt D mesons without PID selection (dashed lines).}
\label{fig:Deff}
\end{center}
\end{figure}

The efficiencies were evaluated from simulated events that have the same average charged-particle multiplicity, 
corresponding to the same detector occupancy, as observed in data in the centrality classes 0--10\% and  30--50\%. 
Figure~\ref{fig:Deff} shows the $\Dzero$, $\Dplus$ and $\Dstar$
acceptance-times-efficiency $({\rm Acc}\times\varepsilon)$ for primary and feed-down D mesons
with rapidity $|y| < y_{\rm fid}(\pt)$ in the centrality class 0--10\%. 
The efficiencies range from about
0.1$\%$ at low $\pt$ to 10--30$\%$ at high $\pt$, because of the
momentum dependence of the D-meson decay length and of the topological
selections applied in the different momentum intervals.
Also shown in the figure are the $({\rm Acc}\times\varepsilon)$ values for the case where no PID is applied. 
The relative difference with respect to the $({\rm
  Acc}\times\varepsilon)$ obtained using the PID selection is
about 5\%, illustrating the high efficiency of the PID criteria. The
$({\rm Acc}\times\varepsilon)$ for D mesons from B-meson decays is larger than for prompt D mesons by a factor of about 1.5, because the decay vertices of the feed-down D mesons are more separated from the primary vertex and are, therefore, more efficiently selected by the analysis cuts.

The $f_{\rm prompt}$ factor was obtained, following the procedure introduced in~\cite{ALICE:2012ab}, as
\begin{equation}
  \label{eq:fcNbMethod}
\begin{split}
f_{\rm prompt} &= 1-\frac{ N^{\rm D+\overline D\,\textnormal{feed-down}}_{\rm raw}} { N^{\rm D+\overline D}_{\rm raw} }\\
 &= 1 -    \RAA^{\textnormal{feed-down}} \cdot    \langle \TAA \rangle 
	 \cdot \left( \frac{{\rm d} \sigma}{{\rm d}\pt }
         \right)^{\textnormal{FONLL,\,EvtGen}} _{{\textnormal{
               feed-down},\,|y|<0.5}} \cdot
         \frac{\Delta\pt\cdot\alpha_y\cdot({\rm
             Acc}\times\epsilon)_{\textnormal{feed-down}}\cdot {\rm
             BR} \cdot N_{\rm events}  }{ \frac{1}{2} N^{\rm D+\overline D}_{\rm raw}  } \, .
\end{split}
\end{equation}
In this expression, the symbols denoting the $\pt$ dependence have
  been omitted for brevity, $N^{\rm D+\overline D}_{\rm raw}$ is the measured raw yields
and $N^{{\rm D+\overline D}\,\textnormal{feed-down}}_{\rm raw}$ is the
estimated raw yields of $\rm D$ mesons from B-meson decays. In
detail, the B-meson production cross section in pp collisions at $\sqrt{s}=2.76~\tev$, estimated with FONLL calculations~\cite{Cacciari:2012ny}, was folded with 
the ${\rm B\rightarrow D}+X$ decay kinematics using the EvtGen package~\cite{Lange2001152} and multiplied by
$\langle \TAA \rangle$ in each centrality class, by the $({\rm
  Acc}\times\varepsilon)$ for feed-down $\rm D$ mesons, and by the other factors introduced in Eq.~(\ref{eq:dNdpt}).
In addition, 
the nuclear modification factor of D mesons from B-meson decays was accounted for.
The comparison of the $\RAA$ of prompt D
mesons ($\RAA^{\rm prompt}$)~\cite{Adam:2015nna} with that of $\rm
J/\psi$ from B-meson decays~\cite{CMS:2012vxa} measured in the CMS experiment indicates
that charmed hadrons are more suppressed than beauty hadrons. 
The value $\RAA^{\textnormal{feed-down}}=2\cdot\RAA^{\rm prompt}$ was
used to compute the correction, and the variation over the range
$1<\RAA^{\textnormal{feed-down}}/\RAA^{\rm prompt}<3$ was considered
for the evaluation of the systematic uncertainties, in order to take into account possible centrality and $\pt$ dependences.
Assuming $\RAA^{\textnormal{feed-down}}=2\cdot\RAA^{\rm prompt}$, the resulting $f_{\rm prompt}$
ranges from about $0.65$ to 0.85, depending on the D-meson
species and on the $\pt$ interval. 

\subsection{Proton--proton reference for $\RAA$}
\label{sec:ppref}

 The $\pt$-differential cross section of prompt D mesons with $|y|<0.5$ in pp collisions at $\sqrt s=2.76~\tev$,
  used as reference for the nuclear modification factor, was obtained as follows:
  \begin{itemize}
  \item in the interval $2<\pt<16\,(24)~\gev/c$ for $\Dzero$ ($\Dplus$ and $\Dstar$), the measurement at $\sqrt{s}=7~\TeV$~\cite{ALICE:2011aa} scaled to $\sqrts=2.76~\tev$ with FONLL calculations~\cite{Cacciari:2012ny} was used;
  \item in the interval $1<\pt<2~\gev/c$ for $\Dzero$,  an average of the aforementioned $\sqrts=7~\tev$ scaled measurement and of the measurement at $\sqrt s=2.76~\tev$~\cite{Abelev:2012vra} was used; 
  \item in the interval $16\,(24)<\pt<24\,(36)~\gev/c$ for $\Dzero$
    ($\Dplus$ and $\Dstar$), where their cross sections were not
    measured in pp collisions, the FONLL calculation at
    $\sqrt s=2.76~\tev$~\cite{Cacciari:2012ny} was
    used as a reference, after scaling it to match the central value
    of the data at lower $\pt$.
  \end{itemize}

The $\pt$-dependent scaling factor from $\sqrts=7~\tev$ to $\sqrts=2.76~\tev$ was determined
with FONLL calculations and its uncertainties were determined by varying
the parameters (charm-quark mass, factorisation and renormalisation scales) as described in~\cite{ConesadelValle:2011fw}.
The uncertainties on the scaling factor range from $^{+57}_{-11}\%$ for $1<\pt<2~\gev/c$ to about $\pm5\%$ for $\pt>10~\gev/c$.
The result of the scaling of the $\sqrts=7~\tev$ $\pt$-differential cross sections to $\sqrts=2.76~\tev$ was validated with measurements from a
smaller data sample in pp collisions at $\sqrt s= 2.76~\tev$~\cite{Abelev:2012vra}. These measurements cover a reduced $\pt$ interval with a statistical uncertainty of 
20--25\% and therefore they were not used as a pp reference for $\pt >2~\gev/c$.

For the lowest $\pt$ interval for the $\Dzero$ meson,
the two references (obtained from the
measurement in pp collisions at $\sqrts=2.76~\tev$ and from the
$\sqrts=7~\tev$ scaled measurement)
have comparable uncertainties. Therefore, in this
interval, the two values were averaged using the inverse of the
squared relative uncertainties as weights. 
In particular, the statistical uncertainties and the uncorrelated part
of the systematic uncertainties, i.e.\,the systematic uncertainty from
data analysis (yield extraction, efficiency corrections)  and the
scaling uncertainty, were used in the weight. 
The uncertainties on the feed-down subtraction were considered as
fully correlated among the two measurements, and were propagated linearly.

The cross section measurements for D mesons in pp collisions 
at $\sqrt s=7~\TeV$ are limited to $\pt \leq 16~\gev/c$ for $\Dzero$ and to
$\pt \leq 24~\gev/c$ for $\Dplus$ and $\Dstar$. 
Beyond these limits the pp reference was obtained using the cross section from the FONLL
calculation at $\sqrt s=
2.76~\tev$~\cite{Cacciari:2012ny}. 
Since the central value of the FONLL calculation underestimates the
measurement for $\pt>5~\gev/c$ at both
$\sqrts=2.76~\tev$ and
$\sqrts=7~\tev$~\cite{Abelev:2012vra,ALICE:2011aa}, the FONLL cross
section was multiplied by a scaling factor ($\kappa$)
\begin{equation}
\label{eq:highptextr}
\frac{{\rm d}\sigma}{{\rm d}\pt} = \kappa \cdot \Bigg(\frac{{\rm d}\sigma}{{\rm d}\pt}\Bigg)^{\textnormal {FONLL}}_{\sqrt s=2.76\,{\rm TeV},\,|y|<0.5}\,.
\end{equation}
 The factor $\kappa$ was determined by fitting with a constant the data-to-theory ratio at $\sqrts=7~\tev$ in the interval $5<\pt<16~\gev/c$. 
 Since the measurements at $\sqrts=2.76~\tev$ are less precise, they do not constrain further the scaling factor. 
 Depending on the D-meson species, the factor $\kappa$ ranges from 1.4
 to 1.5, for the central values of the 
 FONLL calculation parameters~\cite{Cacciari:2012ny}.
  The statistical uncertainty of the extrapolated cross section was
  determined by propagating the statistical uncertainties of the
  measurement in the determination of $\kappa$ and it amounts to about $5\%$. 
 The systematic uncertainties were evaluated under the conservative
 assumption that the systematic uncertainties of the measurement are fully
 correlated over $\pt$, i.e.\ by repeating the calculation of $\kappa$
 after shifting all data points consistently within their systematic uncertainties.
 In addition, the calculation in Eq.~(\ref{eq:highptextr}) was performed considering the FONLL cross sections obtained from
combinations of the renormalisation and factorisation scales with values $(0.5,1,2)\cdot \sqrt{m_{\rm c}^{2}+p_{\rm T,c}^{2}}$~\cite{Cacciari:2012ny}, as
well as the upper and lower limits of their envelope\footnote{Where
  $m_{\rm c}$ and $p_{\rm T,c}$ are respectively the mass and the transverse momentum
  of the charm quark considered in the calculations.}. 
This resulted in a total systematic uncertainty on the $\pt$-extrapolated cross section of about $^{+50}_{-35}\%$.


\section{Systematic uncertainties}
\label{sec:Syst}
\subsection{Systematic uncertainties on the D-meson $\pt$ spectra} 
\label{systPbPbSpectra}

\begin{table}[!t]
\centering
\renewcommand{\arraystretch}{1.3}
\begin{tabular}{|l|ccc|ccc|ccc|}
\hline 
\multicolumn{1}{|l|}{Particle} & \multicolumn{3}{c|}{$\Dzero$} 
 & \multicolumn{3}{c|}{$\Dplus$}& \multicolumn{3}{c|}{$\Dstar$} \\
\hline
\multicolumn{10}{|c|}{0--10\% centrality class} \\
\hline
\multicolumn{1}{|l|}{$\pt$ interval ($\gev/c$)} 
& 1--2 & 6--8 & 16--24 & 3--4 & 6--8 & 24--36 & 3--4 & 6--8 & 24--36\\
\cline{1-10} 
Yield extraction  & 15\%	& 5\%  &15\%
                      & 10\% & 8\%  & 8\%  
                      & 12\% & 5\%  & 10\% \\
 Tracking efficiency & 10\% & 10\% & 10\% 
                      & 15\% & 15\%  & 15\% 
                      & 15\% & 15\%  & 15\%  \\
 Selection cuts      &   15\% & 5\%  & 5\% 
                      &   10\% & 10\%  & 10\% 
                      &   10\% & 10\%  & 10\%  \\
 PID efficiency      &   \phantom{0}5\% & \phantom{0}5\% & \phantom{0}5\%
                      &   \phantom{0}5\% & \phantom{0}5\% &
                                                            \phantom{0}5\% 
                      &   \phantom{0}5\% & \phantom{0}5\%
                              & \phantom{0}5\% \\
 MC $\pt$ shape      & 15\% & \phantom{0}1\%  & \phantom{0}1\%
                      & \phantom{0}6\% & \phantom{0}1\% & \phantom{0}1\%
                      & \phantom{0}4\% & \phantom{0}1\% & \phantom{0}1\% \\
 FONLL feed-down corr.  & $_{-45}^{+\phantom{0}5}\%$ & $_{-13}^{+\phantom{0}8}\%$  & $_{-16}^{+10}\%$ 
                      & $_{-12}^{+\phantom{0}4}\%$ & $_{-11}^{+\phantom{0}6}\%$ & $_{-14}^{+\phantom{0}8}\%$ 
                      & $_{-12}^{+\phantom{0}3}\%$ &
                                                               $_{-\phantom{0}7}^{+\phantom{0}4}\%$  & $_{-\phantom{0}8}^{+\phantom{0}3}\%$  \\
 $R_{{\rm AA}}^{\rm feed-down} / R_{{\rm AA}}^{\rm prompt}$ (Eq.~(\ref{eq:fcNbMethod}))	
                      &  $^{+\phantom{0}5}_{-\phantom{0}5} \%$ & $^{+11}_{-10} \%$ & $^{+16}_{-13} \%$ 
                      &	  $^{+\phantom{0}6}_{-\phantom{0}5} \%$  &  $^{+\phantom{0}9}_{-\phantom{0}7} \%$ &  $^{+14}_{-11} \%$ 
                      &	  $^{+\phantom{0}4}_{-\phantom{0}4} \%$  &
                                                                   $^{+\phantom{0}6}_{-\phantom{0}6}\%$
                             &  $^{+\phantom{0}6}_{-\phantom{0}6}\%$\\

 BR         & \multicolumn{3}{c|}{ 1.3\%} & \multicolumn{3}{c|}{2.1\%}& \multicolumn{3}{c|}{ 1.5\%}\\
 Centrality class definition         & \multicolumn{3}{c|}{$<1\%$} & \multicolumn{3}{c|}{$<1\%$}& \multicolumn{3}{c|}{$<1\%$}\\
\hline
\multicolumn{10}{|c|}{30--50\% centrality class} \\
\hline
\multicolumn{1}{|l|}{$\pt$ interval ($\gev/c$)} 
& 1--2 & 6--8 & 12--16 & 2--3 & 6--8 & 12--16 & 2--3 & 6--8& 12--16\\
\cline{1-10}
 Yield extraction      & 10\%	& \phantom{0}8\% & \phantom{0}8\%
                      & 10\%	&  10\%  & 12\%
                      & 12\%    & \phantom{0}8\% & \phantom{0}5\%\\   

 Tracking efficiency & 10\% & 10\% &10\% 
                      & 15\% & 15\%  &15\% 
                      & 15\% & 15\%  &15\% \\
Selection cuts      &   10\% & 10\%  &15\% 
                      &   10\% & 10\%  & 15\%
                      &   15\% & 10\%  & 5\% \\
 PID efficiency      &  \phantom{0}5\%  & \phantom{0}5\% & \phantom{0}5\%
                      &   \phantom{0}5\% & \phantom{0}5\% & \phantom{0}5\%
                      &   \phantom{0}5\%  & \phantom{0}5\%
  & \phantom{0}5\%\\
 MC $\pt$ shape      &\phantom{0}5\% & \phantom{0}1\% & \phantom{0}3\%  
                      & 10\% & \phantom{0}2\%  & \phantom{0}2\%
                      & 10\% & \phantom{0}1\%   &
                                                            \phantom{0}1\%  \\
 FONLL feed-down corr.  & $_{-45}^{+\phantom{0}5}\%$ & $_{-12}^{+\phantom{0}7}\%$  & $_{-11}^{+\phantom{0}8}\%$ 
                      & $_{-21}^{+\phantom{0}6}\%$ & $_{-12}^{+\phantom{0}6}\%$ & $_{-13}^{+11}\%$ 
                      & $_{-19}^{+\phantom{0}3}\%$ &
                                                               $_{-\phantom{0}8}^{+\phantom{0}5}\%$  & $_{-\phantom{0}8}^{+\phantom{0}4}\%$  \\
 $R_{{\rm AA}}^{\rm feed-down} / R_{{\rm AA}}^{\rm prompt}$ (Eq.~(\ref{eq:fcNbMethod}))	
                      &  $^{+\phantom{0}6}_{-\phantom{0}5} \%$ & $^{+11}_{-\phantom{0}9} \%$ & $^{+14}_{-11} \%$ 
                      &	  $^{+\phantom{0}7}_{-\phantom{0}6} \%$  &  $^{+\phantom{0}9}_{-\phantom{0}8} \%$ &  $^{+16}_{-12} \%$ 
                      &	  $^{+\phantom{0}4}_{-\phantom{0}4} \%$  &  $^{+\phantom{0}7}_{-\phantom{0}6} \%$ &  $^{+\phantom{0}6}_{-\phantom{0}6} \%$\\
 BR         & \multicolumn{3}{c|}{1.3\%} & \multicolumn{3}{c|}{2.1\%}& \multicolumn{3}{c|}{1.5\%}\\
 Centrality class definition         & \multicolumn{3}{c|}{$2\%$} & \multicolumn{3}{c|}{2\%}& \multicolumn{3}{c|}{2\%}\\
\hline
\end{tabular}
\caption{Relative systematic uncertainties on the prompt D-meson
  production yields in Pb--Pb collisions for three selected $\pt$ 
intervals, in the two centrality classes.}
\label{tab:SystPbPbSpectra}
\end{table}

The systematic uncertainties were estimated as a
function of transverse momentum for the two centrality classes. 
Table~\ref{tab:SystPbPbSpectra} lists the uncertainties
for three $\pt$ intervals for each meson species.

The systematic uncertainty on the raw yield extraction was evaluated by repeating the fit of the invariant-mass distributions while varying the fit range; 
by fixing the mean and sigma of the Gaussian term to the world-average value and the expectations
from Monte Carlo simulations, respectively; and by using different fit
functions for the background. Specifically, first- and second-order polynomials
were used for $\Dzero$ and $\Dplus$, and a power law
multiplied by an exponential or a threshold function for $\Dstar$. A
method based on bin counting of the signal after background subtraction was also used.
This method does not assume any particular shape for the invariant-mass
distribution of the signal. The estimated uncertainties depend on the centrality class
and on the $\pt$ interval, ranging from 5\%  to 15\% for $\Dzero$, 8\%  to 10\% for $\Dplus$ and
 5\%  to 10\% for $\Dstar$, typically with larger values in the lowest and highest
$\pt$ intervals.

For $\Dzero$ mesons, the systematic uncertainty due to signal reflections 
in the invariant-mass distribution
was estimated by changing by $\pm50\%$ 
the ratio of the integral of the reflections over the integral of the signal (obtained from the simulation) used in the invariant-mass fit with the reflections template. 
In addition, the shape of the
template was varied using a polynomial parameterisation (of third or sixth order) 
of the simulated distribution, instead of a double-Gaussian parameterisation.
A test was carried out using, in the fit, a template histogram of the
reflections obtained directly from the simulation, rather than a functional form. 
The variation observed in the 
raw yields, ranging from 3\% to 7\% from low to high $\pt$, was added in quadrature as an
independent contribution to the yield extraction systematic uncertainty.

The systematic uncertainty on the tracking efficiency was estimated by comparing the probability to match the TPC tracks to the ITS hits in data and simulation, and by varying the track quality
selection criteria (for example, the minimum number of associated hits
in the TPC and in the ITS, and the maximum $\chi^2$/ndf of the momentum
fit). The efficiency of the track matching and the association of hits
in the silicon pixel layers was found to be well reproduced by the
simulation with maximal deviations on the level of 5\% in the $\pt$
range relevant for this analysis (0.5--25~$\gev/c$)~\cite{Abelev:2014ffa}. The effect of
mis-associating ITS hits to tracks was studied using simulations.
The mis-association probability is about 5\%, for central collisions, in the transverse
momentum interval $1 < \pt < 3~\gev/c$ and drops rapidly to zero at larger $\pt$.
It was verified that the signal selection efficiencies are the same for D mesons with and without wrong hit associations. 
The total systematic uncertainty on the track reconstruction procedure amounts to 5\% for single tracks, which results in a 10\% uncertainty for $\Dzero$ mesons (two-tracks decay) and 15\% for $\Dplus$ and $\Dstar$ mesons (three-tracks decay).

The uncertainty on the D-meson selection efficiency reflects a
possible non-exact description of the D-meson kinematic properties
and of the detector resolutions and alignments in the simulation. This effect was
estimated by repeating the analysis with different values of the selection cuts, significantly modifying the efficiencies, raw yield
and background values. As expected, larger deviations in the 
corrected yields were observed at low $\pt$, where the efficiencies are low and 
vary steeply with $\pt$, because of the tighter selections. Due to this,
the systematic uncertainties are slightly larger in these $\pt$
intervals. The assigned systematic uncertainty
varies from 5\% to 15\% for $\Dzero$, equals 10\%  for $\Dplus$, and varies from 10\% to 15\%  for $\Dstar$. 

A 5\% systematic uncertainty related to the PID selection was
evaluated by comparing the ratio of the corrected yields extracted with and without particle
identification.

The uncertainty on the efficiencies arising from the difference
between the real and simulated D-meson transverse momentum distributions depends on the width of the $\pt$ intervals and on the variation of the efficiencies within them. 
This uncertainty also includes the effect of the $\pt$ dependence of the nuclear modification factor. 
As explained in Section~\ref{Dcorr}, for the centrality 
class 0--10\%, the D-meson transverse momentum distribution from the PYTHIA simulation 
was re-weighted in order to reproduce the $\Dzero$ spectrum shape observed in data, while
for the 30--50\% centrality class, the weights were defined in order to match the $\pt$ distributions 
from FONLL calculations multiplied by the $\RAA$ from the BAMPS model.
A systematic uncertainty was estimated by using two alternative
D-meson $\pt$ distributions in both centrality classes:
i) FONLL $\pt$ distributions, ii) FONLL $\pt$ distributions multiplied 
by $\RAA$ from the BAMPS model. In addition, for the most central
events, a different parameterisation of the measured $\pt$ spectrum
was used. The resulting uncertainties decrease with increasing $\pt$,
varying from 5--6\% to 1\% in the interval $2<\pt< 36~\gev/c$.
For $\Dzero$ mesons, efficiencies increase by more than a factor five
within the interval $1<\pt<2~\gev/c$ in the most central
collisions. 
As a consequence, a larger uncertainty of 15\% 
resulted from a detailed study of the stability of the corrected
yields when changing the $\pt$ spectrum in the simulation.

\begin{figure}[!t]
 \begin{center}
 \includegraphics[angle=0, width=9cm]{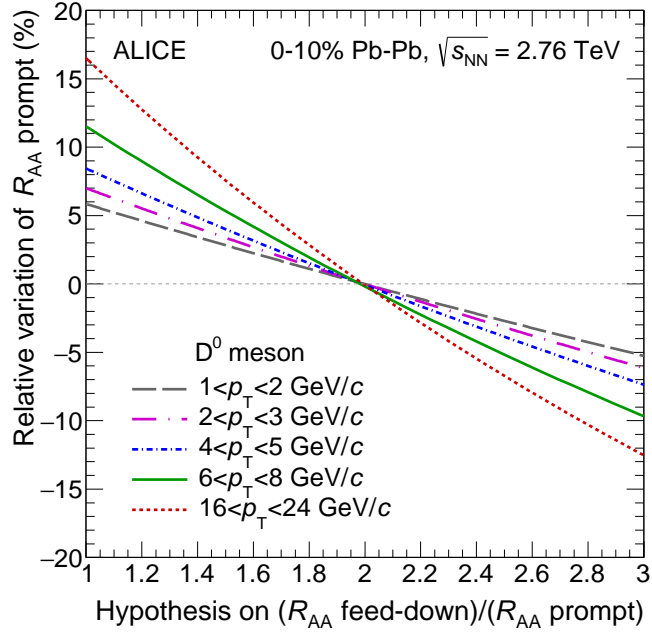}
 \end{center}
 \caption{Relative variation of the prompt $\Dzero$ yields as a
   function of the hypothesis on $\Raa^{\rm
  feed-down} / \Raa^{\rm prompt}$ for the B-meson feed-down subtraction.}
 \label{BEnegyLossHypo} 
\end{figure}

The systematic uncertainty on the subtraction of feed-down from B decays (i.e.\,the calculation of 
the $f_{\rm prompt}$ fraction) was estimated i)
by varying the $\pt$-differential feed-down D-meson cross section from the FONLL calculation within the
theoretical uncertainties, ii) by varying the hypothesis on the ratio
of the prompt and feed-down D-meson $\Raa$ in the range $1 < \RAA^{\rm
  feed-down}/\RAA^{\rm prompt} < 3$, and iii) by applying an
alternative method to compute $f_{\rm prompt}$. This second method is
based on the ratio of charm and beauty FONLL cross sections, instead
of the absolute beauty cross section.
The procedure is the same used for previous measurements of D-meson
production with ALICE~\cite{ALICE:2011aa,ALICE:2012ab,Abelev:2014ipa}.
The resulting uncertainty 
ranges between $_{-45}^{+\phantom{0}5}$\% at low $\pt$ and  $_{-8}^{+3}$\%
at high $\pt$ for the 0--10\% centrality class, and between
$_{-45}^{+\phantom{0}5}$\% at low $\pt$ and $_{-8}^{+4}$\%
at high $\pt$ for the 30--50\% centrality class. 
The uncertainty from the variation of the feed-down D-meson $\RAA$
hypothesis ranges from 6 to 16\%,  
as shown in Fig.~\ref{BEnegyLossHypo}, where the relative variation of
the prompt $\Dzero$ yield is shown as a function of the hypothesis on $\Raa^{\rm
  feed-down} / \Raa^{\rm prompt}$ for four $\pt$ intervals.

The uncertainties on the branching ratios were also considered~\cite{Agashe:2014kda} as well as the contribution due to the 1.1\% uncertainty on the fraction of the hadronic cross section used in the 
Glauber fit to determine the centrality
classes~\cite{Abelev:2013qoq}. The latter was estimated from the variation of the D-meson 
${\rm d}N/{\rm d}\pt$ when the limits of the centrality classes are shifted by $\pm 1.1\%$ 
(e.g. shifted from 30--50\% to 30.3--50.6\% and 29.7--49.5\%)~\cite{ALICE:2012ab}.
The resulting uncertainty, common to all $\pt$ intervals, is smaller than 1\% for the 0--10\% centrality class and about 
2\% for the 30--50\% centrality class. 

\subsection{Systematic uncertainties on $\Raa$}

\begin{table}[!t]
\centering
\renewcommand{\arraystretch}{1.3}
\begin{tabular}{|l|ccc|ccc|ccc|}
\hline 
\multicolumn{1}{|l|}{Particle} & \multicolumn{3}{c|}{$\Dzero$} 
 & \multicolumn{3}{c|}{$\Dplus$}& \multicolumn{3}{c|}{$\Dstar$} \\
\hline
\multicolumn{10}{|c|}{0--10\% centrality class} \\
\hline
\multicolumn{1}{|l|}{$\pt$ interval ($\gev/c$)} 
& 1--2 & 6--8 & 16--24 & 3--4 & 6--8 & 24--36 & 3--4 & 6--8 & 24--36\\
\cline{1-10} 
d$N_{\rm Pb-Pb}/$d$\pt $ (excl. feed-down)   & 28\% & 14\%  &22\%
                      & 22\% & 20\%  & 22\%  
                      & 24\% & 20\%  & 21\% \\
d$N_{\rm pp}/$d$\pt $ (excl. feed-down) & 21\%* & 16\% & 17\% 
                      & 20\% & 19\%  & 20\% 
                      & 17\% & 17\%  & 18\%  \\
 $\sqrt{s}-\rm{scaling}$ of the pp ref.   & $_{-30}^{+\phantom{0}6}\%$* & $_{-10}^{+\phantom{0}6}\%$  & -- 
                      &   $_{-19}^{+\phantom{0}8}\%$  & 
                                                       $_{-10}^{+\phantom{0}6}\% $  & --
                      &   $_{-20}^{+\phantom{0}9}\%$
                       &$_{-10}^{+\phantom{0}6}\% $ & --\\
High-$\pt$ extrapolation & -- & -- & $^{+34}_{-51}\%$ &
						 -- & -- & $^{+37}_{-56}\%$ &
						  -- & -- & $^{+34}_{-53}\%$ \\
 FONLL feed-down corr.   & $_{-\phantom{0}4}^{+\phantom{0}1}\%$ & $_{-\phantom{0}4}^{+\phantom{0}2}\%$ & $_{-16}^{+10}\%$
                          & $_{-\phantom{0}1}^{+\phantom{0}2}\%$ & $_{-\phantom{0}2}^{+\phantom{0}1}\%$  & $_{-14}^{+\phantom{0}8}\%$ 
                          & $_{-\phantom{0}4}^{+\phantom{0}1}\%$ & $_{-\phantom{0}4}^{+\phantom{0}2}\%$& $_{-\phantom{0}8}^{+\phantom{0}3}\%$  \\ 
 $R_{{\rm AA}}^{\rm feed-down} / R_{{\rm AA}}^{\rm prompt}$ (Eq.~(\ref{eq:fcNbMethod}))
      		      & $^{+12}_{-\phantom{0}9} \%$  & $^{+14}_{-11}
                                                       \%$ & $^{+19}_{-13} \%$
                      &	$^{+\phantom{0}8}_{-\phantom{0}7} \%$  & $^{+12}_{-\phantom{0}9}
                                                       \%$  & $^{+16}_{-12} \%$
                      &	$^{+\phantom{0}6}_{-\phantom{0}6} \%$  &
                                                      $^{+\phantom{0}8}_{-\phantom{0}7} \%$ & $^{+\phantom{0}8}_{-\phantom{0}7} \%$ \\
 Normalisation       & \multicolumn{3}{c|}{ 4.8\%} & \multicolumn{3}{c|}{4.8\%}& \multicolumn{3}{c|}{ 4.8\%}\\
\hline
\multicolumn{10}{|c|}{30--50\% centrality class} \\
\hline
\multicolumn{1}{|l|}{$\pt$ interval ($\gev/c$)} 
& 1--2 & 6--8 & 12--16 & 2--3 & 6--8 & 12--16 & 2--3 & 6--8& 12--16\\
\cline{1-10}
 d$N_{\rm Pb-Pb}/$d$\pt $ (excl. feed-down) & 20\% & 20\%  & 22\%
                      & 25\% & 21\%  & 22\%  
                      & 29\% & 19\%  & 18\% \\
d$N_{\rm pp}/$d$\pt $ (excl. feed-down) & 21\%* & 16\% & 17\% 
                      & 20\% & 19\%  & 20\% 
                      & 17\% & 17\%  & 18\%  \\
 $\sqrt{s}-\rm{scaling}$ of the pp ref.    & $_{-30}^{+\phantom{0}6}\%$* & $_{-10}^{+\phantom{0}6}\%$  & $_{-6}^{+5}\%$
                      &   $_{-19}^{+\phantom{0}8}\%$  &
                                                       $_{-10}^{+\phantom{0}6}\% $  & $_{-6}^{+5}\%$
                      &   $_{-20}^{+\phantom{0}9}\%$
                       &$_{-10}^{+\phantom{0}6}\% $ & $_{-6}^{+5}\%$ \\ 
 FONLL feed-down corr.   & $_{-\phantom{0}5}^{+\phantom{0}1}\%$ & $_{-\phantom{0}3}^{+\phantom{0}2}\%$ & $_{-\phantom{0}4}^{+\phantom{0}3}\%$
                          & $_{-\phantom{0}2}^{+\phantom{0}1}\%$ & $_{-\phantom{0}3}^{+\phantom{0}1}\%$  & $_{-\phantom{0}4}^{+\phantom{0}3}\%$ 
                          & $_{-\phantom{0}2}^{+\phantom{0}1}\%$ & $_{-\phantom{0}5}^{+\phantom{0}3}\%$& $_{-\phantom{0}3}^{+\phantom{0}2}\%$  \\ 
 $R_{{\rm AA}}^{\rm feed-down} / R_{{\rm AA}}^{\rm prompt}$ (Eq.~(\ref{eq:fcNbMethod}))
      		      & $^{+12}_{-\phantom{0}9} \%$  & $^{+14}_{-11}
                                                       \%$ & $^{+15}_{-11} \%$
                      &	$^{+\phantom{0}9}_{-\phantom{0}7} \%$  & $^{+13}_{-10}
                                                       \%$  & $^{+17}_{-13} \%$
                      &	$^{+\phantom{0}7}_{-\phantom{0}6} \%$  &
                                                      $^{+10}_{-\phantom{0}8} \%$ & $^{+\phantom{0}9}_{-\phantom{0}8}\%$ \\
 Normalisation    & \multicolumn{3}{c|}{6.2\%} & \multicolumn{3}{c|}{6.2\%}& \multicolumn{3}{c|}{6.2\%}\\
\hline
\end{tabular}
\caption{Relative systematic uncertainties on the prompt D-meson
  $\Raa$ for three $\pt$ intervals, in the two centrality
  classes. Uncertainties marked with a * were obtained as the average of the
  measurement at $\sqrts=2.76~\TeV$ and the measurement at
  $\sqrts=7~\TeV$, scaled using FONLL~\cite{Cacciari:2012ny}, as described in Section~\ref{sec:ppref}.
    } 
\label{tab:SystRaa}
\end{table}

The systematic uncertainties on the $\Raa$ measurement include those 
on the D-meson corrected yields, those on the
proton--proton cross section reference, and the uncertainties on the
average nuclear overlap function. 

The systematic uncertainties on the D-meson corrected yields are
obtained considering as uncorrelated the different contributions described in the previous section.

The uncertainty on the pp reference used for the calculation of $\RAA$ 
has two contributions. The first is the systematic uncertainty on the measured 
$\pt$-differential D-meson cross section at $\sqrt s=7$~TeV.
This uncertainty is about 25\% at the lowest $\pt$ and 17\% at the highest $\pt$ for $\Dzero$ mesons, 
excluding the uncertainty for feed-down corrections, and few percent larger for $\Dplus$ and $\Dstar$ mesons~\cite{ALICE:2011aa}. 
The systematic uncertainty on the feed-down subtraction deriving from
the variation of the parameters of the FONLL calculation 
and from the use of the alternative method to compute $f_{\rm prompt}$
was considered to be
correlated in the $\PbPb$ and pp measurements.
These variations were carried out simultaneously for the numerator and denominator of $\RAA$, 
so only the residual effect was attributed as a systematic uncertainty.
Therefore, the variation of the value of $\RAA^{\rm
  feed-down}/\RAA^{\rm prompt}$ between 1 and 3 is the
main contribution to the feed-down uncertainty on $\RAA$.

The second contribution to the pp reference uncertainty is the scaling to $\sqrt s = 2.76$~TeV. It ranges from $^{+27}_{-10}$\% in the interval $2<\pt<3~\gev/c$ to about 5\% for $\pt>10~\gev/c$~\cite{ConesadelValle:2011fw}.
Note that the upper/lower uncertainties are reversed when considering $\RAA$, where the pp reference is in the denominator.
In the interval 1--2~GeV/$c$, this scaling uncertainty is much larger
($^{+57}_{-11}$\%), but
its impact on the pp reference was reduced by about a factor of two by
using a weighted average of the cross section scaled from 7~TeV and
the measured cross section at 2.76~TeV (see Section~\ref{Dcorr}).

The extrapolation of the pp reference to the intervals $16<\pt<24~\gev/c$ for $\Dzero$ mesons and 
 $24<\pt<36~\gev/c$ for $\Dplus$ and $\Dstar$ mesons resulted in a total systematic uncertainty of about 
$_{-50}^{+35}\%$, as described in Section~\ref{Dcorr}.

The uncertainties on $\Raa$ are listed in Tab.~\ref{tab:SystRaa}. The
uncertainties on the normalisation are the quadratic sum of the pp
normalisation uncertainty  (3.5\%) and the uncertainty on $\langle
T_{\rm AA} \rangle$, which is 3.2\% and 4.7\% in the 0--10\% and 30--50\% centrality classes, respectively. 

All the uncertainties described in this Section that result from detector
effects are considered to be largely correlated over transverse
momentum, with the exception of the yield extraction uncertainty that
depends on the S/B in each $\pt$ interval. The uncertainties related
to the feed-down assumptions and to the $\sqrts$-scaled pp
reference are fully correlated over $\pt$, with the exception of that
for the hypothesis on the ratio of the prompt and feed-down D-meson
$\Raa$ that might not be constant as a function of $\pt$.
\section{Results and discussion}
\label{sec:DRAA}
\subsection{D-meson $\pt$ spectra and $\Raa$} 

\begin{figure}[!t]
 \begin{center}
\subfigure[]{
\includegraphics[angle=0, width=7.5cm]{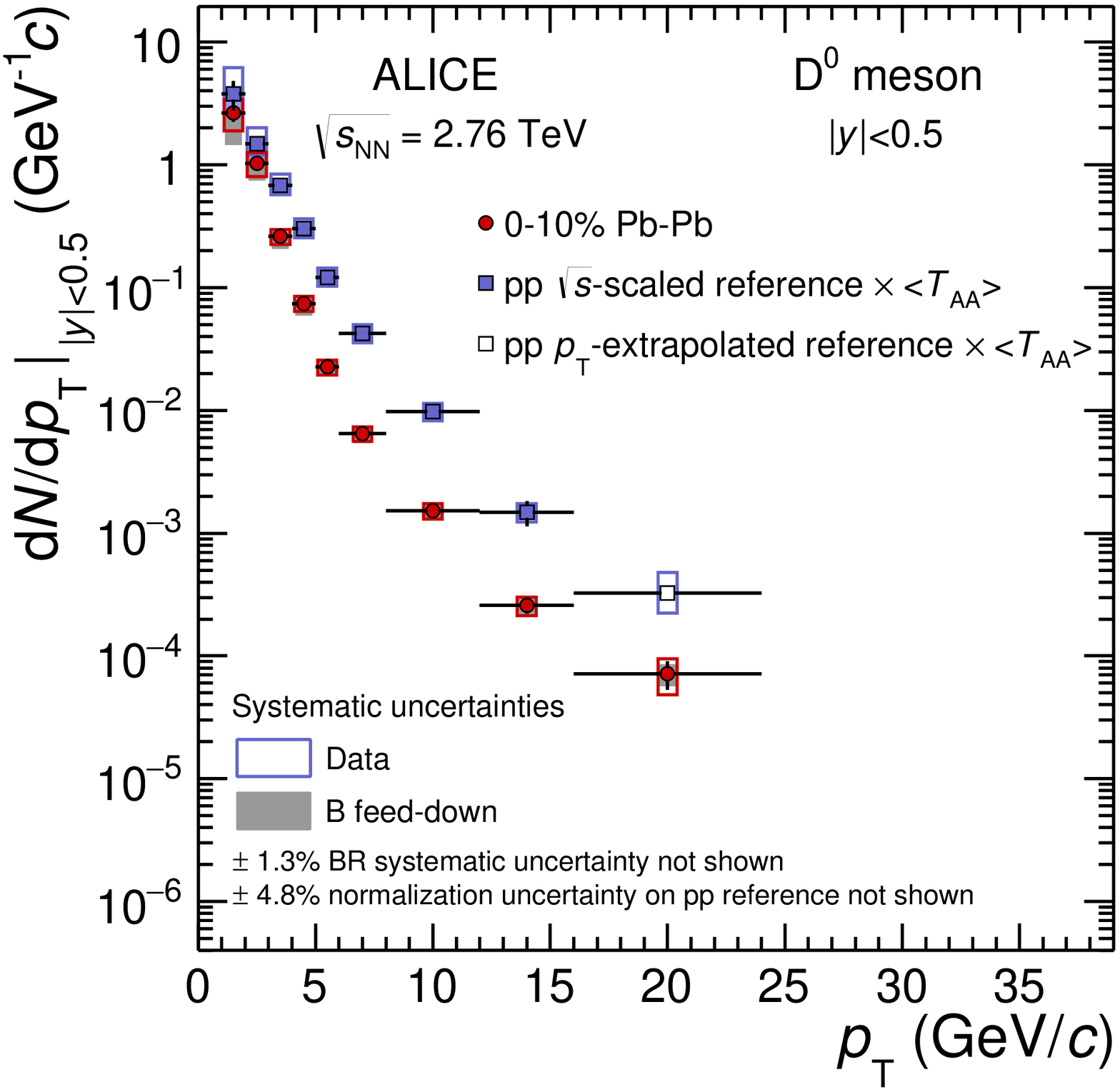}
\label{D0yield}
}
\subfigure[]{
\includegraphics[angle=0, width=7.5cm]{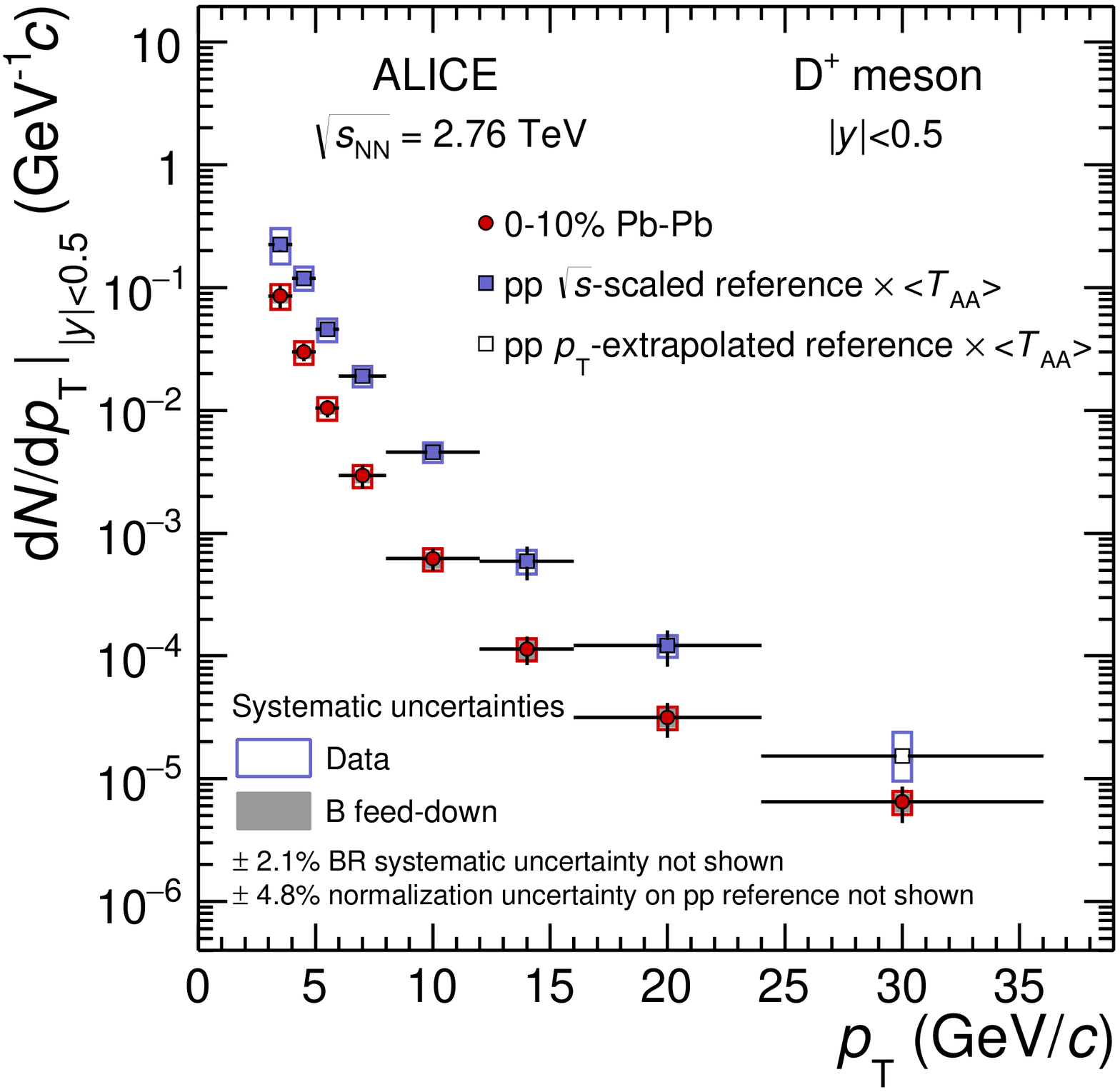}
\label{Dplusyield}
} 
\subfigure[]{
\includegraphics[angle=0, width=7.5cm]{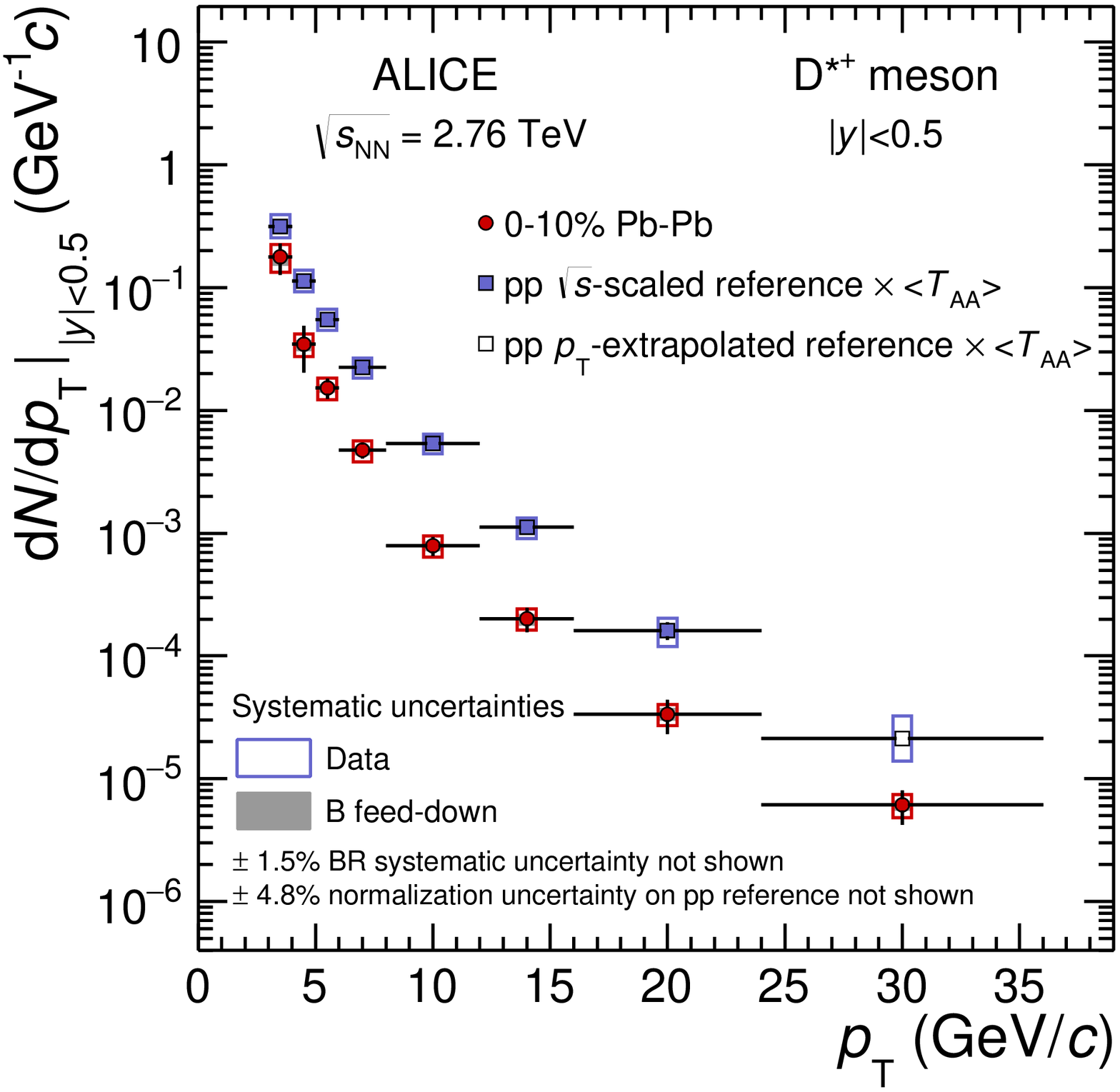}
\label{Dstaryield}
} 
\subfigure[]{
 \includegraphics[angle=0,
 width=7.5cm]{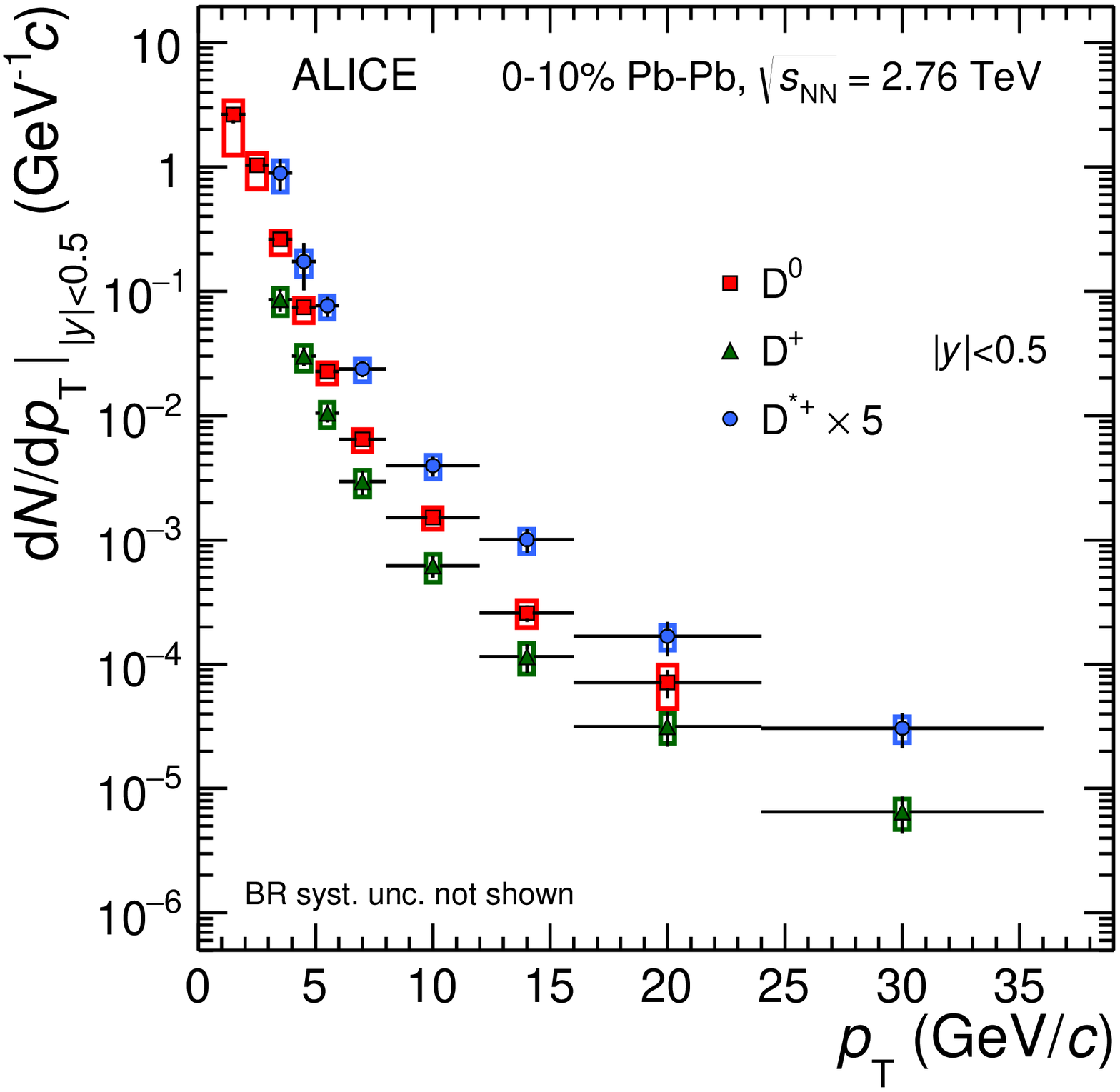}
 \label{DmesYield010Comparison}
}
 \end{center}
 \caption{Transverse momentum distributions $\d N/\d\pt$ of 
prompt $\Dzero$ (a), $\Dplus$ (b) and $\Dstar$ (c) mesons in the 0--10\% 
centrality class in $\PbPb$ collisions 
at $\sqrtsNN=2.76~\tev$. 
The pp reference distributions $\av{\TAA}\,\d \sigma/\d\pt$ are shown as well. 
Statistical uncertainties (bars) and systematic uncertainties from data 
analysis (empty boxes) and from feed-down subtraction 
(shaded boxes) are shown. 
Horizontal bars represent bin widths, symbols are placed at the centre of 
the bin. 
The $\d N/\d\pt$ distributions of the three D-meson species in the 10\% most
central $\PbPb$ collisions are compared to each other in panel (d), where
the $\Dstar$ production yields are scaled by a factor of five for visibility.
}
 \label{DmesCorrYields010} 
\end{figure}

The transverse momentum distributions $\d N/\d \pt$ of prompt $\Dzero$,
$\Dplus$ and $\Dstar$ mesons are shown in Figs.~\ref{D0yield}, 
~\ref{Dplusyield} and~\ref{Dstaryield} for the 10\% most central $\PbPb$ 
collisions. 
The results are presented in the interval $1<\pt<24~\gev/c$ 
for the $\Dzero$ mesons and $3<\pt< 36~\gev/c$ for $\Dplus$ and $\Dstar$ mesons.
They are compared to the corresponding pp cross section reference multiplied by 
$\langle \TAA  \rangle$. 
The vertical bars represent the statistical uncertainties, the empty boxes
the systematic uncertainties from the data analysis, and the shaded boxes
the systematic uncertainty due to the subtraction of the feed-down from 
B-hadron decays. Uncertainties on the pp cross section
normalisation and on the branching ratios are quoted separately.
A clear suppression of the D-meson yields is observed at intermediate ($3<\pt<8~\gev/c$) and high transverse 
momenta ($\pt>8~\gev/c$) in central $\PbPb$ collisions as compared 
to the binary-scaled pp reference.
In Fig.~\ref{DmesYield010Comparison} the transverse momentum distributions of 
prompt $\Dzero$, $\Dplus$ and $\Dstar$ mesons in the 10\% most central
collisions are compared to each other. 
The $\d N/\d \pt$ values of $\Dstar$ mesons are scaled by a factor of five
for visibility.

\begin{figure}[!htpb]
 \begin{center}
\subfigure[]{
\includegraphics[angle=0, width=7.5cm]{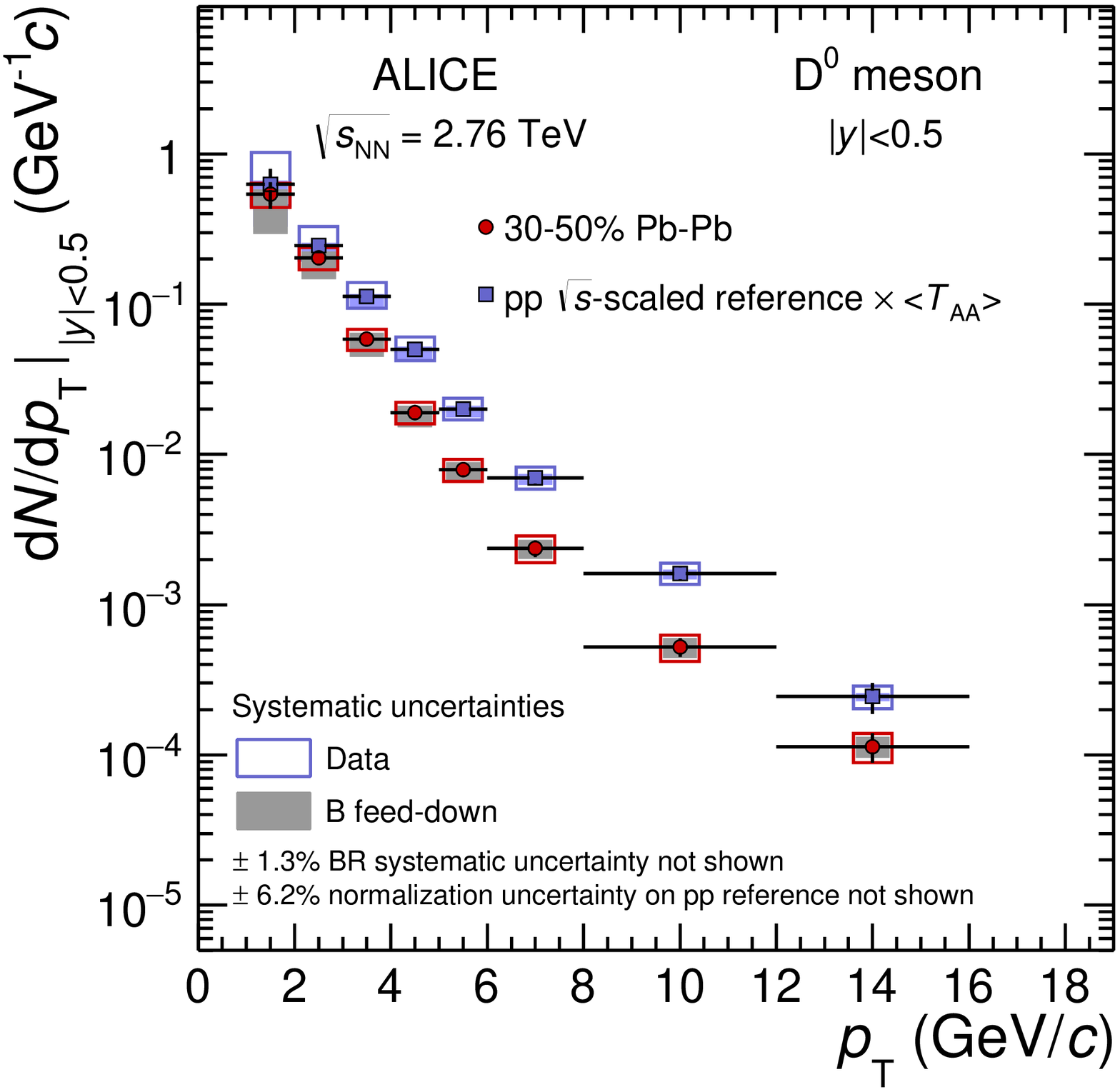}
\label{D0yield3050}
}
\subfigure[]{
\includegraphics[angle=0, width=7.5cm]{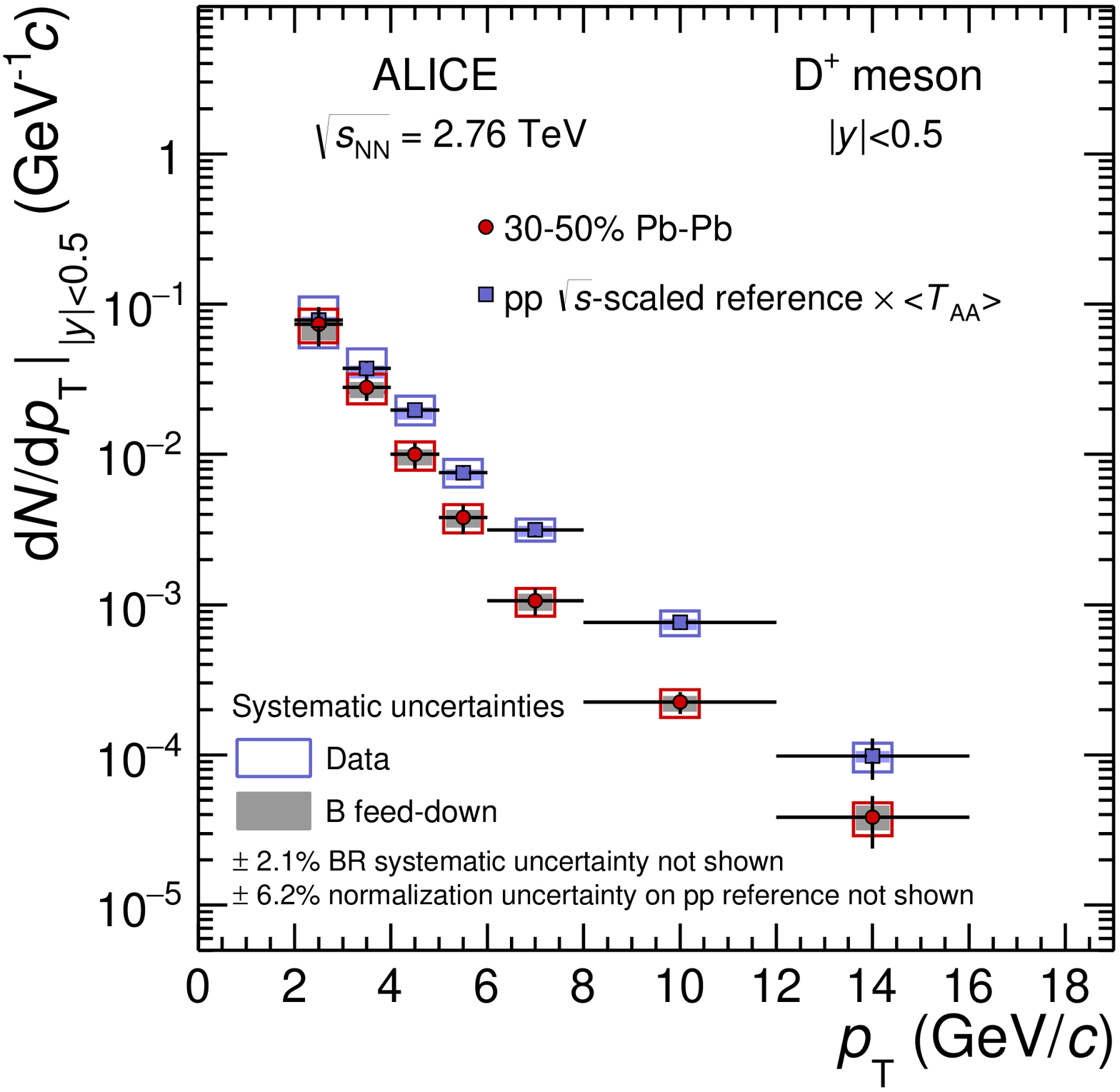}
\label{Dplusyield3050}
} 
\subfigure[]{
\includegraphics[angle=0, width=7.5cm]{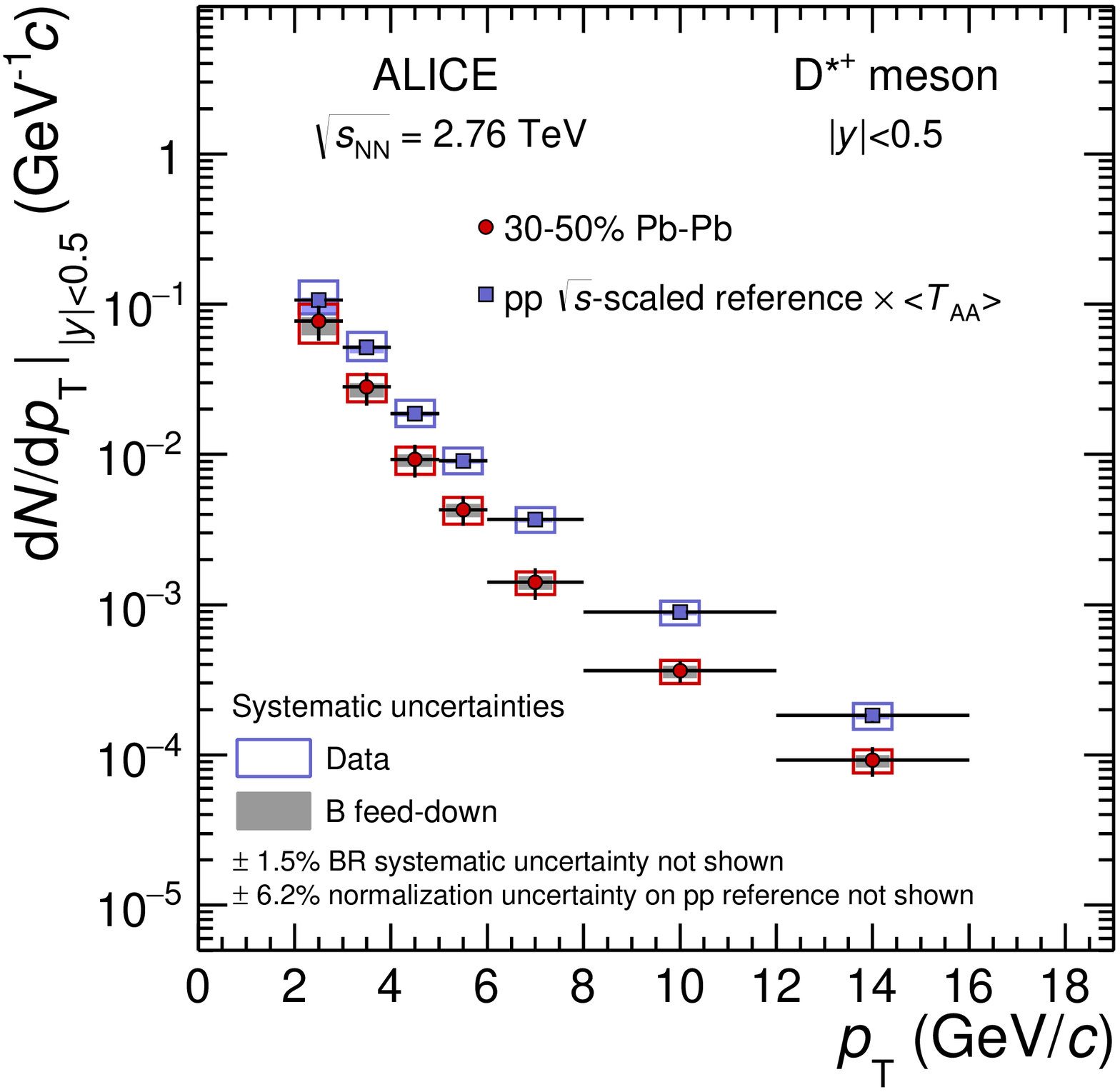}
\label{Dstaryield3050}
} 
\subfigure[]{
 \includegraphics[angle=0,
 width=7.5cm]{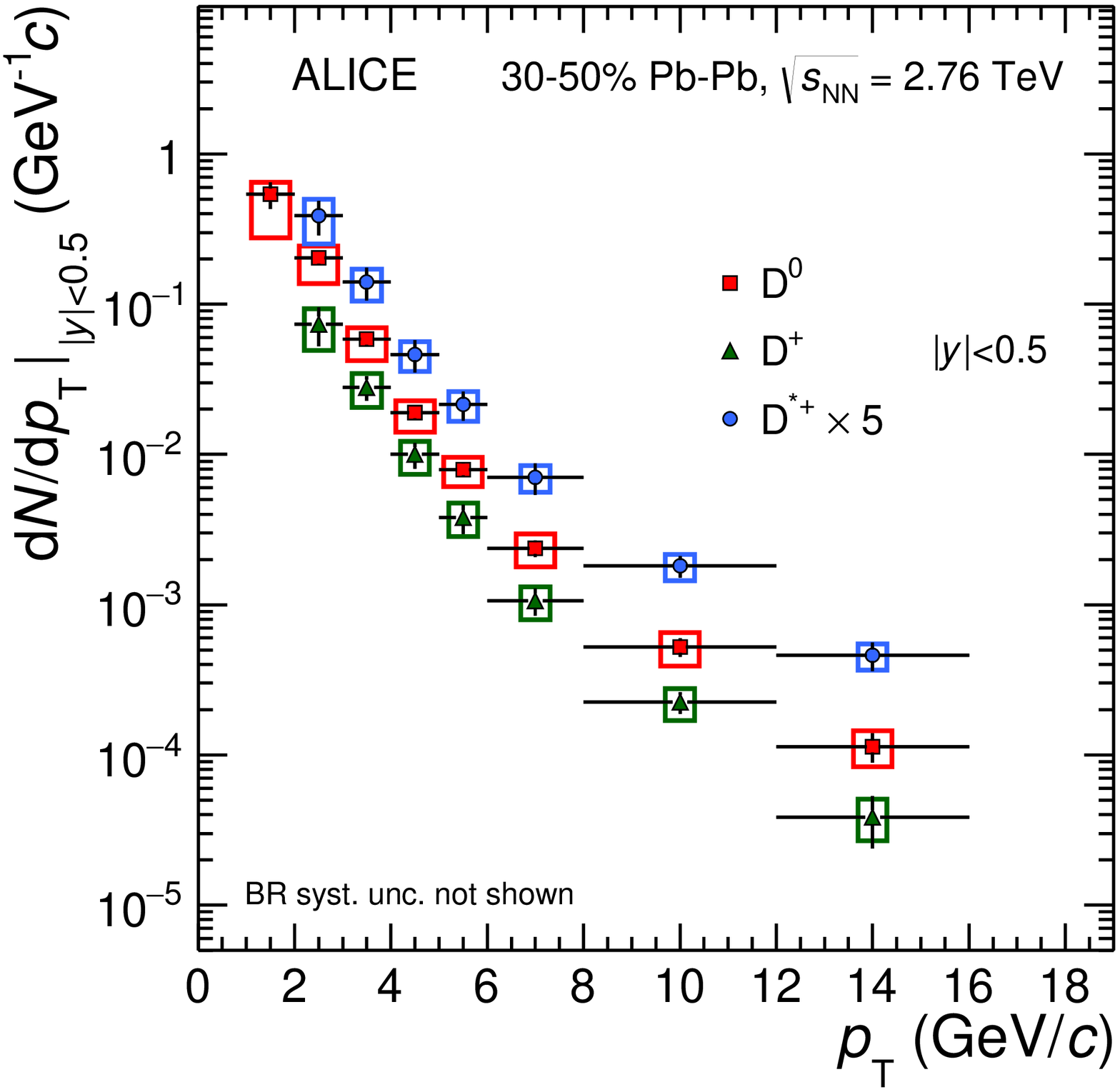}
 \label{DmesYield3050Comparison}
}
 \end{center}
\caption{Transverse momentum distributions $\d N/\d\pt$ of 
prompt $\Dzero$ (a), $\Dplus$ (b) and $\Dstar$ (c) mesons in the 30--50\% 
centrality class in $\PbPb$ collisions at $\sqrtsNN=2.76~\tev$. 
The pp reference distributions $\av{\TAA}\,\d \sigma/\d\pt$ are shown as well. 
Statistical uncertainties (bars) and systematic uncertainties from data 
analysis (empty boxes) and from feed-down subtraction 
(shaded boxes) are shown. 
Horizontal bars represent bin widths, symbols are placed at the centre of 
the bin.
The $\d N/\d\pt$ distributions of the three D-meson species in $\PbPb$
collisions in the 30--50\% centrality class are compared to each other in 
panel (d), where the $\Dstar$ production yields are scaled by a factor of five 
for visibility.
}
\label{DmesCorrYields3050} 
\end{figure}

The D-meson $\d N/\d \pt$ distributions measured in the 30--50\% centrality 
class are shown in Fig.~\ref{DmesCorrYields3050}.
Also for this centrality class, a clear suppression of the D-meson yields as
compared to the expectation based on binary scaling of the pp yields
is observed for $\pt>3~\gev/c$.
In Fig.~\ref{DmesYield3050Comparison}, the $\d N/\d \pt$
of  prompt $\Dzero$, $\Dplus$ and $\Dstar$ (the latter scaled by a factor of 
five) are compared to each other.

\begin{figure}[!t]
 \begin{center}
\includegraphics[angle=0, width=7.5cm]{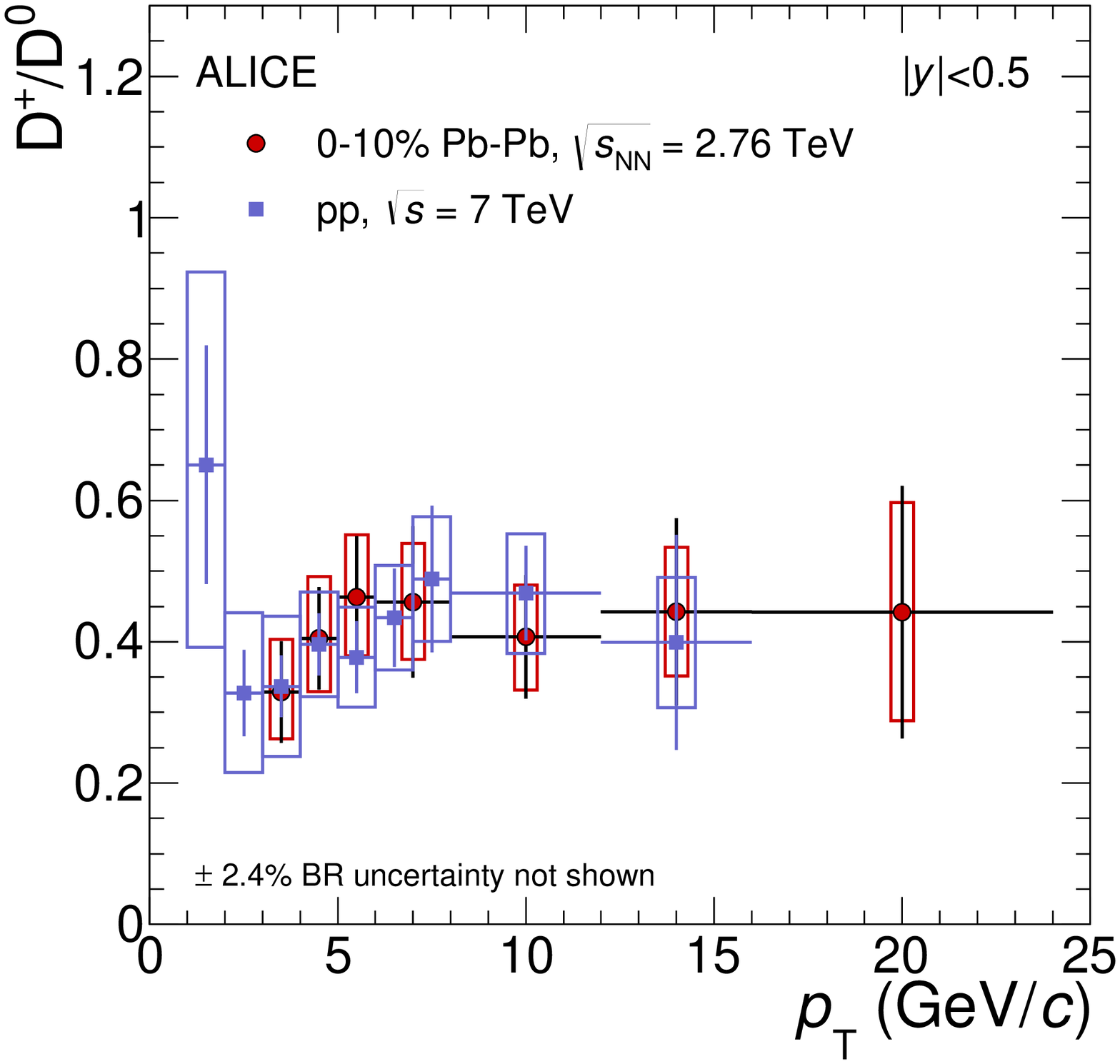}
\includegraphics[angle=0, width=7.5cm]{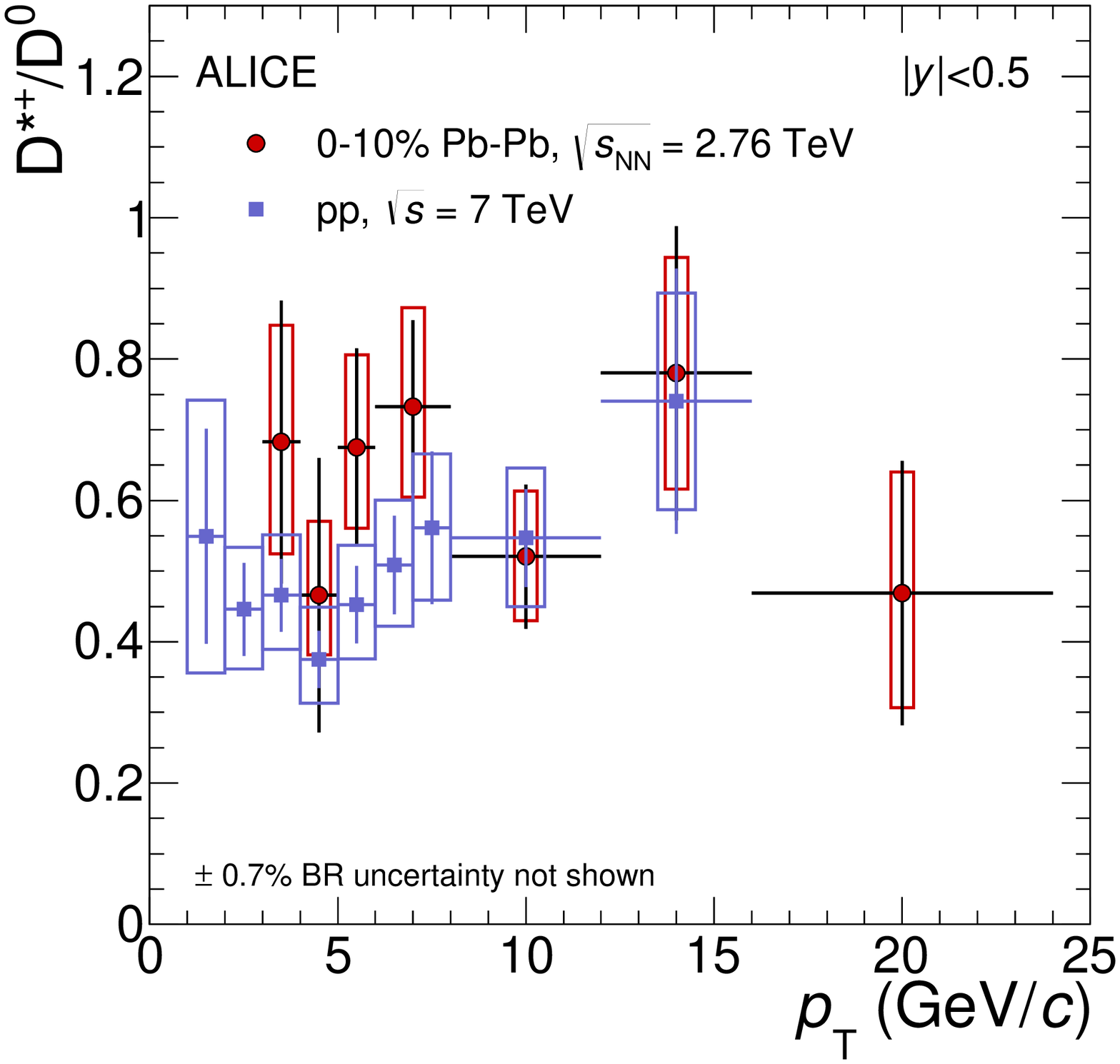}
 \end{center}
 \caption{Ratio of prompt D-meson yields ($\Dplus$/$\Dzero$ and
   $\Dstar$/$\Dzero$) as a function of $\pt$ in the 10\% most central
   $\PbPb$ collisions at $\sqrtsNN=2.76~\TeV$ compared to the results
   in pp collisions at $\sqrts=7~\TeV$. 
Statistical (bars) and systematic (boxes) uncertainties are shown.}
 \label{DmesRatio} 
\end{figure}

Figure~\ref{DmesRatio} shows the $\pt$-dependent ratios of
$\Dplus$/$\Dzero$ and $\Dstar$/$\Dzero$ for
central $\PbPb$ collisions. They are 
found to be compatible within uncertainties with those measured in pp 
collisions at $\sqrts=7~\tev$~\cite{Abelev:2012tca}.
Similar results were also found for the 30--50\% centrality class. 
Therefore, no modification of the relative abundances of these three D-meson species is 
observed within the current uncertainties in central and semi-central 
$\PbPb$ collisions relative to the pp ones at LHC energies.

\begin{figure}[!t]
 \begin{center}
\includegraphics[angle=0, width=7.5cm]{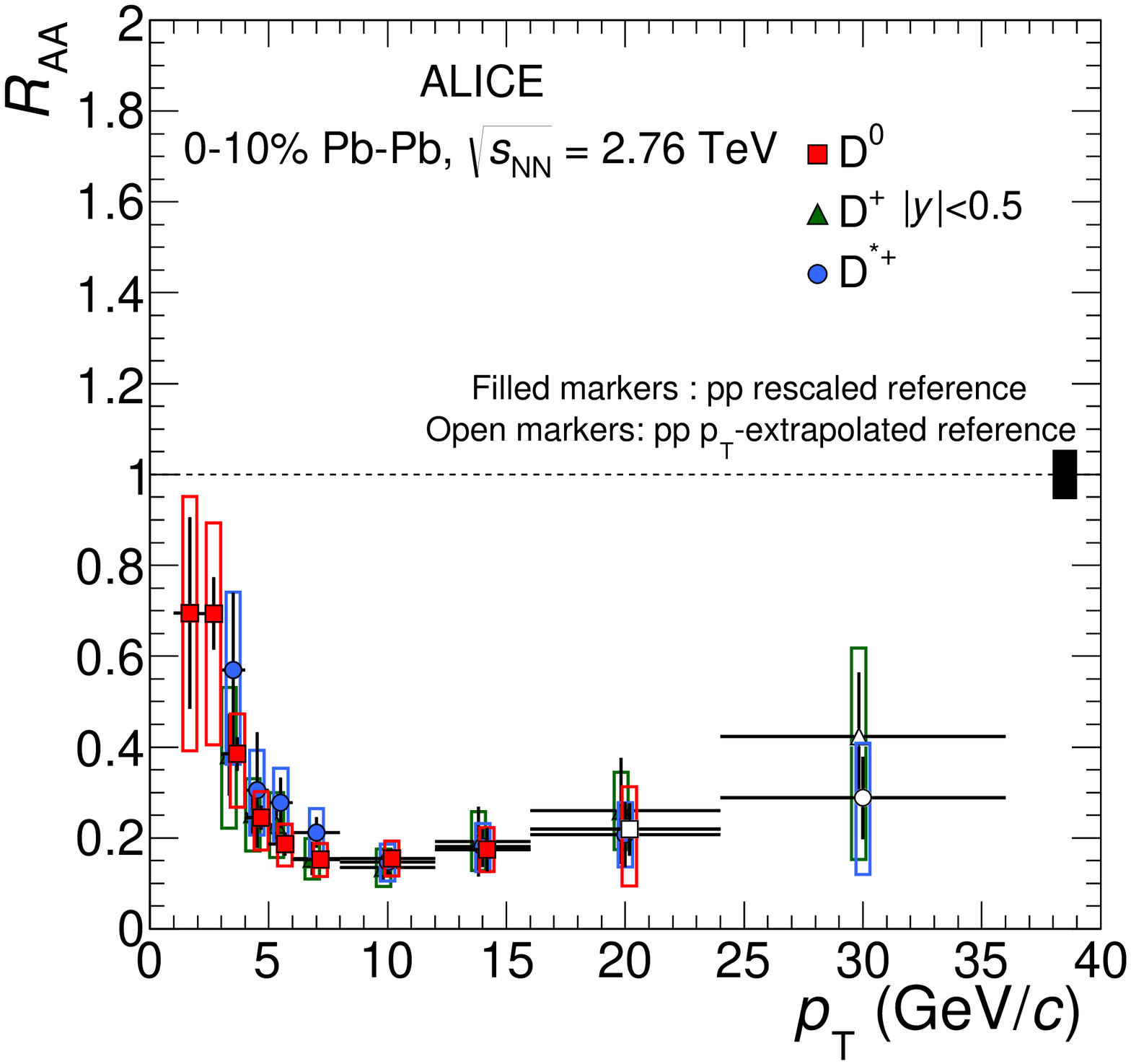}
\includegraphics[angle=0, width=7.5cm]{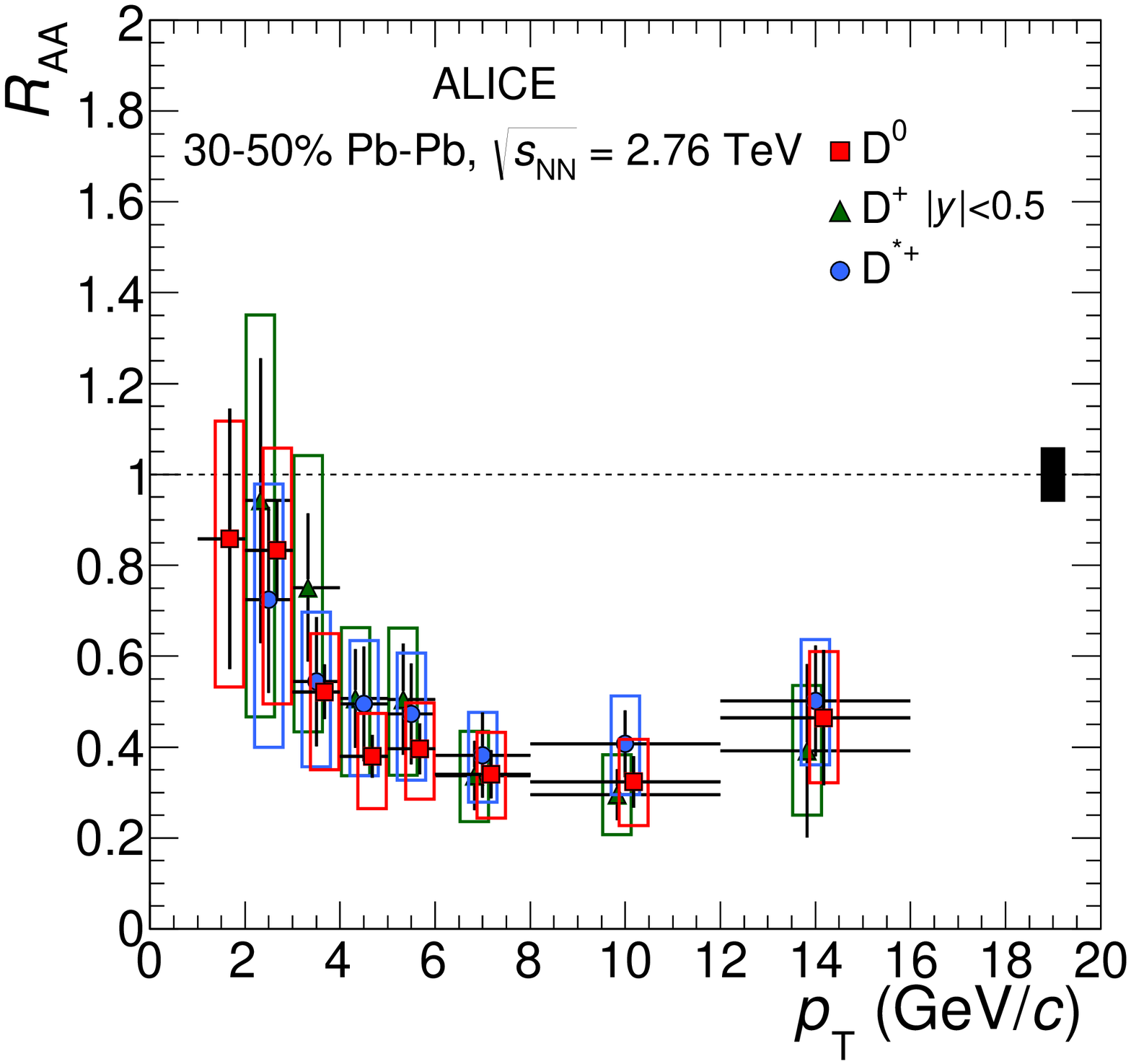}
 \end{center}
 \caption{$\RAA$ 
  of prompt $\Dzero$, $\Dplus$, and $\Dstar$ mesons for the 0--10\% (left) and 
30--50\% (right) centrality classes. 
Statistical (bars),  systematic (empty boxes), and normalisation (shaded box) 
uncertainties are shown.
Horizontal bars represent bin widths. $\Dzero$ symbols are placed at the centre 
of the bin. $\Dplus$, and $\Dstar$ are shifted for visibility.}
 \label{DmesRaa} 
\end{figure}

The $\Raa$ of prompt $\Dzero$, $\Dplus$
and $\Dstar$ mesons is shown in Fig.~\ref{DmesRaa}
for the 0--10\% (left panel) and 30--50\% (right panel)
centrality classes. 
The statistical uncertainties, represented by the vertical error bars,
range from 10\% in the intermediate $\pt$ range 
up to about 25--30\% in the lowest and highest $\pt$ intervals, for the 10\% 
most central collisions. The statistical uncertainty on the reference
measurement at $\sqrts=7~\tev$ dominates this uncertainty in the
interval $2<\pt<16~\gev/c$.
For the 30--50\% centrality class, the statistical uncertainties at low and 
intermediate $\pt$ are similar in magnitude to those of central collisions and 
are about 20\% in the interval $12 < \pt < 16~\gev/c$. 
The total $\pt$-dependent systematic uncertainties, described
in the previous Section, are shown as empty boxes. 
The normalisation uncertainty is represented by a filled box at $\Raa =1$. 
The nuclear modification factors of the three D-meson species are compatible 
within statistical uncertainties for both centrality classes. For the 10\% most
central collisions, the measured $\Raa$ shows a suppression that is
maximal at around $\pt=10~\gev/c$, where a reduction of the yields by
a factor of 5--6 with respect to the binary-scaled pp reference is observed.
The suppression decreases with decreasing $\pt$ for $\pt<10~\GeV/c$, and it 
is of the order of a factor of 3 in the interval $3<\pt<4~\gev/c$,
while the $\Raa$ ranges from about 0.35 to 1 in the first two $\pt$
intervals.
For $\pt> 10~\GeV/c$, the suppression appears to decrease with increasing $\pt$, 
but the large statistical uncertainties do not allow us to determine the 
trend of the $\RAA$. A suppression ($\Raa
< 0.5$) is still observed for D mesons with $\pt > 25~\GeV/c$.
For the 30--50\% centrality class, the suppression amounts to about a
factor of 3 at $\pt=10~\gev/c$, which indicates
that the suppression of the high-$\pt$ D-meson yields is smaller than in the
0--10\% centrality class.
As for the central collisions, the suppression reduces at lower
momenta, with $\Raa$ increasing with decreasing $\pt$ up to a value of about 
0.6 in the interval $3<\pt<4~\gev/c$. For lower $\pt$ the suppression is
further reduced and $\Raa$ is compatible with unity.

The average nuclear modification factor of $\Dzero$, $\Dplus$ and $\Dstar$ 
mesons was computed using the inverse of the squared relative statistical uncertainties as weights. 
The systematic uncertainties were propagated through the averaging procedure,
considering the contributions from the tracking efficiency, the B-meson feed-down subtraction 
and the FONLL-based $\sqrts$-scaling of the pp cross section from
$\sqrts = 7~\TeV$  to 
$\sqrts = 2.76~\TeV$ as fully correlated uncertainties among the three D-meson species. 
The average D-meson $\Raa$ for the two centrality classes is shown in the
left panel of Fig.~\ref{DmesRaaAverAndStar}.
A larger suppression, by about a factor of two, is observed in the 10\% most
central collisions compared to the 30--50\% centrality class for
$\pt>5~\gev/c$.
The stronger suppression observed in central collisions can be understood
as resulting from to the increasing medium density, size and lifetime
from peripheral to central collisions.
The $\Raa$ values measured for the 0--10\% centrality class are slightly lower,
although compatible within uncertainties, than those reported in 
Ref.~\cite{ALICE:2012ab} for the 20\% most central collisions,
measured with the 2010 data sample.
As a consistency check, the analysis on the 2011 data sample was also
performed in the 0-20\% centrality class and the resulting $\Raa$ value
was found to be compatible with the one measured with the 2010 sample
within statistical and systematic uncertainties, considering that the
pp reference uncertainties are the same in the two measurements.
In addition, the larger sample of central $\PbPb$ collisions used in this analysis,
compared to that used in the previous publication, enables the measurement
of the D-meson $\Raa$ in a wider $\pt$ range (the intervals $1<\pt<2~\gev/c$ and 
$\pt>16~\gev/c$ were not accessible with the previous sample), 
with a substantial reduction (by a factor of about 2--3)
of the statistical uncertainties.

Figure~\ref{DmesRaaAverAndStar} (left) also shows the average D-meson
nuclear modification factor measured in minimum-bias p--Pb collisions at $\sqrtsNN=5.02~\tev$~\cite{Abelev:2014hha}. 
Since no significant modification of the D-meson production is observed in p--Pb 
collisions for $\pt>2~\GeV/c$, the strong
suppression of the D-meson yields for $\pt>3~\gev/c$  observed in central and semi-central $\PbPb$ collisions cannot be explained in terms of cold nuclear matter effects and is predominantly due to final-state 
effects induced by the hot and dense medium created in the 
collisions.

\subsection{Comparison with results at lower collision energy}

\begin{figure}[!t]
 \begin{center}
\includegraphics[angle=0, width=7.5cm]{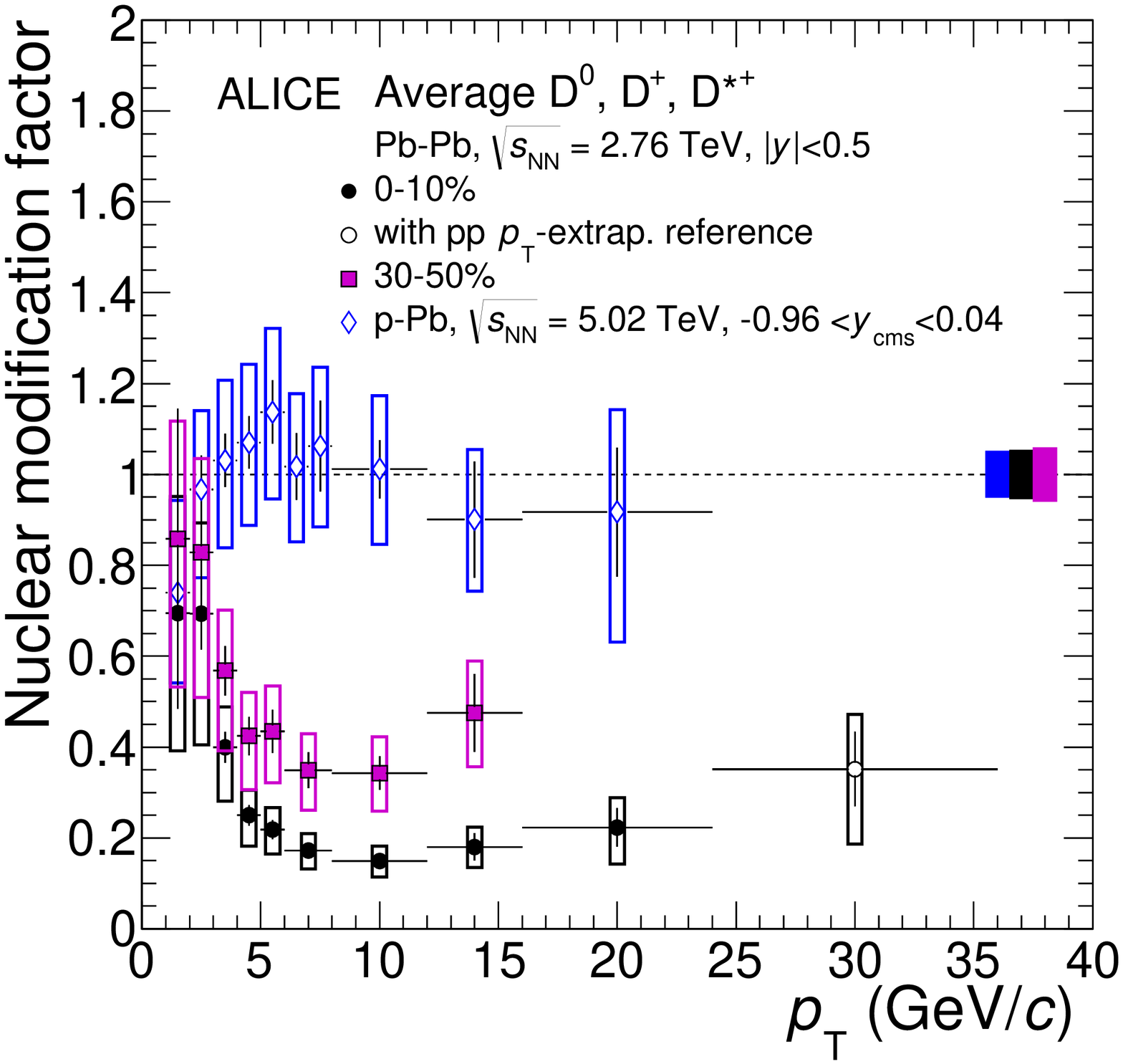}
 \includegraphics[angle=0, width=7.5cm]{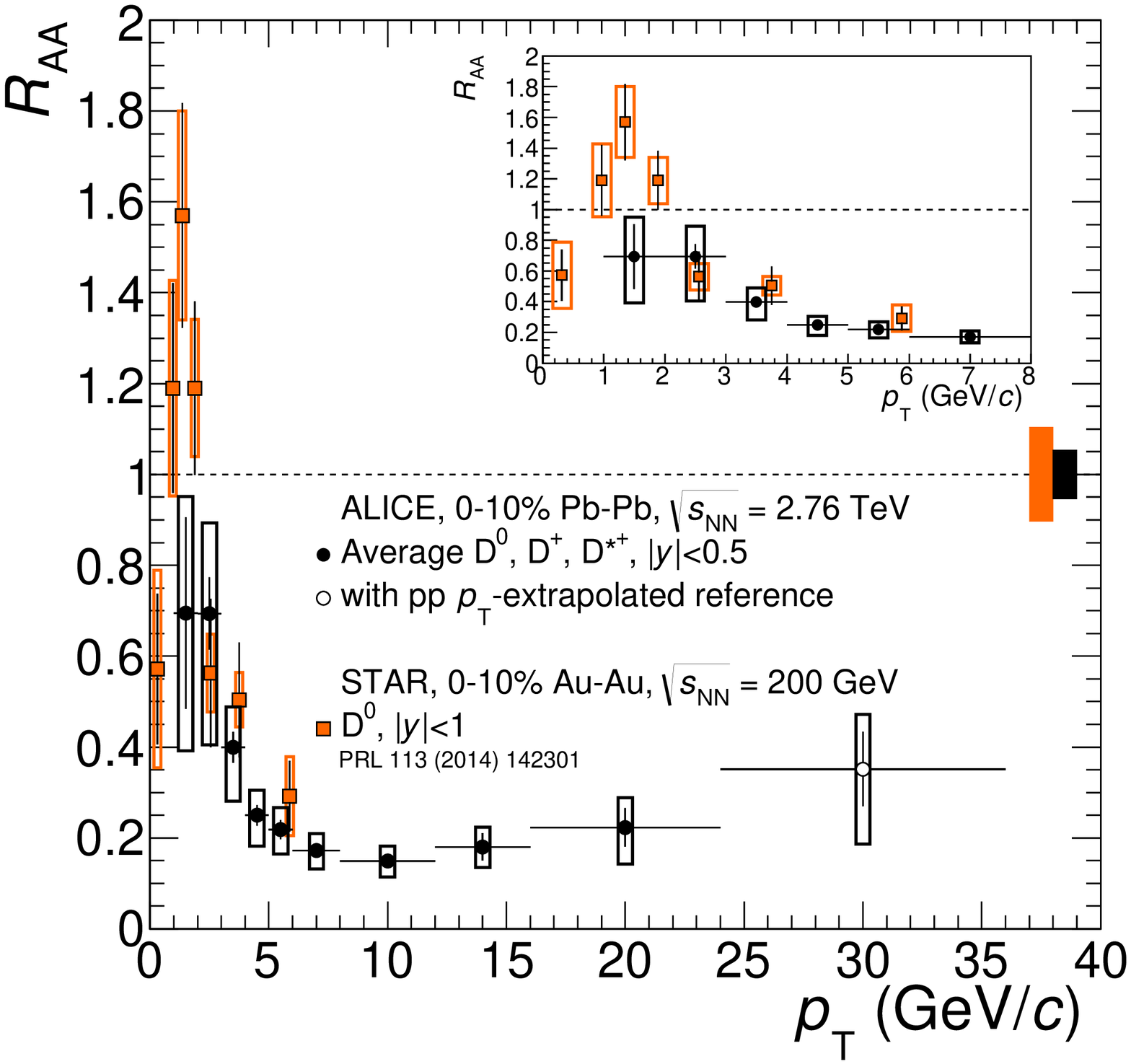}
 \end{center}
 \caption{Left: prompt D-meson $\Raa$ (average of $\Dzero$, $\Dplus$ and 
$\Dstar$) as a function of $\pt$ in $\PbPb$ collisions at 
 $\sqrtsNN=2.76~\tev$ in the 0--10\% and 30--50\% centrality
 classes. Prompt D-meson nuclear modification factor (average of $\Dzero$, $\Dplus$ and 
$\Dstar$) as a function of $\pt$ in $\pPb$ collisions at 
 $\sqrtsNN=5.02~\tev$~\cite{Abelev:2014hha}.
Right: prompt D-meson $\Raa$ (average of $\Dzero$, $\Dplus$ and 
$\Dstar$) as a function of $\pt$ in the 10\% most central $\PbPb$ collisions at 
 $\sqrtsNN=2.76~\tev$ compared to $\Dzero$ $\Raa$ measured
by the STAR Collaboration in $\AuAu$ collisions at RHIC at 
$\sqrtsNN=200~\gev$~\cite{Adamczyk:2014uip}. A zoomed-in plot of the
interval $0<\pt<8~\gev/c$ is shown in the inset. Statistical (bars),
systematic (empty boxes), and normalisation (shaded boxes at $\Raa=1$) uncertainties are shown.
Horizontal bars represent bin widths. Symbols are placed at the centre 
of the bin.
}
 \label{DmesRaaAverAndStar} 
\end{figure}

In the right panel of Fig.~\ref{DmesRaaAverAndStar}, the average 
D-meson $\Raa$ for the 10\% most central $\PbPb$ collisions is
compared to the $\Dzero$ nuclear modification factor measured by the
STAR Collaboration for the 10\% most central Au--Au collisions at 
$\sqrtsNN = 200~\GeV$~\cite{Adamczyk:2014uip}.
The D-meson $\Raa$ measured at the two energies are compatible
within uncertainties for $\pt > 2~\GeV/c$.
It should be noted that the similar $\Raa$ of D mesons with high momentum, 
$\pt>5~\gev/c$, i.e. in the range where the nuclear modification factor is 
expected to be dominated by the effect of in-medium parton energy 
loss, does not necessarily imply a similar charm-quark energy loss 
at the two collision energies.
Since the nuclear modification factor is also sensitive to the 
slope of the $\pt$ spectra in pp collisions, the combined effect of a denser 
medium and of the harder $\pt$ spectra at the LHC 
could result in similar values of $\RAA$ as at lower collision 
energies (see e.g.\,Ref.~\cite{Baier:2002tc}).

At low momentum ($1<\pt<2~\gev/c$), the 
$\Raa$ measured by STAR shows a maximum. This effect can be described by
models including parton energy loss, collective radial flow and the 
contribution of the recombination mechanism to charm-quark 
hadronisation~\cite{Abelev:2006db}.
The ALICE results at higher $\sqrtsNN$ do not show a
maximum. However, the large uncertainties and the coarser binning at low 
$\pt$ prevent a firm conclusion from being drawn.
A different pattern could be explained by the different role of initial-state 
effects or of radial flow at the two collision energies.
In the initial state, the modification of the parton distribution
functions in a nuclear environment is predicted to lead to a stronger 
suppression of the heavy-quark production yields at low $\pt$ with increasing 
$\sqrtsNN$~\cite{Eskola:2009uj}, because of the smaller values of Bjorken-$x$ probed.
In addition, the momentum ($k_{\rm T}$) broadening effect, which gives rise to an 
enhancement of the $\Raa$ at intermediate $\pt$ (Cronin peak), is known to
be more pronounced at lower collision energies~\cite{Wang:1998ww,Vogt:2001nh}.
In the final state, in addition to energy loss, the collective expansion 
of the medium is also predicted to affect the momentum distribution of charmed 
hadrons in heavy-ion collisions.
Indeed, the interactions with the medium constituents are expected to transfer 
momentum to low-$\pt$ charm quarks, which could take part in the collective
radial flow of the medium.
This effect could be enhanced by hadronisation via recombination, which is 
predicted in some models to contribute significantly to hadron formation at 
low and intermediate $\pt$~\cite{Andronic:2015wma}.
The momentum distributions of identified light-flavour hadrons at the 
LHC~\cite{Abelev:2012wca, Abelev:2013vea} indicate that the 
radial flow of the medium at LHC energies is about 10\% higher than at 
RHIC~\cite{Abelev:2008ab}. 
However, this stronger radial flow does not necessarily give rise
to a more pronounced bump-like structure in the $\Raa$ at low $\pt$ with 
increasing collision energy, because its effect can be counterbalanced
by the different shape of the momentum spectra in pp collisions at different
$\sqrts$~\cite{He:2014cla, He:2011qa}.

\begin{figure}[!t]
 \begin{center}
\includegraphics[angle=0, width=7.5cm]{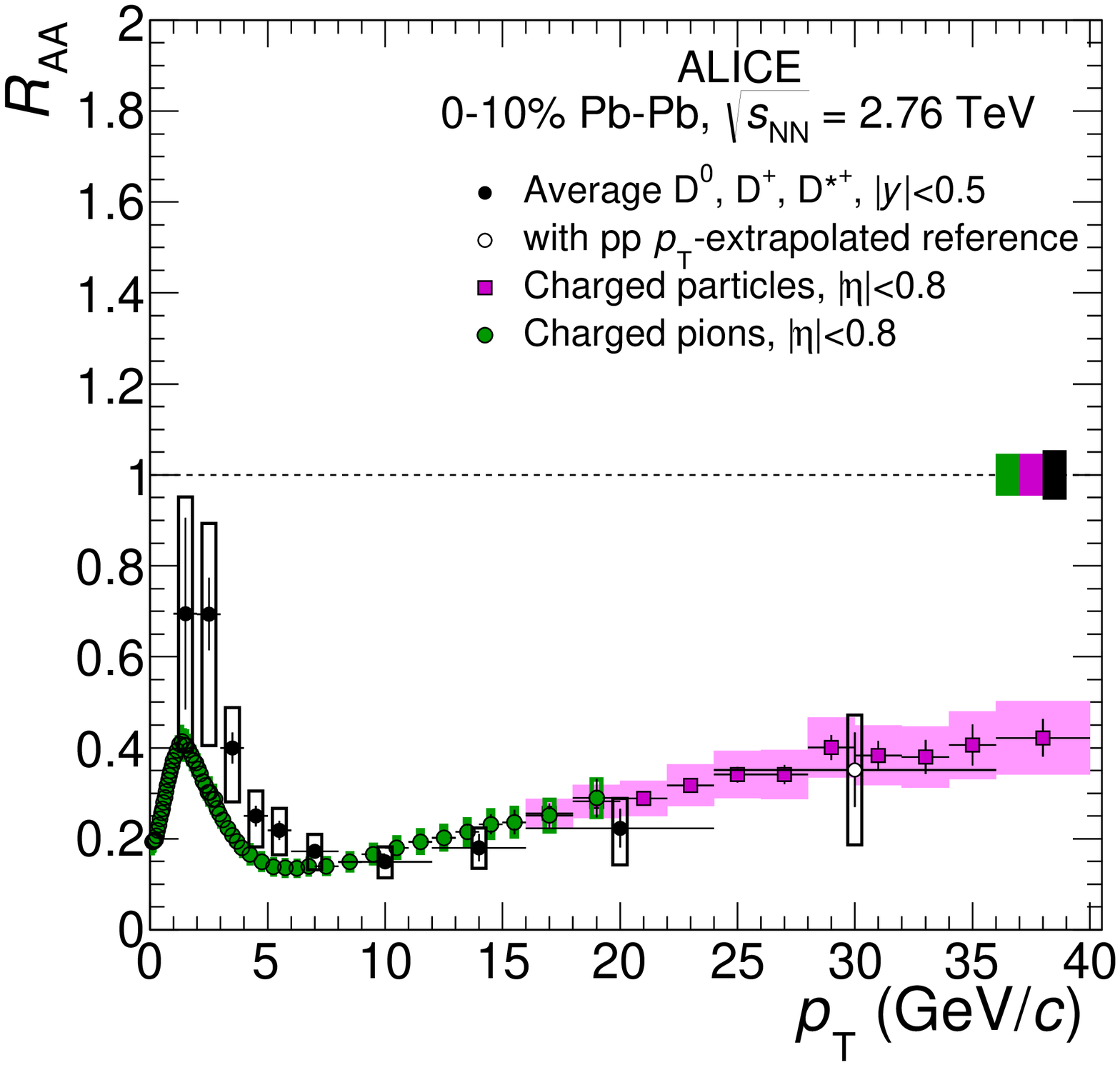}
 \includegraphics[angle=0, width=7.5cm]{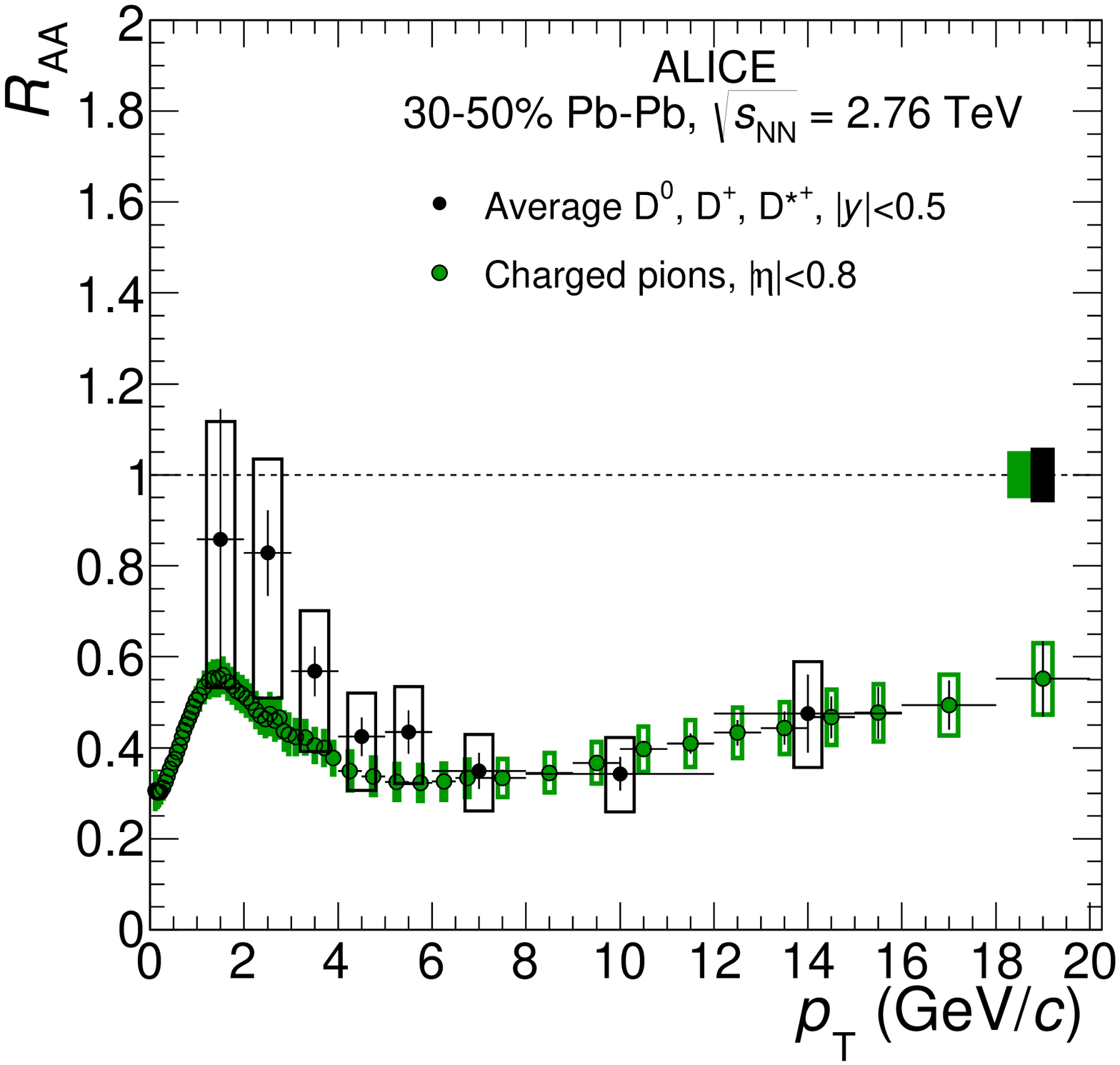}
 \end{center}
 \caption{Prompt D-meson $\Raa$ (average of $\Dzero$, $\Dplus$ and 
$\Dstar$) as a function of $\pt$ compared to the nuclear modification factors 
of pions~\cite{Abelev:2014laa} and charged particles~\cite{Abelev:2012hxa} in 
the 0--10\% (left) and 30--50\% (right) centrality
classes. Statistical (bars),
systematic (empty boxes), and normalisation (shaded box at $\Raa=1$) uncertainties are shown.
Horizontal bars represent bin widths. Symbols are placed at the centre 
of the bin.}
 \label{DmesRaaPions} 
\end{figure} 

\subsection{Comparison with pion and charged-hadron $\Raa$}

As described in Section~\ref{sec:intro}, the colour-charge and
quark-mass dependence of the energy loss can be tested with the comparison of 
D-meson and pion nuclear modification factors. 
In the left panel of Fig.~\ref{DmesRaaPions}, the D-meson $\Raa$ 
(average of $\Dzero$, $\Dplus$ and $\Dstar$) measured for the 10\% most central 
$\PbPb$ collisions is compared with the pion $\Raa$ in the interval
$1<\pt<20~\GeV/c$ and with the $\Raa$ of charged particles in
$16<\pt<40~\GeV/c$.
The charged-particle $\Raa$ is shown in order to extend the comparison up to the
higher $\pt$ interval in which the D-meson yield was measured.
The comparison of D mesons with charged hadrons at high-$\pt$ is
relevant because the $\Raa$ of different light-flavour hadron species
are consistent with one another for $\pt >
8~\GeV/c$~\cite{Abelev:2014laa}. Moreover the contribution of pions dominates
the charged-hadron yields at $\pt$ of about $20~\gev/c$ with respect to other hadron species (about
65\%)~\cite{Adam:2015kca}.
A similar comparison is performed in the right panel of 
Fig.~\ref{DmesRaaPions} for the 30--50\% centrality class. 

The $\RAA$ of D mesons and light-flavour hadrons are consistent for
$\pt>6~\gev/c$ for both centrality classes.
For $\pt < 6~\GeV/c$, the $\Raa$ of D mesons tends to be slightly higher than
that of pions. This can be also observed from the ratio of
nuclear modification factors, presented in Fig.~\ref{fig:DandpionRAAcmpModels}.
Considering that the systematic uncertainties of D-meson yields are mainly
correlated with $\pt$, we observe $\RAA^{\rm D}>\RAA^{\pi}$ at low
$\pt$ with a significance of about 1\,$\sigma$ in four $\pt$ intervals, in
the most central events. In the 30--50\% centrality class, the significance of the effect is smaller than in central collisions. 

A direct interpretation of a possible 
difference between the D-meson and pion $\Raa$ at low $\pt$ is not straightforward.
In the presence of a colour-charge and 
quark-mass dependent energy loss, the harder $\pt$ distribution and the harder fragmentation function of 
charm quarks compared to those of light quarks and gluons could lead to similar 
values of D-meson and pion $\Raa$, as discussed in 
Ref.~\cite{Djordjevic:2013pba}.
In addition, it should be considered that the pion yield could have a 
substantial contribution from soft production processes up to transverse 
momenta of about 2--3~$\gev/c$ due to the strong radial flow at LHC energies. 
This soft contribution, which is not present in the D-meson yield, does not 
scale with the number of binary nucleon--nucleon collisions.
Finally, the effects of radial flow and hadronisation via recombination, 
as well as initial-state effects, could affect D-meson and pion 
(light-flavour particle) yields differently at a given $\pt$, thus introducing an additional
complication in interpreting the magnitude of the $\Raa$ in terms
of different in-medium parton energy loss of charm quarks, light quarks and 
gluons.

\subsection{Comparison with models} 
\begin{figure}[!t]
 \begin{center}
\subfigure[]{
\includegraphics[angle=0, width=7.5cm]{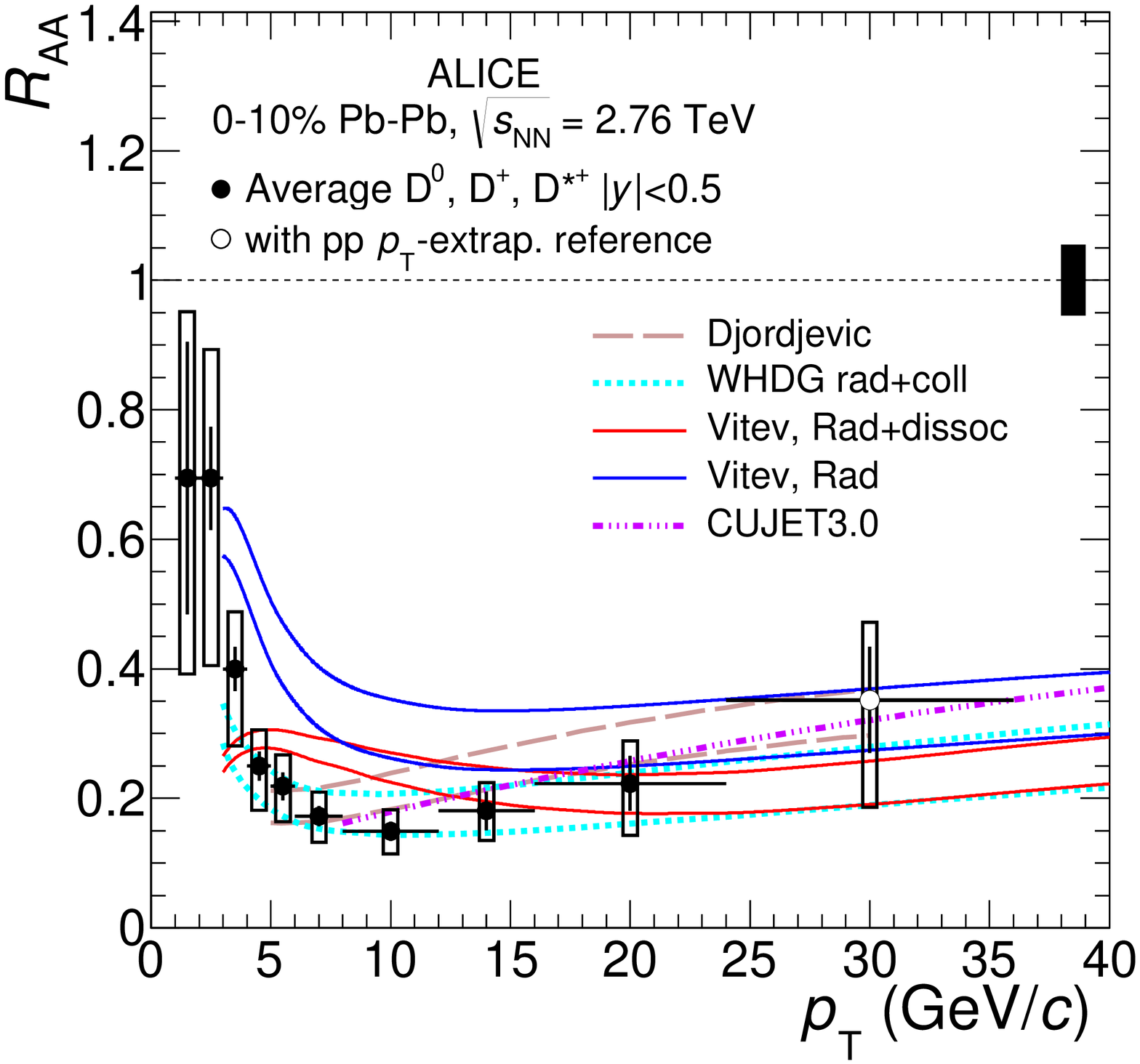}
\label{Raa010ModelsPions}
}
\subfigure[]{
\includegraphics[angle=0, width=7.5cm]{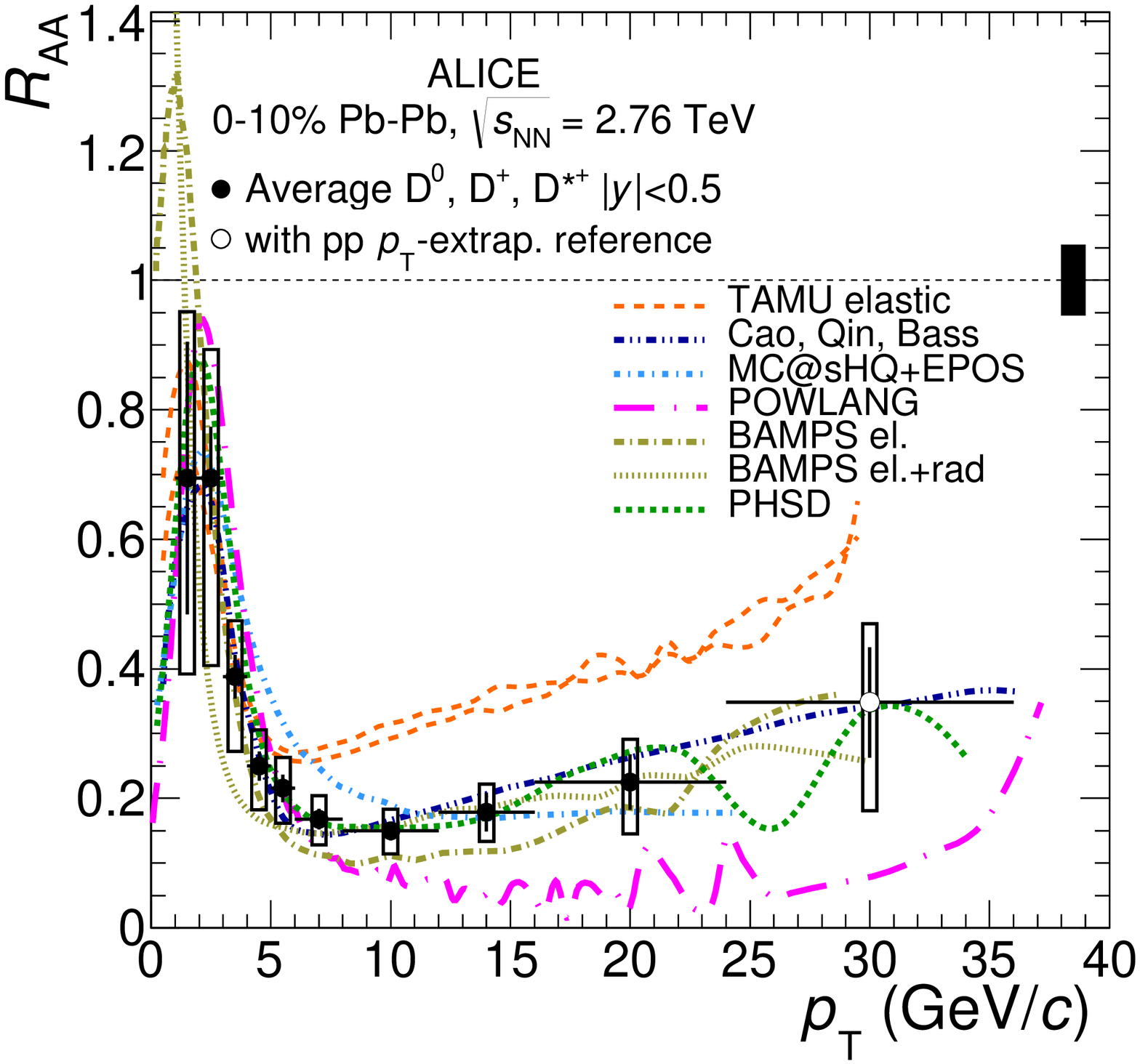}
\label{Raa010ModelsNoPions}
} 
\subfigure[]{
\includegraphics[angle=0, width=7.5cm]{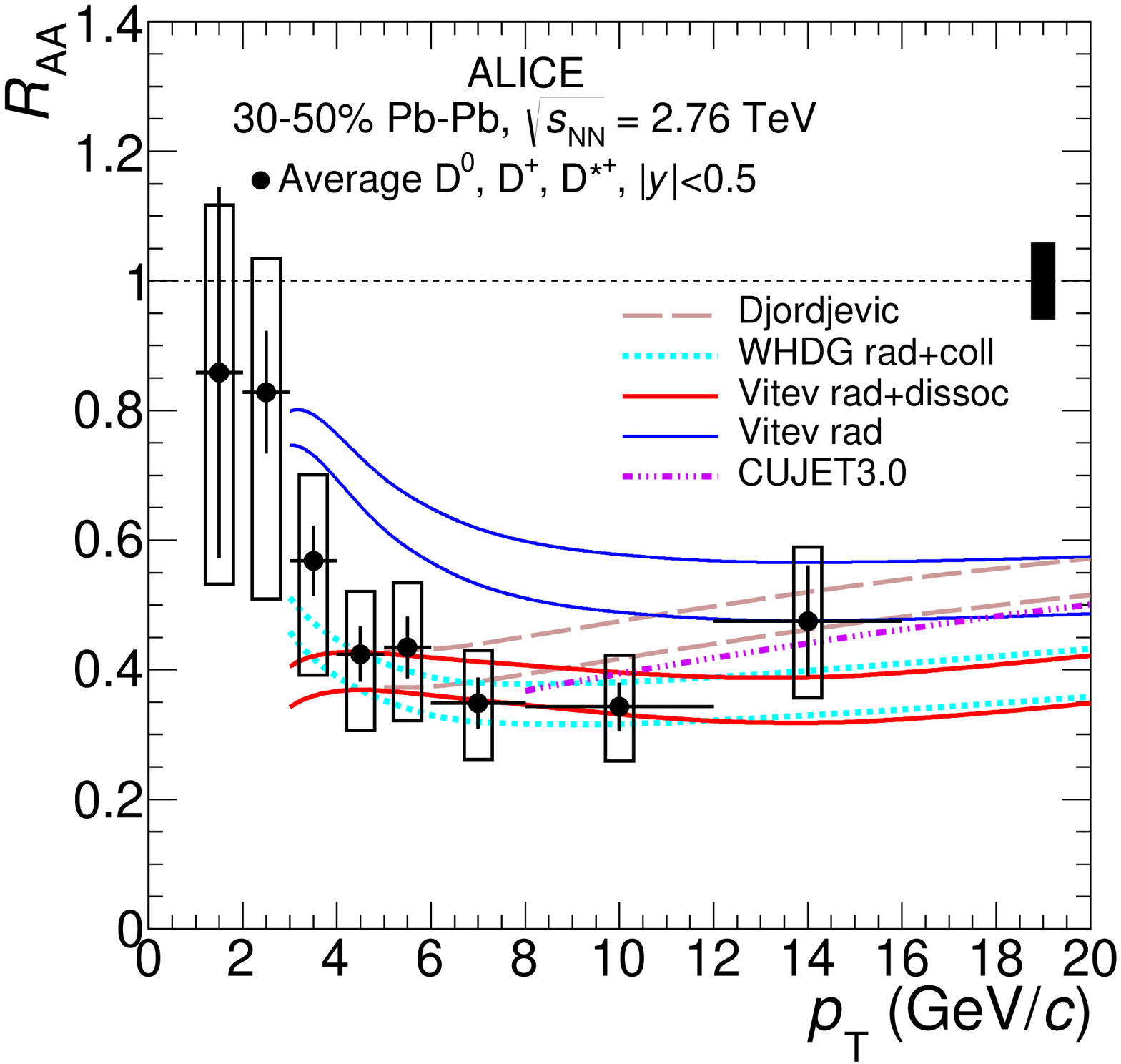}
\label{Raa3050ModelsPions}
} 
\subfigure[]{
 \includegraphics[angle=0, width=7.5cm]{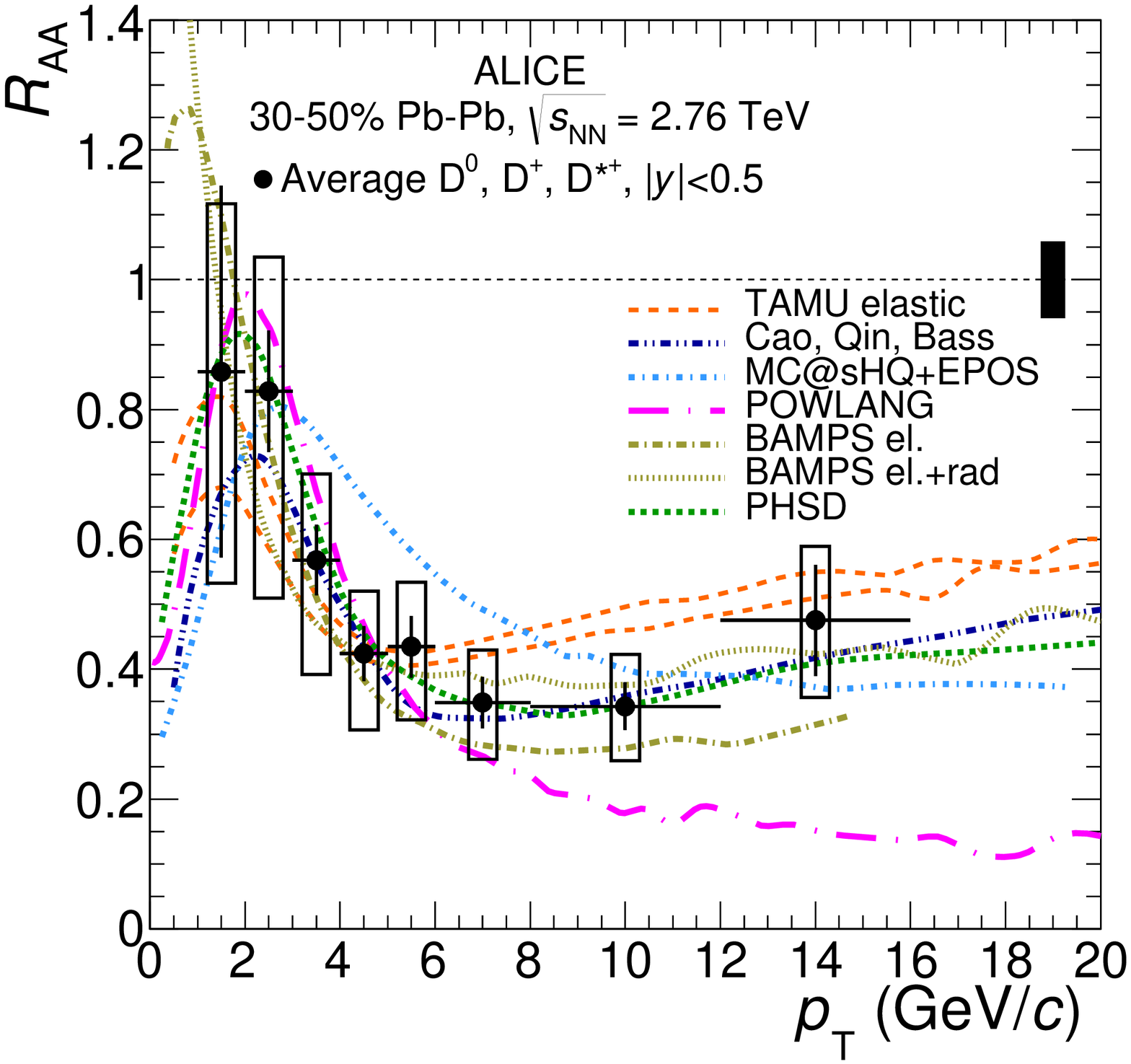}
 \label{Raa3050ModelsNoPions}
}
 \end{center}
 \caption{Average of prompt $\rm D^0$, $\rm D^+$ and $\rm D^{*+}$ $\RAA$ in the centrality classes 0--10\% (a and b) and 30--50\% (c and d) compared with model calculations:
{\it Djordjevic}~\cite{Djordjevic:2014tka},  {\it CUJET3.0}~\cite{Xu:2014tda, Xu:2015bbz}, {\it  WHDG}~\cite{Wicks:2005gt,Horowitz:2011gd,Horowitz:2011wm},  {\it Vitev}~\cite{Sharma:2009hn} (a and c), 
{\it  TAMU elastic}~\cite{He:2014cla}, {\it Cao, Qin and Bass}~\cite{Cao:2013ita},
{\it MC@sHQ+EPOS, Coll+Rad(LPM)}~\cite{Nahrgang:2013xaa},
 {\it POWLANG}~\cite{Alberico:2011zy,Beraudo:2014boa},  
 {\it BAMPS}~\cite{Uphoff:2011ad,Fochler:2011en,Uphoff:2012gb}, {\it PHSD}~\cite{Song:2015ykw} (b and d).
Some of the model calculations are shown by two lines to represent their uncertainties.}
 \label{fig:DmesRaaModels} 
\end{figure}

Figure~\ref{fig:DmesRaaModels} shows the comparison of the average D-meson $\RAA$ for the two centrality classes 0--10\% (a and b) and 30--50\% (c and d) with most of the available model calculations.
The model calculations are described and compared in a recent review~\cite{Andronic:2015wma}.
A concise summary is given in the following paragraphs.

The interaction of heavy quarks with the medium constituents is computed considering radiative and collisional processes
in the calculations indicated as {\it
  Djordjevic}~\cite{Djordjevic:2014tka}, {\it
  WHDG}~\cite{Wicks:2005gt,Horowitz:2011gd,Horowitz:2011wm}, {\it
  CUJET3.0}~\cite{Xu:2014tda, Xu:2015bbz}, {\it
  MC@sHQ+EPOS}~\cite{Nahrgang:2013xaa}, {\it
  BAMPS}~\cite{Uphoff:2011ad,Fochler:2011en,Uphoff:2012gb}, and {\it
  Cao, Qin , Bass}~\cite{Cao:2013ita}. Only collisional interactions
are considered in the model calculations {\it
  POWLANG}~\cite{Alberico:2011zy,Beraudo:2014boa}, {\it TAMU
  elastic}~\cite{He:2014cla} and PHSD~\cite{Song:2015ykw}. In {\it BAMPS}, two different options are considered: including only collisional energy loss but introducing a scaling factor to match RHIC high-$\pt$ data (where radiative energy loss is expected to be dominant) or including both collisional and radiative energy loss.
Also for the {\it Vitev} model~\cite{Sharma:2009hn} two different options are considered: including only radiative energy loss ({\it Vitev} rad) or also considering the in-medium dissociation of heavy-flavour hadrons ({\it Vitev} rad+dissoc). 

The medium is described  using an underlying hydrodynamical model in
{\it CUJET3.0}, {\it Cao, Qin, Bass}, {\it MC@sHQ+EPOS}, {\it BAMPS},
{\it POWLANG}, {\it TAMU elastic} and {\it PHSD},
while {\it Djordjevic}, {\it WHDG} and {\it Vitev} use a Glauber model nuclear overlap without radial expansion. 

The initial heavy-quark $\pt$ distributions are based on next-to-leading order (NLO) or FONLL perturbative QCD calculations in all model calculations, except for {\it Cao, Qin, Bass}, 
which uses the PYTHIA event generator~\cite{Sjostrand:2006za}. The EPS09 NLO parameterisation~\cite{Eskola:2009uj} of the nuclear parton distribution functions is included by 
{\it POWLANG}, {\it MC@sHQ+EPOS}, {\it TAMU elastic}, {\it PHSD}
and {\it Cao, Qin, Bass}.

All model calculations use in-vacuum fragmentation of heavy quarks for the high-momentum region. At low momentum this is supplemented by 
hadronisation via recombination in the {\it MC@sHQ+EPOS}, {\it POWLANG} \footnote{Note that 
recombination was not included in the version of the {\it POWLANG} model used for the comparison with the D-meson $v_2$ measurement
in~\cite{Abelev:2014ipa}.},  {\it Cao, Qin, Bass}, {\it TAMU elastic}
and {\it PHSD} models. The two last models also include scattering of
D mesons in the hadronic phase of the medium. Also for the  {\it Cao,
  Qin, Bass} model, the hadronic-rescattering effects have been studied in a
recent publication~\cite{Cao:2015hia} and no large differences in the $\Raa$ are observed, when these processes are considered.

Several model calculations provide a good description of the measured $\RAA$ for both centrality classes.
The {\it MC@sHQ+EPOS} model has recently improved the description of the $\RAA$ in the $\pt$ interval 2--8~$\gev/c$ including the EPS09 shadowing parameterisation in addition to in-medium energy loss, the {\it TAMU elastic} model overestimates the $\RAA$ in central collisions in the $\pt$ interval 6--30~$\gev/c$ and the {\it POWLANG} model underestimates 
it in the interval 5--36 (8--16)~$\gev/c$ in the 0--10\% (30--50\%) centrality class. Interestingly, these model calculations provide a fair description of the 
D-meson $v_2$ measured at LHC~\cite{Abelev:2014ipa} and of the D-meson $\RAA$ measured at RHIC~\cite{Adamczyk:2014uip}. 
On the other hand, the model calculations that do not include a hydrodynamical medium expansion and hadronisation via recombination, namely 
{\it Djordjevic}, {\it Vitev},  {\it WHDG}  ---and as a consequence do not describe
the features observed for the $v_2$ at the LHC and the $\RAA$ at RHIC in the momentum region up to about 3--5~$\gev/c$--- provide a good description of the $\RAA$ in the full ``high $\pt$ interval", above $5~\gev/c$. The {\it Vitev} model shows a better agreement when including the D-meson in-medium dissociation mechanism. 
The BAMPS model with collisional energy loss describes the data better for the low-$\pt$ interval, as is the case for the D-meson $v_2$~\cite{Abelev:2014ipa}. The inclusion of radiative energy loss improves the agreement at high $\pt$. 
The {\it Cao, Qin, Bass} model describes the $\Raa$ in both centrality classes, but underestimates the D-meson $v_2$~\cite{Abelev:2014ipa}. The {\it PHSD} model describes the $\Raa$ in both centrality classes.

\begin{figure}[!th]
 \begin{center}
\includegraphics[angle=0, width=7.5cm]{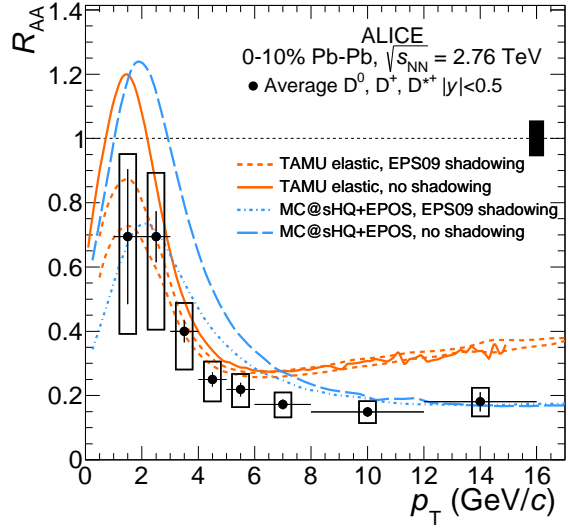}
\end{center}
 \caption{Average of prompt $\rm D^0$, $\rm D^+$ and $\rm D^{*+}$ $\RAA$ in the centrality classes 0--10\% compared with {\it TAMU elastic} and {\it MC@sHQ+EPOS} models calculations with and without including EPS09 shadowing parameterisations~\cite{Eskola:2009uj}.}
 \label{fig:DmesRaaWWOshadow} 
\end{figure}

Figure~\ref{fig:DmesRaaWWOshadow} shows the {\it TAMU elastic} and {\it MC@sHQ+EPOS} calculations of the nuclear modification factor, for the 10\% most central events, with and without including the EPS09 shadowing parameterisation. 
For both models the inclusion of shadowing reduces the $\Raa$ by up to about 30--40\% 
in the interval $\pt<5~\gev/c$, resulting 
in a better description of the data.

\begin{figure}[!th]
 \begin{center}
\includegraphics[width=0.49\textwidth]{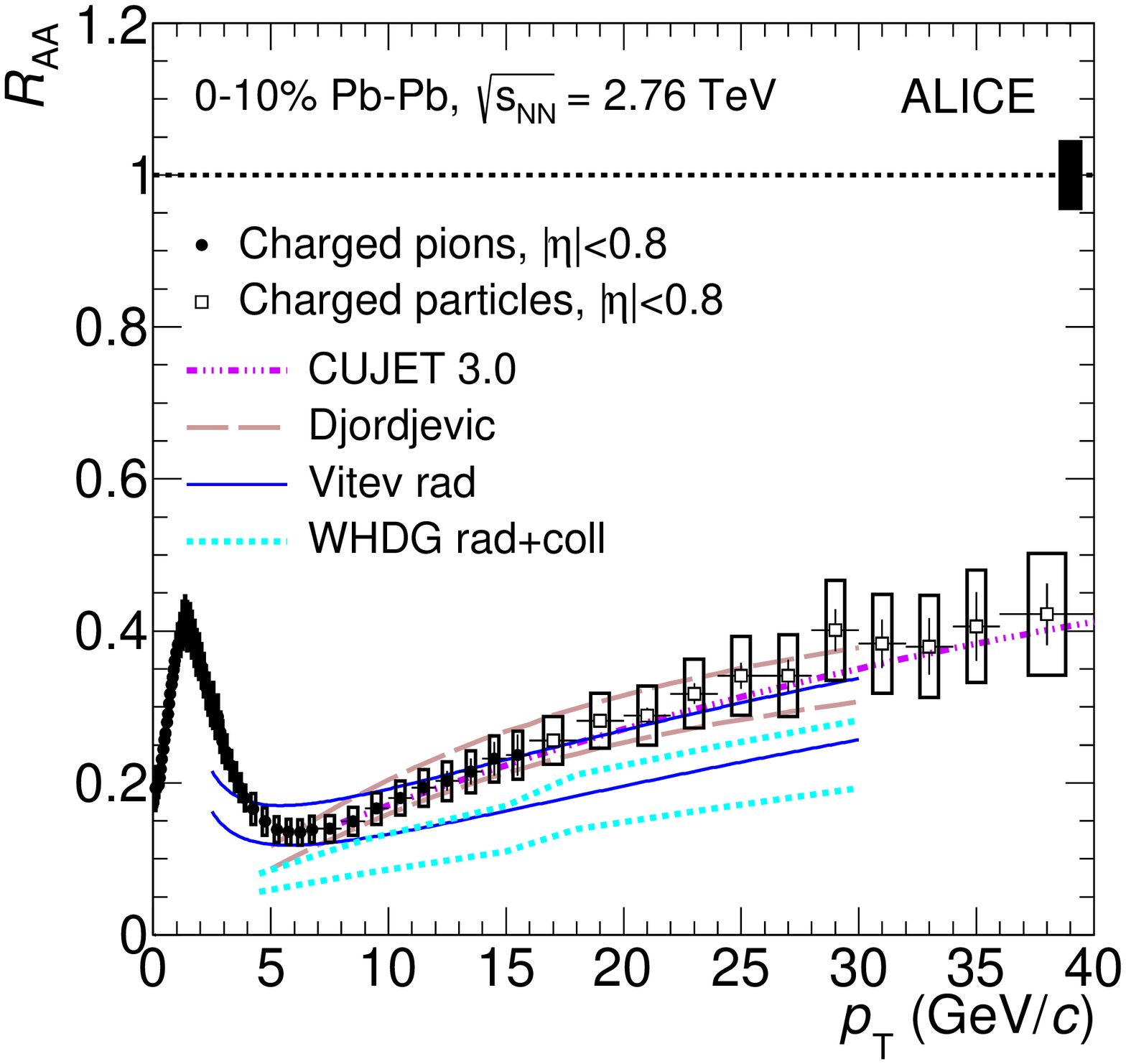}
\includegraphics[width=0.49\textwidth]{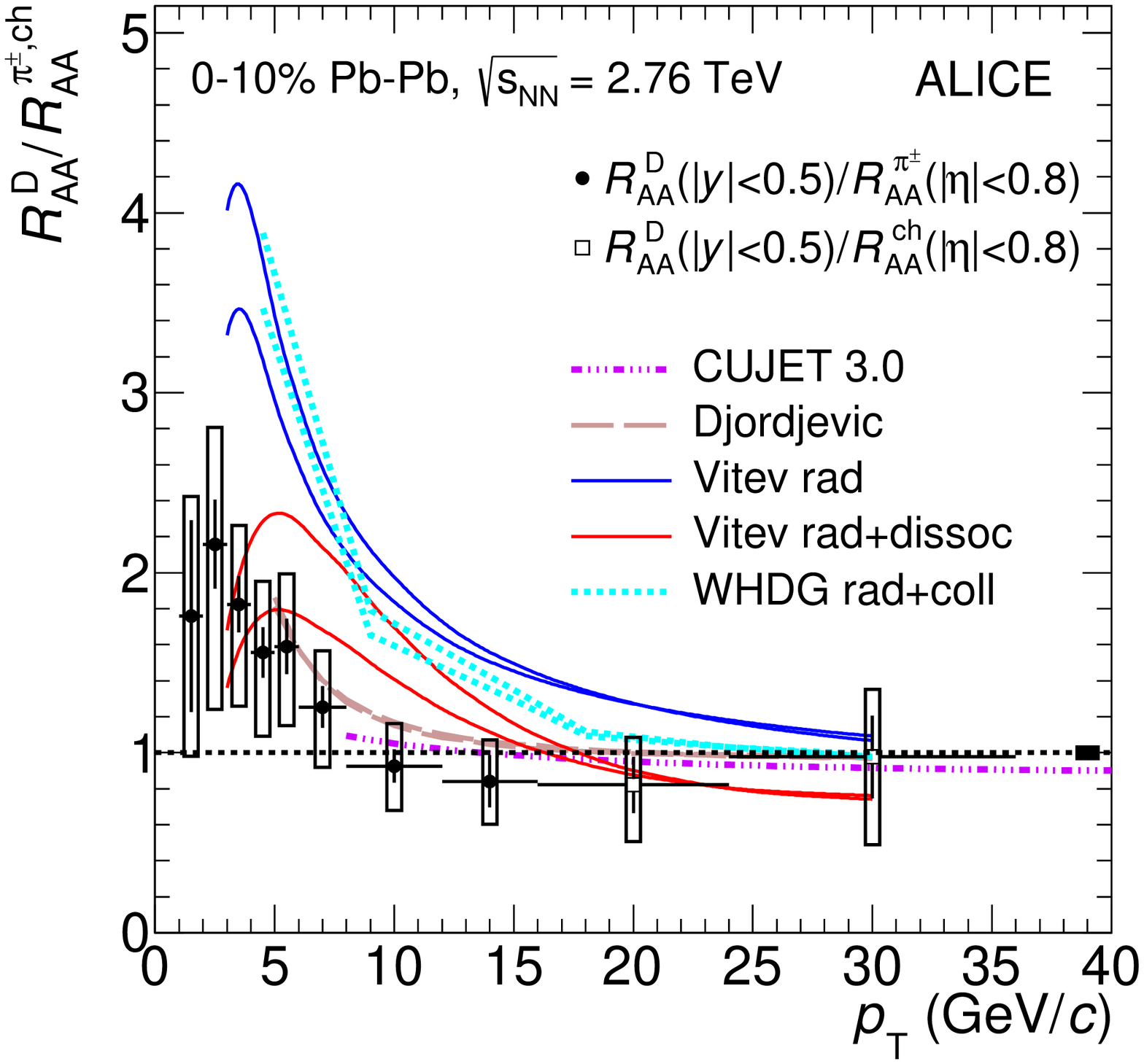}
 \end{center}
 \caption{Left: $\RAA$ of charged pions ($\pt<16~\gev/c$)~\cite{Abelev:2014laa} and of charged particles ($\pt>16~\gev/c$)~\cite{Abelev:2012hxa} compared with model calculations that compute also the D-meson $\RAA$. Right: ratio of the 
 $\RAA$ of prompt D mesons (average of $\rm D^0$, $\rm D^+$ and $\rm D^{*+}$ as shown in Fig.~\ref{fig:DmesRaaModels})
 and the $\RAA$ of charged pions (for $\pt<20~\gev/c$) or charged particles (for $\pt>20~\gev/c$), compared 
 with the same model calculations shown in the left panel.}
 \label{fig:DandpionRAAcmpModels} 
\end{figure}

Four of the model calculations also provide the nuclear modification factor of pions and charged particles ({\it Djordjevic},  {\it CUJET3.0},
{\it  WHDG} and 
 {\it Vitev}). All these calculations include radiative and collisional energy loss\footnote{The in-medium formation and dissociation process, included by {\it Vitev} for D mesons, is not relevant for pions, which have a much larger formation time.}.
 The left panel of Fig.~\ref{fig:DandpionRAAcmpModels} shows the comparison with the measured charged-pion $\RAA^{\pi}$ ($\pt<16~\gev/c$)~\cite{Abelev:2014laa}
 and charged-particle $\RAA^{ch}$ ($\pt>16~\gev/c$)~\cite{Abelev:2012hxa}.  The model calculations provide a reasonable description of the measurements, 
 with {\it WHDG} generally showing smaller $\RAA$ values than seen in data, although consistent within experimental and theoretical uncertainties. 
 
 The right panel of Fig.~\ref{fig:DandpionRAAcmpModels} shows the $\RAA^{\rm D}/\RAA^{\pi}$ ($\pt<16~\gev/c$) and $\RAA^{\rm D}/\RAA^{ch}$ ($\pt>16~\gev/c$) ratios for data and for these four model calculations. In the case of data, the uncertainties of D-meson and charged-pion (or charged-particle) measurements were propagated as uncorrelated uncertainties, except for the uncertainty on $\av{\TAA}$, which cancels in the ratio\footnote{The uncertainty on the normalisation (integrated luminosity) of the pp reference cross sections for D mesons and pions (charged particles) does not cancel in the ratio, because the two cross sections were measured in two data samples at different centre-of-mass energies.}. In the case of model calculations, the theoretical uncertainty, when provided, was propagated assuming full correlation between 
 D mesons and pions (charged particles), since it accounts for a variation of the medium density (or temperature).
 Only the {\it Djordjevic} and {\it CUJET3.0} models, which use radiative and collisional energy loss, can describe the two $\RAA$ results and their ratio over the full $\pt$ interval in which they provide the calculations ($\pt>5$ and 8~GeV/$c$, respectively).
The  {\it Vitev} model can describe the data at the lowest $\pt$ (2--6~GeV/$c$) only if the dissociation mechanism is included, suggesting that the effect is relevant in this model. However, the model overestimates the data in the interval 6--12~GeV/$c$.


\section{Conclusions}
\label{sec:conclusions}
We have presented the measurements of the production of
prompt $\Dzero$, $\Dplus$ and $\Dstar$ mesons at central rapidity 
in Pb--Pb collisions at a centre-of-mass energy per nucleon pair $\sqrtsNN=2.76~\tev$, as well as their nuclear modification factor $\RAA$.
The measurements cover the interval $1<\pt<36~\gev/c$ for the 0--10\% centrality class and $1<\pt<16~\gev/c$ for 
the 30--50\% centrality class. 

The nuclear modification factor shows a maximum reduction of the yields with respect to binary scaling by a factor 5--6, for transverse momenta of about $10~\gevc$ for the 10\% most central $\PbPb$ collisions. 
A suppression of a factor about 2--3 persists 
in the highest $\pt$ interval covered by the measurements (24--36~GeV/$c$). 
At low $\pt$ (1--3~GeV/$c$), the $\Raa$ has large uncertainties, that span the range from 0.35 (factor of three suppression) to 1 (no suppression).
In all $\pt$ intervals above $5~\gev/c$, the $\Raa$ for the 30--50\% centrality class is about twice that for the 0--10\% centrality class.
The suppression observed for $\pt>3~\gev/c$
is interpreted to be due to interactions of the charm quarks within the high-energy density medium formed in the
final-state of Pb--Pb collisions. 
This is demonstrated by the nuclear modification factor measurements in p--Pb collisions at $\sqrtsNN=5.02~\tev$, which indicate that D-meson production is consistent with binary collision scaling~\cite{Abelev:2014hha}. 

The D-meson $\Raa$ was compared with that of charged pions in the interval $1<\pt<16~\gev/c$,
also in terms of the ratio $\Raa^{\rm D}/\Raa^\pi$, and with that of charged particles up to $\pt=36~\gev/c$ ($\Raa^{\rm D}/\Raa^{ch}$).
In the interval $1 < \pt < 6~\gev/c$, the $\RAA$ values of D mesons are higher than those of pions, although consistent within uncertainties. For the 10\% most central collisions, the ratio $\RAA^{\rm D}/\RAA^{\pi,\,ch}$ is larger than unity by about $1\,\sigma$ of the total uncertainties, which are to some extent correlated among $\pt$ intervals. 
For $\pt>8~\gev/c$, the $\RAA$ values are compatible with those of pions and charged particles up to $\pt=36~\gev/c$.

Several models provide a good description of the $\RAA$ for both centrality classes. 
Interestingly, the models that show larger deviation from the data, especially in the high-$\pt$ region, are among those that provide a good description of the D-meson $v_2$ measured at the LHC and of the D-meson $\RAA$ measured at RHIC, in the low-$\pt$ region. 
On the other hand, the models that do not include a hydrodynamical medium expansion and recombination, and as a consequence do not describe $v_2$ in the momentum region up to about 3--5~GeV/$c$, provide a good description of the $\RAA$ at the LHC in the full high-$\pt$ interval, above 5~GeV/$c$.

Only two out of the four models that compute $\RAA^{\rm D}/\RAA^{\pi,\,ch}$ can describe this measurement over the full $\pt$ interval for which they provide the calculations.
In these models, the nuclear modification factors of D mesons and pions turn out to be very similar as a consequence of 
a compensation among the larger energy loss of gluons with respect to that of charm quarks (mainly due to the larger colour coupling factor), the different amount of gluon and light quark yields on the pion $\Raa$ and the harder $\pt$ distribution and fragmentation of charm quarks with respect to those of gluons and of light quarks.


%
%
\newpage
\section*{Acknowledgements}

The ALICE Collaboration would like to thank all its engineers and technicians for their invaluable contributions to the construction of the experiment and the CERN accelerator teams for the outstanding performance of the LHC complex.
The ALICE Collaboration gratefully acknowledges the resources and support provided by all Grid centres and the Worldwide LHC Computing Grid (WLCG) collaboration.
The ALICE Collaboration acknowledges the following funding agencies for their support in building and
running the ALICE detector:
State Committee of Science,  World Federation of Scientists (WFS)
and Swiss Fonds Kidagan, Armenia;
Conselho Nacional de Desenvolvimento Cient\'{\i}fico e Tecnol\'{o}gico (CNPq), Financiadora de Estudos e Projetos (FINEP),
Funda\c{c}\~{a}o de Amparo \`{a} Pesquisa do Estado de S\~{a}o Paulo (FAPESP);
National Natural Science Foundation of China (NSFC), the Chinese Ministry of Education (CMOE)
and the Ministry of Science and Technology of China (MSTC);
Ministry of Education and Youth of the Czech Republic;
Danish Natural Science Research Council, the Carlsberg Foundation and the Danish National Research Foundation;
The European Research Council under the European Community's Seventh Framework Programme;
Helsinki Institute of Physics and the Academy of Finland;
French CNRS-IN2P3, the `Region Pays de Loire', `Region Alsace', `Region Auvergne' and CEA, France;
German Bundesministerium fur Bildung, Wissenschaft, Forschung und Technologie (BMBF) and the Helmholtz Association;
General Secretariat for Research and Technology, Ministry of Development, Greece;
Hungarian Orszagos Tudomanyos Kutatasi Alappgrammok (OTKA) and National Office for Research and Technology (NKTH);
Department of Atomic Energy and Department of Science and Technology of the Government of India;
Istituto Nazionale di Fisica Nucleare (INFN) and Centro Fermi -
Museo Storico della Fisica e Centro Studi e Ricerche ``Enrico Fermi'', Italy;
MEXT Grant-in-Aid for Specially Promoted Research, Ja\-pan;
Joint Institute for Nuclear Research, Dubna;
National Research Foundation of Korea (NRF);
Consejo Nacional de Cienca y Tecnologia (CONACYT), Direccion General de Asuntos del Personal Academico(DGAPA), M\'{e}xico, Amerique Latine Formation academique - European Commission~(ALFA-EC) and the EPLANET Program~(European Particle Physics Latin American Network);
Stichting voor Fundamenteel Onderzoek der Materie (FOM) and the Nederlandse Organisatie voor Wetenschappelijk Onderzoek (NWO), Netherlands;
Research Council of Norway (NFR);
National Science Centre, Poland;
Ministry of National Education/Institute for Atomic Physics and National Council of Scientific Research in Higher Education~(CNCSI-UEFISCDI), Romania;
Ministry of Education and Science of Russian Federation, Russian
Academy of Sciences, Russian Federal Agency of Atomic Energy,
Russian Federal Agency for Science and Innovations and The Russian
Foundation for Basic Research;
Ministry of Education of Slovakia;
Department of Science and Technology, South Africa;
Centro de Investigaciones Energeticas, Medioambientales y Tecnologicas (CIEMAT), E-Infrastructure shared between Europe and Latin America (EELA), Ministerio de Econom\'{i}a y Competitividad (MINECO) of Spain, Xunta de Galicia (Conseller\'{\i}a de Educaci\'{o}n),
Centro de Aplicaciones Tecnológicas y Desarrollo Nuclear (CEA\-DEN), Cubaenerg\'{\i}a, Cuba, and IAEA (International Atomic Energy Agency);
Swedish Research Council (VR) and Knut $\&$ Alice Wallenberg
Foundation (KAW);
Ukraine Ministry of Education and Science;
United Kingdom Science and Technology Facilities Council (STFC);
The United States Department of Energy, the United States National
Science Foundation, the State of Texas, and the State of Ohio;
Ministry of Science, Education and Sports of Croatia and  Unity through Knowledge Fund, Croatia;
Council of Scientific and Industrial Research (CSIR), New Delhi, India;
Pontificia Universidad Cat\'{o}lica del Per\'{u}.

\newpage
\bibliographystyle{utphys}
\bibliography{DmesonRaavspT.bib}

\newpage
\appendix

\section{The ALICE Collaboration}
\label{app:collab} 



\begingroup
\small
\begin{flushleft}
J.~Adam\Irefn{org40}\And
D.~Adamov\'{a}\Irefn{org83}\And
M.M.~Aggarwal\Irefn{org87}\And
G.~Aglieri Rinella\Irefn{org36}\And
M.~Agnello\Irefn{org110}\And
N.~Agrawal\Irefn{org48}\And
Z.~Ahammed\Irefn{org132}\And
S.U.~Ahn\Irefn{org68}\And
S.~Aiola\Irefn{org136}\And
A.~Akindinov\Irefn{org58}
% ...
\renewcommand\labelenumi{\textsuperscript{\theenumi}~}

\section*{Affiliation notes}
\renewcommand\theenumi{\roman{enumi}}
\begin{Authlist}
\item \Adef{0}Deceased
\item \Adef{idp1745888}{Also at: Georgia State University, Atlanta, Georgia, United States}
\item \Adef{idp3782096}{Also at: M.V. Lomonosov Moscow State University, D.V. Skobeltsyn Institute of Nuclear, Physics, Moscow, Russia}
\end{Authlist}

\section*{Collaboration Institutes}
\renewcommand\theenumi{\arabic{enumi}~}
\begin{Authlist}

\item \Idef{org1}A.I. Alikhanyan National Science Laboratory (Yerevan Physics Institute) Foundation, Yerevan, Armenia
\item \Idef{org2}Benem\'{e}rita Universidad Aut\'{o}noma de Puebla, Puebla, Mexico
\item \Idef{org3}Bogolyubov Institute for Theoretical Physics, Kiev, Ukraine
\item \Idef{org4}Bose Institute, Department of Physics and Centre for Astroparticle Physics and Space Science (CAPSS), Kolkata, India
\item \Idef{org5}Budker Institute for Nuclear Physics, Novosibirsk, Russia
\item \Idef{org6}California Polytechnic State University, San Luis Obispo, California, United States
\item \Idef{org7}Central China Normal University, Wuhan, China
\item \Idef{org8}Centre de Calcul de l'IN2P3, Villeurbanne, France
\item \Idef{org9}Centro de Aplicaciones Tecnol\'{o}gicas y Desarrollo Nuclear (CEADEN), Havana, Cuba
\item \Idef{org10}Centro de Investigaciones Energ\'{e}ticas Medioambientales y Tecnol\'{o}gicas (CIEMAT), Madrid, Spain
\item \Idef{org11}Centro de Investigaci\'{o}n y de Estudios Avanzados (CINVESTAV), Mexico City and M\'{e}rida, Mexico
\item \Idef{org12}Centro Fermi - Museo Storico della Fisica e Centro Studi e Ricerche ``Enrico Fermi'', Rome, Italy
\item \Idef{org13}Chicago State University, Chicago, Illinois, USA
\item \Idef{org14}China Institute of Atomic Energy, Beijing, China
\item \Idef{org15}Commissariat \`{a} l'Energie Atomique, IRFU, Saclay, France
\item \Idef{org16}COMSATS Institute of Information Technology (CIIT), Islamabad, Pakistan
\item \Idef{org17}Departamento de F\'{\i}sica de Part\'{\i}culas and IGFAE, Universidad de Santiago de Compostela, Santiago de Compostela, Spain
\item \Idef{org18}Department of Physics and Technology, University of Bergen, Bergen, Norway
\item \Idef{org19}Department of Physics, Aligarh Muslim University, Aligarh, India
\item \Idef{org20}Department of Physics, Ohio State University, Columbus, Ohio, United States
\item \Idef{org21}Department of Physics, Sejong University, Seoul, South Korea
\item \Idef{org22}Department of Physics, University of Oslo, Oslo, Norway
\item \Idef{org23}Dipartimento di Elettrotecnica ed Elettronica del Politecnico, Bari, Italy
\item \Idef{org24}Dipartimento di Fisica dell'Universit\`{a} 'La Sapienza' and Sezione INFN Rome, Italy
\item \Idef{org25}Dipartimento di Fisica dell'Universit\`{a} and Sezione INFN, Cagliari, Italy
\item \Idef{org26}Dipartimento di Fisica dell'Universit\`{a} and Sezione INFN, Trieste, Italy
\item \Idef{org27}Dipartimento di Fisica dell'Universit\`{a} and Sezione INFN, Turin, Italy
\item \Idef{org28}Dipartimento di Fisica e Astronomia dell'Universit\`{a} and Sezione INFN, Bologna, Italy
\item \Idef{org29}Dipartimento di Fisica e Astronomia dell'Universit\`{a} and Sezione INFN, Catania, Italy
\item \Idef{org30}Dipartimento di Fisica e Astronomia dell'Universit\`{a} and Sezione INFN, Padova, Italy
\item \Idef{org31}Dipartimento di Fisica `E.R.~Caianiello' dell'Universit\`{a} and Gruppo Collegato INFN, Salerno, Italy
\item \Idef{org32}Dipartimento di Scienze e Innovazione Tecnologica dell'Universit\`{a} del  Piemonte Orientale and Gruppo Collegato INFN, Alessandria, Italy
\item \Idef{org33}Dipartimento Interateneo di Fisica `M.~Merlin' and Sezione INFN, Bari, Italy
\item \Idef{org34}Division of Experimental High Energy Physics, University of Lund, Lund, Sweden
\item \Idef{org35}Eberhard Karls Universit\"{a}t T\"{u}bingen, T\"{u}bingen, Germany
\item \Idef{org36}European Organization for Nuclear Research (CERN), Geneva, Switzerland
\item \Idef{org37}Excellence Cluster Universe, Technische Universit\"{a}t M\"{u}nchen, Munich, Germany
\item \Idef{org38}Faculty of Engineering, Bergen University College, Bergen, Norway
\item \Idef{org39}Faculty of Mathematics, Physics and Informatics, Comenius University, Bratislava, Slovakia
\item \Idef{org40}Faculty of Nuclear Sciences and Physical Engineering, Czech Technical University in Prague, Prague, Czech Republic
\item \Idef{org41}Faculty of Science, P.J.~\v{S}af\'{a}rik University, Ko\v{s}ice, Slovakia
\item \Idef{org42}Faculty of Technology, Buskerud and Vestfold University College, Vestfold, Norway
\item \Idef{org43}Frankfurt Institute for Advanced Studies, Johann Wolfgang Goethe-Universit\"{a}t Frankfurt, Frankfurt, Germany
\item \Idef{org44}Gangneung-Wonju National University, Gangneung, South Korea
\item \Idef{org45}Gauhati University, Department of Physics, Guwahati, India
\item \Idef{org46}Helsinki Institute of Physics (HIP), Helsinki, Finland
\item \Idef{org47}Hiroshima University, Hiroshima, Japan
\item \Idef{org48}Indian Institute of Technology Bombay (IIT), Mumbai, India
\item \Idef{org49}Indian Institute of Technology Indore, Indore (IITI), India
\item \Idef{org50}Inha University, Incheon, South Korea
\item \Idef{org51}Institut de Physique Nucl\'eaire d'Orsay (IPNO), Universit\'e Paris-Sud, CNRS-IN2P3, Orsay, France
\item \Idef{org52}Institut f\"{u}r Informatik, Johann Wolfgang Goethe-Universit\"{a}t Frankfurt, Frankfurt, Germany
\item \Idef{org53}Institut f\"{u}r Kernphysik, Johann Wolfgang Goethe-Universit\"{a}t Frankfurt, Frankfurt, Germany
\item \Idef{org54}Institut f\"{u}r Kernphysik, Westf\"{a}lische Wilhelms-Universit\"{a}t M\"{u}nster, M\"{u}nster, Germany
\item \Idef{org55}Institut Pluridisciplinaire Hubert Curien (IPHC), Universit\'{e} de Strasbourg, CNRS-IN2P3, Strasbourg, France
\item \Idef{org56}Institute for Nuclear Research, Academy of Sciences, Moscow, Russia
\item \Idef{org57}Institute for Subatomic Physics of Utrecht University, Utrecht, Netherlands
\item \Idef{org58}Institute for Theoretical and Experimental Physics, Moscow, Russia
\item \Idef{org59}Institute of Experimental Physics, Slovak Academy of Sciences, Ko\v{s}ice, Slovakia
\item \Idef{org60}Institute of Physics, Academy of Sciences of the Czech Republic, Prague, Czech Republic
\item \Idef{org61}Institute of Physics, Bhubaneswar, India
\item \Idef{org62}Institute of Space Science (ISS), Bucharest, Romania
\item \Idef{org63}Instituto de Ciencias Nucleares, Universidad Nacional Aut\'{o}noma de M\'{e}xico, Mexico City, Mexico
\item \Idef{org64}Instituto de F\'{\i}sica, Universidad Nacional Aut\'{o}noma de M\'{e}xico, Mexico City, Mexico
\item \Idef{org65}iThemba LABS, National Research Foundation, Somerset West, South Africa
\item \Idef{org66}Joint Institute for Nuclear Research (JINR), Dubna, Russia
\item \Idef{org67}Konkuk University, Seoul, South Korea
\item \Idef{org68}Korea Institute of Science and Technology Information, Daejeon, South Korea
\item \Idef{org69}KTO Karatay University, Konya, Turkey
\item \Idef{org70}Laboratoire de Physique Corpusculaire (LPC), Clermont Universit\'{e}, Universit\'{e} Blaise Pascal, CNRS--IN2P3, Clermont-Ferrand, France
\item \Idef{org71}Laboratoire de Physique Subatomique et de Cosmologie, Universit\'{e} Grenoble-Alpes, CNRS-IN2P3, Grenoble, France
\item \Idef{org72}Laboratori Nazionali di Frascati, INFN, Frascati, Italy
\item \Idef{org73}Laboratori Nazionali di Legnaro, INFN, Legnaro, Italy
\item \Idef{org74}Lawrence Berkeley National Laboratory, Berkeley, California, United States
\item \Idef{org75}Moscow Engineering Physics Institute, Moscow, Russia
\item \Idef{org76}Nagasaki Institute of Applied Science, Nagasaki, Japan
\item \Idef{org77}National Centre for Nuclear Studies, Warsaw, Poland
\item \Idef{org78}National Institute for Physics and Nuclear Engineering, Bucharest, Romania
\item \Idef{org79}National Institute of Science Education and Research, Bhubaneswar, India
\item \Idef{org80}Niels Bohr Institute, University of Copenhagen, Copenhagen, Denmark
\item \Idef{org81}Nikhef, Nationaal instituut voor subatomaire fysica, Amsterdam, Netherlands
\item \Idef{org82}Nuclear Physics Group, STFC Daresbury Laboratory, Daresbury, United Kingdom
\item \Idef{org83}Nuclear Physics Institute, Academy of Sciences of the Czech Republic, \v{R}e\v{z} u Prahy, Czech Republic
\item \Idef{org84}Oak Ridge National Laboratory, Oak Ridge, Tennessee, United States
\item \Idef{org85}Petersburg Nuclear Physics Institute, Gatchina, Russia
\item \Idef{org86}Physics Department, Creighton University, Omaha, Nebraska, United States
\item \Idef{org87}Physics Department, Panjab University, Chandigarh, India
\item \Idef{org88}Physics Department, University of Athens, Athens, Greece
\item \Idef{org89}Physics Department, University of Cape Town, Cape Town, South Africa
\item \Idef{org90}Physics Department, University of Jammu, Jammu, India
\item \Idef{org91}Physics Department, University of Rajasthan, Jaipur, India
\item \Idef{org92}Physik Department, Technische Universit\"{a}t M\"{u}nchen, Munich, Germany
\item \Idef{org93}Physikalisches Institut, Ruprecht-Karls-Universit\"{a}t Heidelberg, Heidelberg, Germany
\item \Idef{org94}Purdue University, West Lafayette, Indiana, United States
\item \Idef{org95}Pusan National University, Pusan, South Korea
\item \Idef{org96}Research Division and ExtreMe Matter Institute EMMI, GSI Helmholtzzentrum f\"ur Schwerionenforschung, Darmstadt, Germany
\item \Idef{org97}Rudjer Bo\v{s}kovi\'{c} Institute, Zagreb, Croatia
\item \Idef{org98}Russian Federal Nuclear Center (VNIIEF), Sarov, Russia
\item \Idef{org99}Russian Research Centre Kurchatov Institute, Moscow, Russia
\item \Idef{org100}Saha Institute of Nuclear Physics, Kolkata, India
\item \Idef{org101}School of Physics and Astronomy, University of Birmingham, Birmingham, United Kingdom
\item \Idef{org102}Secci\'{o}n F\'{\i}sica, Departamento de Ciencias, Pontificia Universidad Cat\'{o}lica del Per\'{u}, Lima, Peru
\item \Idef{org103}Sezione INFN, Bari, Italy
\item \Idef{org104}Sezione INFN, Bologna, Italy
\item \Idef{org105}Sezione INFN, Cagliari, Italy
\item \Idef{org106}Sezione INFN, Catania, Italy
\item \Idef{org107}Sezione INFN, Padova, Italy
\item \Idef{org108}Sezione INFN, Rome, Italy
\item \Idef{org109}Sezione INFN, Trieste, Italy
\item \Idef{org110}Sezione INFN, Turin, Italy
\item \Idef{org111}SSC IHEP of NRC Kurchatov institute, Protvino, Russia
\item \Idef{org112}Stefan Meyer Institut f\"{u}r Subatomare Physik (SMI), Vienna, Austria
\item \Idef{org113}SUBATECH, Ecole des Mines de Nantes, Universit\'{e} de Nantes, CNRS-IN2P3, Nantes, France
\item \Idef{org114}Suranaree University of Technology, Nakhon Ratchasima, Thailand
\item \Idef{org115}Technical University of Ko\v{s}ice, Ko\v{s}ice, Slovakia
\item \Idef{org116}Technical University of Split FESB, Split, Croatia
\item \Idef{org117}The Henryk Niewodniczanski Institute of Nuclear Physics, Polish Academy of Sciences, Cracow, Poland
\item \Idef{org118}The University of Texas at Austin, Physics Department, Austin, Texas, USA
\item \Idef{org119}Universidad Aut\'{o}noma de Sinaloa, Culiac\'{a}n, Mexico
\item \Idef{org120}Universidade de S\~{a}o Paulo (USP), S\~{a}o Paulo, Brazil
\item \Idef{org121}Universidade Estadual de Campinas (UNICAMP), Campinas, Brazil
\item \Idef{org122}University of Houston, Houston, Texas, United States
\item \Idef{org123}University of Jyv\"{a}skyl\"{a}, Jyv\"{a}skyl\"{a}, Finland
\item \Idef{org124}University of Liverpool, Liverpool, United Kingdom
\item \Idef{org125}University of Tennessee, Knoxville, Tennessee, United States
\item \Idef{org126}University of the Witwatersrand, Johannesburg, South Africa
\item \Idef{org127}University of Tokyo, Tokyo, Japan
\item \Idef{org128}University of Tsukuba, Tsukuba, Japan
\item \Idef{org129}University of Zagreb, Zagreb, Croatia
\item \Idef{org130}Universit\'{e} de Lyon, Universit\'{e} Lyon 1, CNRS/IN2P3, IPN-Lyon, Villeurbanne, France
\item \Idef{org131}V.~Fock Institute for Physics, St. Petersburg State University, St. Petersburg, Russia
\item \Idef{org132}Variable Energy Cyclotron Centre, Kolkata, India
\item \Idef{org133}Warsaw University of Technology, Warsaw, Poland
\item \Idef{org134}Wayne State University, Detroit, Michigan, United States
\item \Idef{org135}Wigner Research Centre for Physics, Hungarian Academy of Sciences, Budapest, Hungary
\item \Idef{org136}Yale University, New Haven, Connecticut, United States
\item \Idef{org137}Yonsei University, Seoul, South Korea
\item \Idef{org138}Zentrum f\"{u}r Technologietransfer und Telekommunikation (ZTT), Fachhochschule Worms, Worms, Germany
\end{Authlist}
\endgroup

\end{document}